\documentclass[a4paper,11pt]{article}
\pdfoutput=1

\usepackage{jcappub}

\usepackage{amsmath}
\usepackage{amsthm}
\usepackage{amssymb}
\usepackage{amsfonts}
\usepackage{mathtools}

 \usepackage[english]{babel}
\usepackage{lineno}
\usepackage{lscape}
\usepackage{soul}
\usepackage{graphicx}
\usepackage{bm}
\usepackage{epstopdf}
\usepackage{url}
\usepackage{booktabs}
\usepackage{mdframed}
\usepackage[colorinlistoftodos]{todonotes}
\usepackage{booktabs}
\usepackage{graphicx}
\usepackage{booktabs}
\usepackage[super]{nth}
\usepackage[math]{cellspace}

\usepackage{fontawesome}
\usepackage{microtype}
\usepackage{xspace}
\usepackage[capitalise]{cleveref}
\crefname{section}{Sec.}{Secs.} \Crefname{section}{Section}{Sections.}
\usepackage{verbatim}
\usepackage{enumitem}
\usepackage{siunitx}
\DeclareSIUnit\parsec{pc}
\DeclareSIUnit\year{yr}
\DeclareSIUnit\solarmass{\textit{M}_\odot}

\definecolor{burgundy}{RGB}{144,0,32}

\newcommand{\LCDM}{$\Lambda$CDM\xspace}
\newcommand{\fPBH}{\ensuremath{f_\mathrm{PBH}}\xspace}
\newcommand{\diff}{\ensuremath{\mathrm{d}}}

\newcommand{\rmi}{\ensuremath{\mathrm{i}}}
\newcommand{\vect}[1]{\boldsymbol{#1}}
\newcommand{\CAMB}{\texttt{CAMB}}

\newcommand{\IFCA}{%
Instituto de F\'isica de Cantabria (IFCA), UC-CSIC,\\
Avenida de Los Castros s/n, 39005 Santander, Spain}

\newcommand{\IFT}{%
Instituto de Física Teórica UAM-CSIC,
Universidad Autónoma de Madrid,\\
Cantoblanco, 28049 Madrid, Spain}

\newcommand{\IFIC}{%
Instituto de F\'isica Corpuscular, Universidad de Valencia and CSIC,\\
Edificio Institutos de Investigac\'ion, Calle Catedr\'atico Jos\'e Beltr\'an 2, 46980 Paterna, Spain}

\newcommand{\INAF}{%
INAF - Osservatorio Astronomico di Roma,\\
via Frascati 33,
00040 Monteporzio Catone (Roma), Italy}

\newcommand{\IPhT}{%
Université Paris-Saclay, CNRS, CEA,\\
Institut de physique théorique,
91191, Gif-sur-Yvette,
France}

\newcommand{\Torino}{%
Dipartimento di Fisica, Universit\`a di Torino,
via P. Giuria 1, I–10125 Torino, Italy}

\preprint{IFT-UAM/CSIC-22-50}

\begin{document}

\title{Dancing in the dark:\\detecting a population\\of distant primordial black holes}

\author[a, b]{Matteo Martinelli,}
\author[b]{Francesca Scarcella,}
\author[c, b]{Natalie B. Hogg,}
\author[d]{Bradley J. Kavanagh,}
\author[e, f, b]{Daniele Gaggero,}
\author[c, b]{and Pierre Fleury}

\affiliation[a]{\INAF}
\affiliation[b]{\IFT}
\affiliation[c]{\IPhT}
\affiliation[d]{\IFCA}
\affiliation[e]{\IFIC}
\affiliation[f]{\Torino}

\emailAdd{matteo.martinelli@inaf.it}
\emailAdd{francesca.scarcella@uam.es}
\emailAdd{natalie.hogg@ipht.fr}
\emailAdd{kavanagh@ifca.unican.es}
\emailAdd{daniele.gaggero@ific.uv.es}
\emailAdd{pierre.fleury@ipht.fr}

\abstract{Primordial black holes (PBHs) are compact objects proposed to have formed in the early Universe from the collapse of small-scale over-densities. Their existence may be detected from the observation of gravitational waves (GWs) emitted by PBH mergers, if the signals can be distinguished from those produced by the merging of astrophysical black holes. In this work, we forecast the capability of the Einstein Telescope, a proposed third-generation GW observatory, to identify and measure the abundance of a subdominant population of distant PBHs, using the difference in the redshift evolution of the merger rate of the two populations as our discriminant. We carefully model the merger rates and generate realistic mock catalogues of the luminosity distances and errors that would be obtained from GW signals observed by the Einstein Telescope. We use two independent statistical methods to analyse the mock data, finding that, with our more powerful, likelihood-based method, PBH abundances as small as  $f_\mathrm{PBH} \approx 7 \times 10^{-6}$ ($f_\mathrm{PBH} \approx 2\times10^{-6}$) would be distinguishable from $f_\mathrm{PBH} = 0$ at the level of $3\sigma$ with a one year (ten year) observing run of the Einstein Telescope.
Our mock data generation code, \texttt{darksirens}, is fast, easily extendable and publicly available on GitLab \href{https://gitlab.com/matmartinelli/darksirens}{\faGitlab}.
}

\maketitle

\section{Introduction}
\label{sec:introduction}

The idea that a population of primordial black holes (PBHs) may exist in the Universe and constitute a significant portion of the dark matter (DM) has been debated in the literature since the pioneering works of the late 1960s and early 1970s \cite{Zeldovich:1967lct,Hawking:1971ei,Chapline:1975ojl} (see e.g.~\cite{Green:2020jor} for a recent review). The discovery of such a population would have profound implications for fundamental physics, even if it were to represent only a subdominant portion of the DM. The formation of PBHs could reveal precious hints about inflationary and early Universe physics \cite{Dolgov1993,Jedamzik:1996mr,Garcia-Bellido:1996mdl,Khlopov:2008qy,Garcia-Bellido:2017mdw,Ballesteros:2017fsr,Pattison:2017mbe}, and their subsequent evolution could impact structure formation \cite{Inman:2019wvr} and solve long-standing puzzles related to the early formation of supermassive black holes \cite{Volonteri:2021sfo}. Moreover, their detection could help to exclude the presence of thermal weakly interacting massive particles and could strongly constrain models of particle physics which invoke new physics at the Weak scale \cite{Lacki:2010zf,Adamek:2019gns,Bertone:2019vsk,Carr:2020mqm}.

PBHs whose masses lie between \SI{1}{\solarmass} and \SI{100}{\solarmass} are particularly promising candidates for study, given the now routine detection of gravitational wave (GW) signals from the mergers of compact objects in this mass range. 
Gravitational waves from the inspiral, merger and ringdown of a pair of BHs were first detected by the LIGO Scientific Collaboration (henceforth LIGO) and the Virgo Collaboration (henceforth Virgo) in 2015 \cite{GWthefirst} and the detection of a binary neutron star merger with an electromagnetic counterpart signal was made in 2017 by LIGO and Virgo \cite{BNSmerger}. In 2020, the KAGRA observatory in Japan \cite{KAGRA:2018plz} joined the global network of GW detectors, reporting its first observations in conjunction with the GEO600 instrument in March 2022 \cite{LIGOScientific:2022myk}.

The analysis of the combined information about the merger rate, mass and spin distribution of low redshift merger events collected by LIGO, Virgo and KAGRA (LVK) since 2015 have allowed upper limits on the PBH abundance $f_\mathrm{PBH} = \Omega_\mathrm{PBH}/\Omega_\mathrm{DM}$ in the aforementioned mass window to be set \cite{Ali-Haimoud:2017rtz,Kavanagh:2018ggo}. This has also triggered a debate about the possibility of identifying a subdominant population of PBHs on top of the (likely dominant) contribution of astrophysical black holes (ABHs) \cite{Franciolini:2021tla,Franciolini:2021xbq,Hutsi:2020sol}, motivated in particular by the detected substructure in the mass distribution of these objects \cite{LIGOScientific:2020kqk,LIGOScientific:2021psn}. 
Definitively detecting such a population of PBHs will become a concrete possibility with the advent of the third generation of GW observatories, such as the Einstein Telescope~(ET)~\cite{Maggiore:2019uih} and Cosmic Explorer (CE)~\cite{Reitze:2019iox}.\footnote{Other proposed third-generation observatories include LISA \cite{Amaro-Seoane2017}, DECIGO \cite{Kawamura:2020pcg}, TianQin and Taiji \cite{Gong:2021gvw}, which will all be space-based rather than terrestrial like the ET and CE, allowing for far longer interferometer arms than is possible on Earth. These instruments will hence be sensitive to a very different GW frequency range to that of ET and CE, which means that the latter will have far better prospects for detecting PBHs in the mass range 1--100  $M_\odot$.} Assessing the potential constraining power of the ET on PBHs is the focus of this work.

The most up-to-date configuration of the planned ET facility is known as ET-D~\cite{Hild:2011np}, which proposes a three-armed observatory consisting of three interferometers arranged in an equilateral triangle. In the ET-D configuration, each detector is in fact made up of a \textit{pair} of detectors -- one sensitive to a lower frequency range and the other to a higher frequency range -- thus greatly increasing the overall sensitivity of the instrument with respect to the current generation of GW observatories. 
Furthermore, the triangular shape of the observatory will enable improved sky localisation of GW events~\cite{Vitale:2016icu}. Lastly, the interferometers and detectors will all be constructed underground, in an effort to reduce seismic noise~\cite{2011GReGr..43..623B}. 

\begin{figure}
    \centering
    \includegraphics[width=0.8\textwidth]{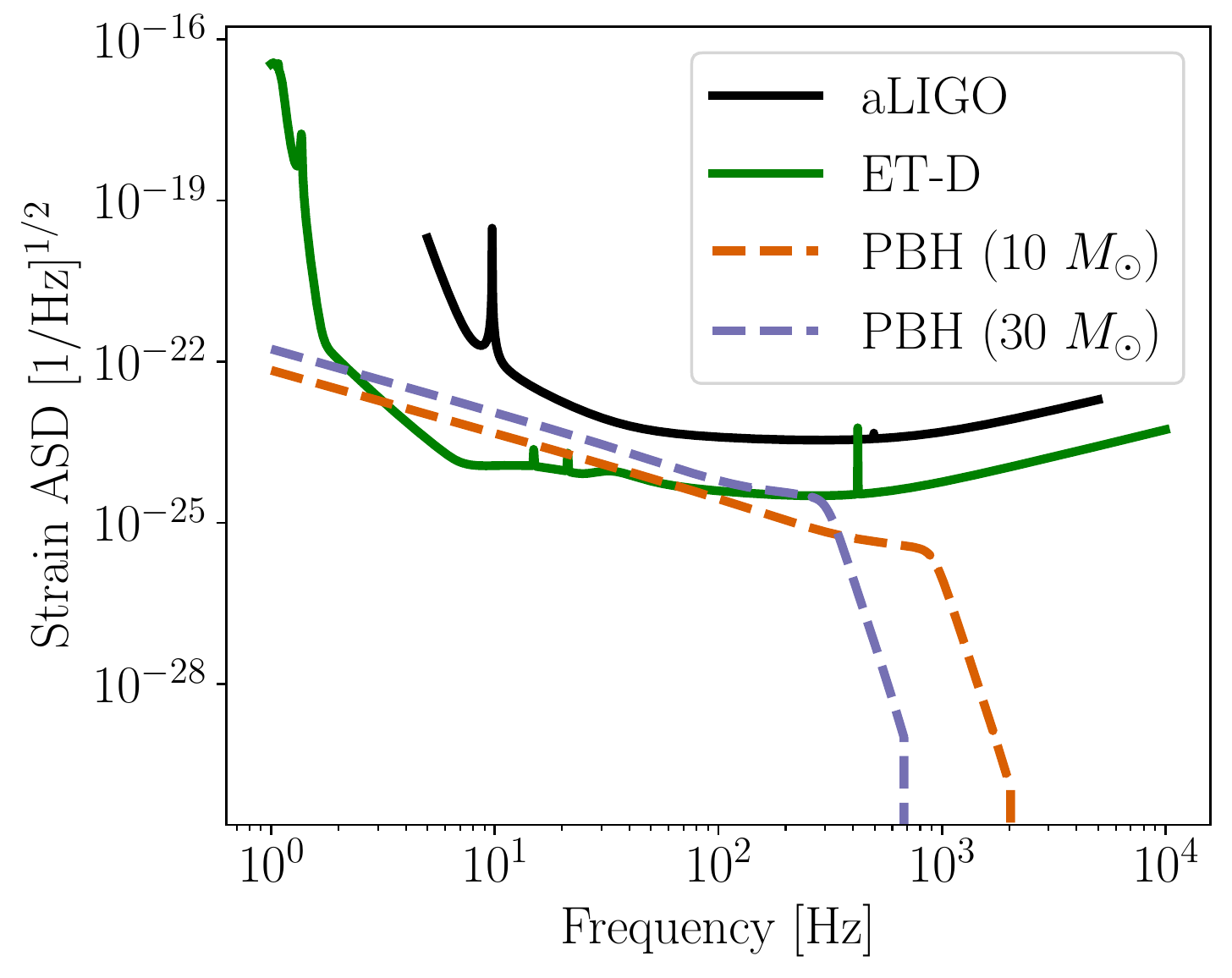}
    \caption{The strain amplitude spectral densities (ASD) of advanced LIGO (black) and the ET-D configuration of the Einstein Telescope (green), along with the characteristic strain of two PBH merger events, with component masses of \SI{10}{\solarmass} (orange) and \SI{30}{\solarmass} (purple) taking place at $z=1$, that is, with a luminosity distance $D\approx\SI{6.8}{\giga\parsec}$.} 
    \label{fig:ET_sensitivity}
\end{figure}

The less noise in the detector, the smaller the amplitude of GWs --~or \textit{strain}~-- that can be detected.  This noise is typically quantified by the strain amplitude spectral density~(ASD), which we show for advanced LIGO\footnote{\url{https://dcc.ligo.org/LIGO-T1800042/public}.} (aLIGO) and ET-D\footnote{\url{http://www.et-gw.eu/index.php/etsensitivities}.} in \cref{fig:ET_sensitivity}. These ASD curves effectively show the lowest GW strain that can be detected and we show for comparison the characteristic strains of two PBH merger events with the example masses we use in the rest of this work: \SI{10}{\solarmass} and \SI{30}{\solarmass}. From this plot, we can see that ET-D will be sensitive to strains around two orders of magnitude smaller than what aLIGO can currently detect, and will also probe a much broader range of frequencies. 

With all these factors taken together, the ET-D design is expected to yield many more observations of GWs at ever-greater cosmic distances (or redshifts) than the current generation of terrestrial detectors. 

A crucial difference between ABH mergers and PBH mergers is the \emph{redshift evolution} of their merger rates. While there is broad consensus on a steeply decreasing rate for ABHs beyond $z \simeq 2$ (see e.g. \cite{Dvorkin:2016wac} and references therein), the PBH merger rate is expected instead to be a \emph{monotonically increasing} function of redshift. In fact, a significant number of PBH binaries are expected to form by gravitational decoupling from the Hubble flow before matter--radiation equality. As shown in \cite{Nakamura:1997sm,Sasaki:2016jop,Ali-Haimoud:2017rtz}, the distribution of the orbital parameters for these primordial binaries peaks at low values of both semi-major axis and angular momentum, and therefore low values of the merger time (compared to the Hubble time scale). Hence, even a small population of PBHs is expected to dominate the merger rate at early times.

This idea naturally suggests a focus on high redshift GWs for the purposes of detecting PBHs~\cite{Koushiappas:2017kqm}. They may manifest as resolved individual events~\cite{Chen:2019irf,DeLuca:2021wjr,DeLuca:2021hde,Ng:2021sqn}, but also as a measurable contribution to the stochastic GW background, as recently pointed out in \cite{Atal:2022zux,Braglia:2022icu}. In this article we focus on the former case: the prospects for detecting an anomalously large number of \emph{individual resolved events} at large cosmic distances. Specifically, we propose (i) a careful theoretical modelling of both the PBH and ABH merger rates; and (ii) a realistic assessment of the ability of future observatories -- specifically, the ET -- to disentangle the two populations.

As far as modelling is concerned, a useful starting point is the formalism for the PBH merger rate developed in \cite{Nakamura:1997sm,Ioka:1998nz,Ali-Haimoud:2017rtz}, based on the assumption of an initially uniform and isotropic distribution of PBHs all having the same mass (in other words, a ``monochromatic'' mass function). In this scenario, neighbouring PBHs may decouple from the Hubble flow before matter--radiation equality, forming binaries which merge with an ever decreasing rate throughout the age of the Universe. We extend this formalism to include the effects of the early-time formation of PBH clusters~\cite{Chisholm:2005vm,Chisholm:2011kn,Raidal:2018bbj} on the evolution of the binaries. The ``background'' to these PBH merger events is provided by the mergers of ABHs. The ABH merger rate traces the binary ABH birth rate, with some delay, which in turn is expected to trace the star formation rate, again with some delay~\cite{Dvorkin:2016wac}. A careful description of PBH and ABH merger rates as a function of redshift is therefore crucial in understanding how well future GW observatories can distinguish these two populations. 
 
In this work, we make the conservative imposition of monochromatic mass functions for both the ABH and PBH populations -- in other words, all the BHs in a given population have the same mass -- and focus on the redshift dependence of the merger rates as the only discriminant between the two. We also focus on PBHs in the \SI{1}{\solarmass} -- \SI{100}{\solarmass} mass range, but we note that several other investigations of the capability of future observatories to detect PBHs selected the sub-solar mass range as an interesting avenue for study \cite{2019ApJ...885...77B,Mukherjee:2021itf,Pujolas:2021yaw}.

The uncertainty in the measured luminosity distances of events at high redshift is also expected to play an important role in separating the two populations of BHs. The authors of~\cite{Ng:2021sqn} simulate the response of different detector networks located in Europe and the USA, and assess whether a single-event-based PBH identification can be unambiguous, if the redshift is large enough. The authors conclude that the typical redshift measurement is not precise enough to conclude with certainty that a single source is of primordial origin.

Motivated by these results, we present a framework that implements a statistically sound assessment of the capability of ET to:
\begin{itemize}
\item {\it Detect} an excess of merger events at high redshift with respect to the astrophysical expectation;
\item {\it Measure} the $f_{\rm PBH}$ associated with the detection, if present, or constrain $f_{\rm PBH}$ to lie within some range.
\end{itemize}

To this aim, we generate a mock data set associated to the null hypothesis of $f_{\rm PBH} = 0$, and several data sets associated to different PBH fractions. We analyse these data using two different methods. 
First, we use an intuitive two-bin approach to determine the minimum abundance of PBHs that can be detected as a significant ``excess'' with respect to the astrophysical background, identifying the optimal binning in redshift.
Next, we develop a parameter estimation pipeline, in which we estimate the posterior distribution of $f_{\rm PBH}$, by comparing the mock data generated for different fiducial values of $f_{\rm PBH}$ with the theoretical distribution of event distances.

The article is structured as follows: in \cref{sec:merger_rates} we discuss our theoretical and phenomenological models for the redshift evolution of the merger rate of primordial and astrophysical BHs; in \cref{sec:generating_mock} we describe how we generate our mock GW catalogues; in \cref{sec:cut_and_count} we present the results from the ``cut-and-count'' method, which assesses ET's ability to detect PBHs; in \cref{sec:likelihood_based_method} we present the results from the likelihood-based method, which further analyses ET's ability to measure the PBH fraction; in \cref{sec:discussion} we discuss the implications and possible shortcomings of these results and finally in \cref{sec:conclusion} we conclude. \Cref{sec:appendix_clustering,sec:appendix_mocks,sec:appendix_lensing} provide more information on the merger rate suppression due to PBH clustering, the method for computing the signal-to-noise ratio of an event in our mock catalogues, and the effect of lensing on GW measurements. The code associated with the article, \texttt{darksirens},\footnote{\url{https://darksirens.readthedocs.io}} is publicly available on GitLab \href{https://gitlab.com/matmartinelli/darksirens}{\faGitlab}.

\paragraph{Conventions, notation, cosmology.} We assume a spatially flat \LCDM cosmology throughout, using $H_0 = \SI{67.4}{\kilo\meter\per\second\per\mega\parsec}$ and $\Omega_{\rm m} = 0.315$ as reported by the \emph{Planck} collaboration~\cite{Aghanim:2018eyx}. A three-bar equality sign ($\equiv$) indicates a definition; an upper-case $P$ indicates a probability, while a lower-case $p$ indicates a probability density function (PDF); bold symbols ($\vect{x}, \vect{\theta}, \vect{\mathcal{A}}$, \ldots) stand for vectors or matrices. The luminosity distance is simply denoted with a $D$ throughout, because we use no other notion of distance. We endeavour with all heroism to keep explicit factors of $G$ and $c$ where relevant.

\section{Modelling the black hole merger rate}
\label{sec:merger_rates}

This section summarises our assumptions regarding the modelling of the merger rate of both ABHs and PBHs, which will then be used to produce mock catalogues of GWs.

\subsection{Preliminary definitions}
\label{subsec:preliminary_definitions}
Correctly interpreting a population of observed merger events relies on our expectations for the distribution of mergers. We therefore need to accurately model this quantity. Physical models for the formation and mergers of BHs typically yield the merger rate density~$\mathcal{R}(z)$; that is, the number of mergers per unit comoving volume and per unit time in the rest frame of the source, as a function of redshift~$z$. However, from an observational point of view,  the relevant quantity is rather the number of mergers that would be observed by an ideal detector per unit redshift and per unit time in the rest frame of the detector. This is what we shall refer to as the \emph{merger rate}~\cite{LIGOScientific:2016ebi,LIGOScientific:2017zid},
\begin{equation}
R(z) = \frac{\mathcal{R}(z)}{1+z}\,\frac{\diff V_{\rm c}}{\diff z} \, .
\label{eq:MergerRate}
\end{equation}
The factor of $(1+z)^{-1}$ serves to convert a source-frame rate to a detector-frame rate, and $V_{\rm c}$ denotes the comoving volume; $\diff V_{\rm c}/\diff z$ then represents the comoving volume of a spherical shell between $z$ and $z+\diff z$ around the detector,
\begin{equation}
\frac{\diff V_{\rm c}}{\diff z} = 4\pi r^2(z)\,\frac{c}{H(z)}\, ,
\qquad \text{with} \quad
r(z) \equiv \int_0^z \diff\zeta \; \frac{c}{H(\zeta)}\,,
\end{equation}
the comoving distance at redshift $z$. Both $r(z)$ and $H(z)$ are computed using the public cosmology solver \CAMB~\cite{Lewis:1999bs,2012JCAP...04..027H}.\footnote{\href{https://camb.info}{\tt https://camb.info.}} Given a merger rate~$R(z)$, the total theoretical number of events that would be detected by an infinitely sensitive instrument during an observation time~$T_{\rm obs}$ reads
\begin{equation}
\label{eq:N_bar_tot}
\bar{N}_{\rm tot} = T_{\rm obs}
\int_{z_\mathrm{min}}^{z_\mathrm{max}} \diff z \; R(z) \ ,
\end{equation}
where we limit ourselves to mergers in a redshift range $z \in [z_\mathrm{min}, z_\mathrm{max}]$. Their redshift distribution, still in the case of an ideal detector, is given by
\begin{equation}
\label{eq:prob_redshift}
p(z)
=
\frac{T_{\rm obs}R(z)}{\bar{N}_{\rm tot}}
=
\frac{R(z)}{\int_{z_\mathrm{min}}^{z_\mathrm{max}}
\diff\zeta \; R(\zeta)} \, .
\end{equation}

\begin{figure}
    \centering
     \includegraphics[width=0.6\textwidth]{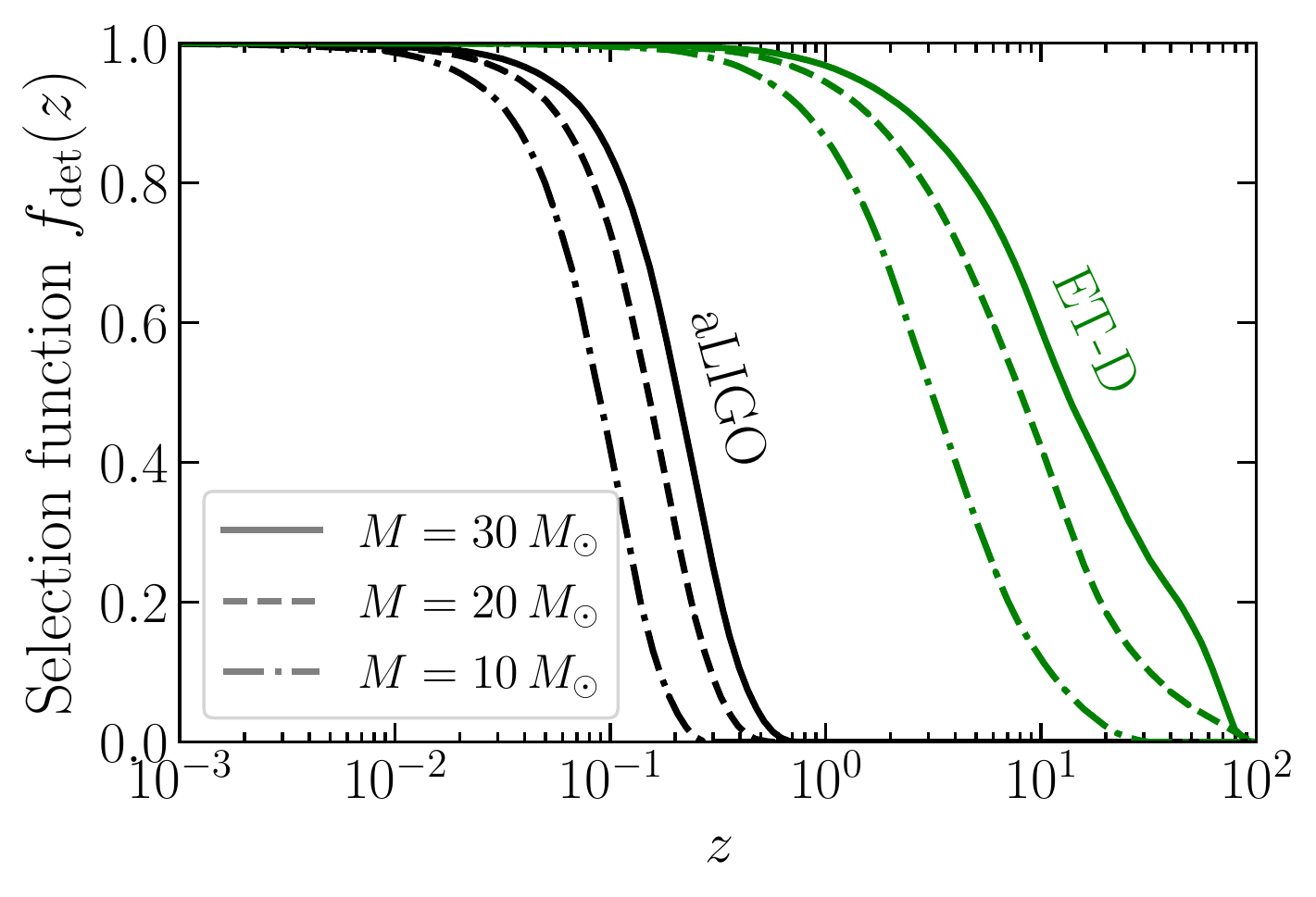}
    \caption{The selection functions $f_{\rm det}(z)$ for aLIGO (black) and ET-D (green) for three different total masses $M$: \SI{10}{\solarmass} (dot-dashed lines), \SI{20}{\solarmass} (dashed lines) and \SI{30}{\solarmass} (solid lines). The selection function is defined as the fraction of mergers which are detectable at a given redshift.}
    \label{fig:ET_selection_functions}
\end{figure}

Importantly, not all of these mergers will have a signal-to-noise ratio (SNR) large enough to be detectable. We therefore define the selection function $f_\mathrm{det}(z)$, which encodes the fraction of mergers that are detectable at a given redshift. We compute this explicitly for aLIGO and ET-D by sampling over binary orientations and sky positions, following the approach of Ref.~\cite{gw-horizon-plot}, and requiring a threshold of $\mathrm{SNR}_\mathrm{min} = 8$ for an event to qualify as detected. We illustrate these selection functions in Fig.~\ref{fig:ET_selection_functions} for a range of BH masses. With this, the rate of \textit{detectable} events is  given by
\begin{equation}
\label{eq:detectable_rate}
R_\mathrm{det}(z) = f_{\rm det}(z) \, R(z) \, .
\end{equation}
The expected number of detected events, $\bar{N}_{\rm det}$, and their redshift distribution, $p_{\rm det}(z)$, are defined analogously to \cref{eq:N_bar_tot,eq:prob_redshift}, respectively, by substituting $R(z)$ with $R_{\rm det}(z)$.

\begin{figure}[tb]
\centering
\includegraphics[width=0.49\textwidth]{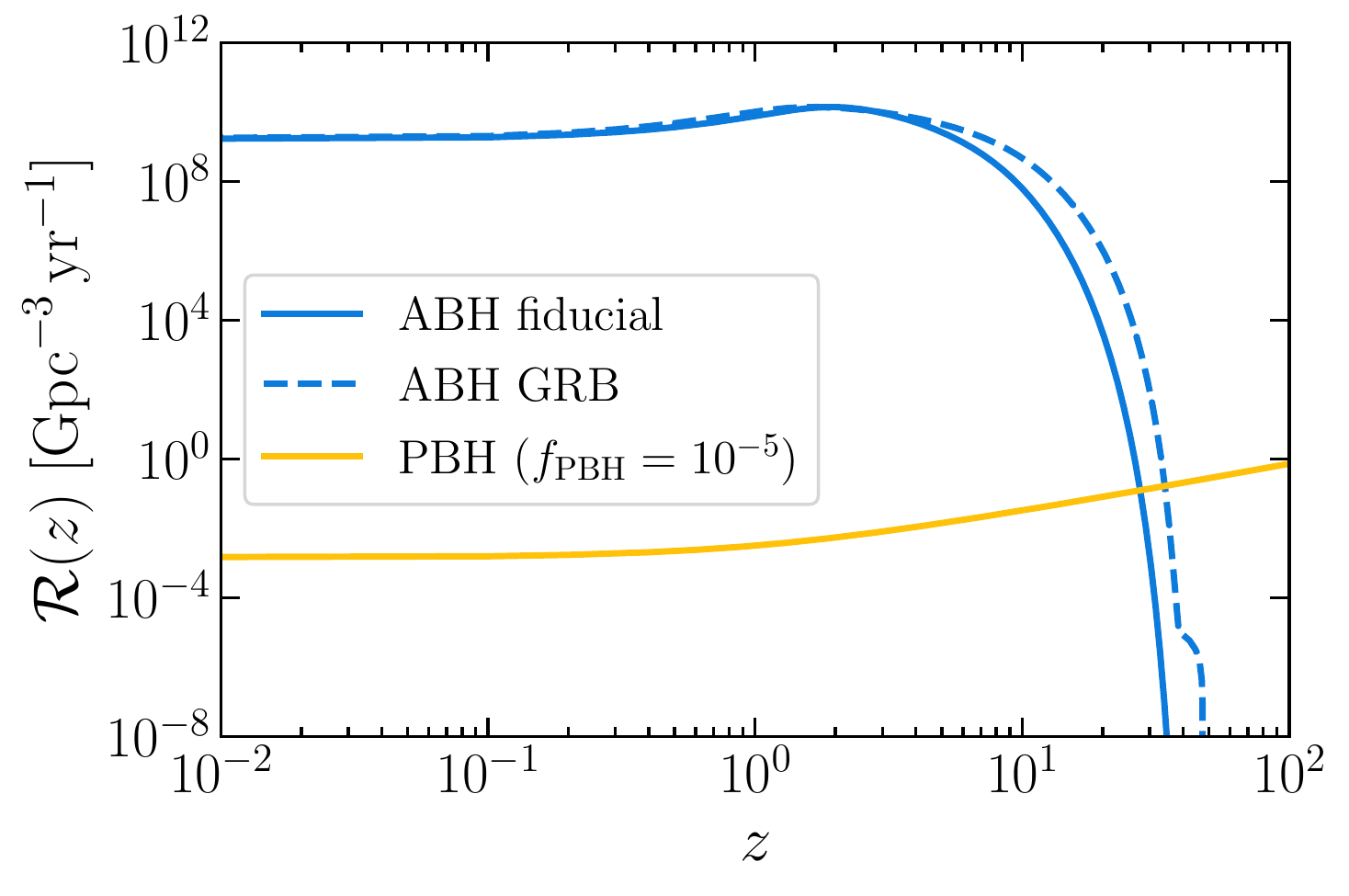}
\hfill
\includegraphics[width=0.48\textwidth]{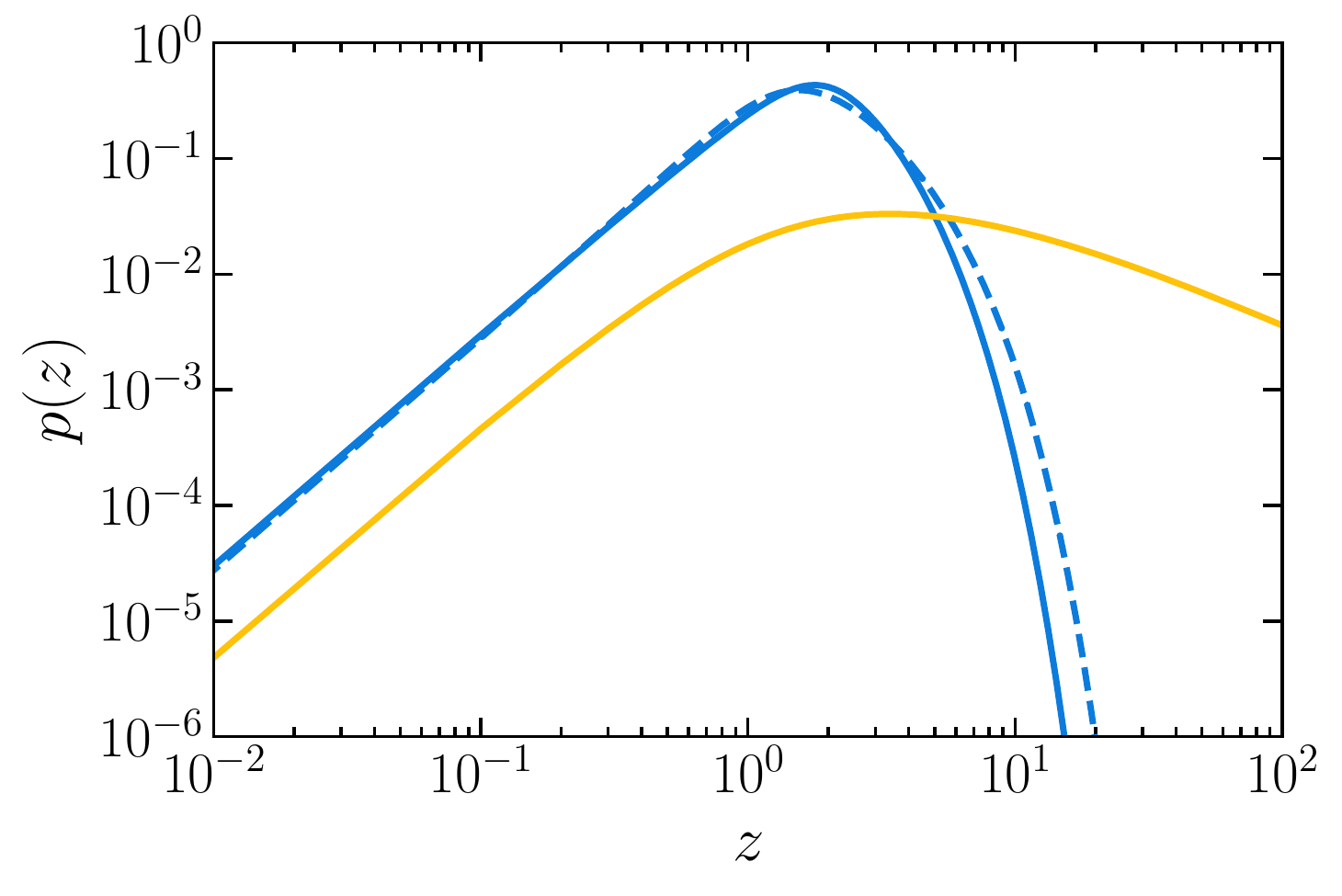}
\caption{Distribution of merger events as a function of redshift for ABHs and PBHs. We show results for our fiducial ABH model (solid blue), the more optimistic ABH model (dashed blue) based on GRB data (\cref{sec:merger_rate_ABH}), and our PBH model (\cref{sec:merger_rate_PBH}) with $f_\mathrm{PBH} = 10^{-5}$ (yellow).
\textbf{Left:} Comoving merger rate density per unit volume per unit time $\mathcal{R}(z)$. \textbf{Right:} Redshift probability distribution $p(z)$, defined in \cref{eq:prob_redshift}. This corresponds to the distribution of observed event redshifts for an ideal detector, ignoring the effects of a selection function $f_\mathrm{det}(z)$.}
\label{fig:MergerRate}
\end{figure}

In the following subsections, we outline our computation of the merger rate density $\mathcal{R}(z)$ for ABH and PBH mergers. In \cref{fig:MergerRate}, we show examples of these merger rate densities (left panel), along with the corresponding redshift distribution $p(z)$ (right panel). These illustrate why high redshift mergers are so crucial to discovering and constraining a PBH population; above $z \sim 30 - 40$ the ABH rate quickly becomes negligible and mergers from a potential PBH population come to rapidly dominate.

\subsection{Primordial black holes}
\label{sec:merger_rate_PBH}

PBH binaries are efficiently formed deep in the radiation era, in what is known as the early-time formation channel. Binaries can also be formed in dense environments through gravitational capture or three-body interactions~\cite{Bird:2016dcv,Franciolini:2022ewd}. These late-time formation channels are expected to become increasingly important for large PBH abundances and low redshifts. In this work, we focus on high redshift observations and our analysis is restricted to $\fPBH \lesssim 10^{-3}$ (in agreement with current constraints, see \cref{subsec:bound}). In this case, we can consider the merger rate to be dominated by the contribution of binaries formed in the early Universe. We compute it following the derivation of \cite{Ali-Haimoud:2017rtz}, which builds on several earlier works~\cite{Nakamura:1997sm,Ioka:1998nz,Sasaki:2016jop}. 

The PBHs are assumed to be initially Poisson distributed (we neglect any possible initial clustering, see e.g.~\cite{Bringmann:2018mxj,Ballesteros:2018swv,Young:2019gfc}) with negligible initial velocity dispersion. Under mutual gravitational attraction, PBH pairs decouple from the Hubble flow and form binary systems. Head-on collisions are avoided thanks to the small angular momentum provided by the presence of surrounding PBHs and smooth density perturbations. After formation, the binaries slowly lose energy through GW emission and eventually merge. The coalescence time (for small values of $j$) is given by~\cite{PhysRev.136.B1224}
\begin{equation}
\label{eq:tmerger}
t_{\mathrm{merger}}
= \frac{3}{170}\frac{c^5}{G^3\,M_\mathrm{PBH}^3} \,
    a^4 j^7 \, ,
\end{equation}
where $M_\mathrm{PBH}$ is the PBH mass, $a$ is the binary's initial semi-major axis and $j$ is the initial dimensionless angular momentum, related to the eccentricity $e$ through $j \equiv \sqrt{1-e^2}$. As the binaries which merge within a Hubble time typically form early in the radiation era~\cite{Ali-Haimoud:2017rtz}, we identify the coalescence time with the cosmic time at which the merger occurs.

Given \cref{eq:tmerger}, we can estimate the merger rate density from the initial distribution of the orbital parameters $(a, j)$ as
\begin{equation}
\label{eq:PBHrate}
\mathcal{R}_\mathrm{PBH}[z(t)]\, = \frac{1}{2} n_\mathrm{PBH}
\int \mathrm{d} a \,  \mathrm{d} j  \;
p(j | a)p(a)  \, \delta \left[  t - t_{\mathrm{merger}}(a, j) \right] ,
\end{equation}
where $n_\mathrm{PBH} = f_\mathrm{PBH} \,\rho_{\mathrm{DM}, 0}/M_\mathrm{PBH}$ is the comoving number density of PBHs, and $\rho_{\mathrm{DM}, 0}$ is the energy density of cold dark matter at redshift zero.
We obtain distributions for the semi-major axis $p(a)$ and for the angular momentum $p(j|a)$ following \cite{Ali-Haimoud:2017rtz,Kavanagh:2018ggo}. The latter distribution is computed taking into account the torques generated by all the surrounding PBHs and from the density perturbations in the matter fluid. 
Most binaries are born with very high eccentricities and small semi-major axes, which correspond to short times to merger, see \cref{eq:tmerger}. It follows that the PBH merger rate increases with redshift, with a larger number of binaries merging at early times, as shown in the left panel of \cref{fig:MergerRate}. 

In estimating the initial distribution of orbital parameters, we do not take into account the early-time disruption of binaries by neighbouring PBHs, which ultimately results in a suppression of the merger rate, as shown in \cite{Raidal:2018bbj}. 
However, in this work we restrict ourselves to densities $\fPBH \lesssim 10^{-3}$ (see \cref{subsec:bound} for details) for which the effect is negligible \cite{Hutsi:2020sol}.

The expression for the merger rate given in \cref{eq:PBHrate} assumes that the orbital parameters evolve exclusively through GW emission. In fact, $a$ and $j$ can be altered via a number of mechanisms. Apart from the early-time disruption by neighbours discussed above, one of these is accretion, which can modify the PBH mass and shrink the binary~\cite{Caputo:2020irr,Ricotti:2007au,Hutsi:2019hlw,DeLuca:2020bjf, DeLuca:2020qqa}.
The effect of accretion is particularly relevant for high PBH masses and results in an enhancement of the high mass tail of the mass function: we leave the study of this effect to a future work where extended mass functions will be considered. Furthermore, if PBHs exist side-by-side with particle DM, the latter forms mini-haloes around the PBHs. In this case, another perturbation to the binary can come from the dynamical friction that the mini-haloes induce on the PBHs when they approach each other \cite{Kavanagh:2018ggo}. While the binaries are dramatically perturbed by this interaction, the shrinking of the binary compensates the increase of angular momentum in such way that the coalescence time is almost unaffected. 

The simplified picture of isolated binaries which we have presented so far fails to capture an important aspect: under the action of gravity, PBHs form bound clusters, where complex $N$-body interactions take place. Various works have been dedicated to studying the effect of clustering on the evolution of binaries~\cite{Raidal:2018bbj,Belotsky:2018wph,Vaskonen:2019jpv,Jedamzik:2020ypm,DeLuca:2020jug}, showing that it can have a significant impact on the merger rate. In the most extreme scenarios, binaries can be easily disrupted in these dense environments. But even if the binary is not completely disrupted, it must be extremely eccentric in order to merge within a Hubble time; given the strong dependence of the time of merger on the angular momentum, $t_{\mathrm{merger}} \propto j^7$, even a small increase in $j$ due to interactions within the cluster is sufficient to delay the merger beyond our time.

In this work, we quantify the suppression of the merger rate due to clustering following the semi-analytical modelling presented in \cite{Vaskonen:2019jpv}. This calculation is based on estimating the fraction of clusters and sub-clusters that undergo core collapse following gravo-thermal instability. It is assumed that all binary systems within these structures do not contribute to the merger rate: they end up being perturbed in the high density cores in such a way that their coalescence time exceeds the age of the Universe. In this sense, the calculation can be considered an over-estimation of the effect (while it is very likely that a binary is perturbed in a dense environment, it is not necessarily so; furthermore, new binaries can be created in the cores). However, this may be partially compensated by the fact that the formalism does not consider perturbations to binaries in stable clusters (i.e.~those not affected by core collapse). The calculation of the suppression factor is detailed in \cref{sec:appendix_clustering}.

\begin{figure}
\centering
\includegraphics[width=0.7\textwidth]{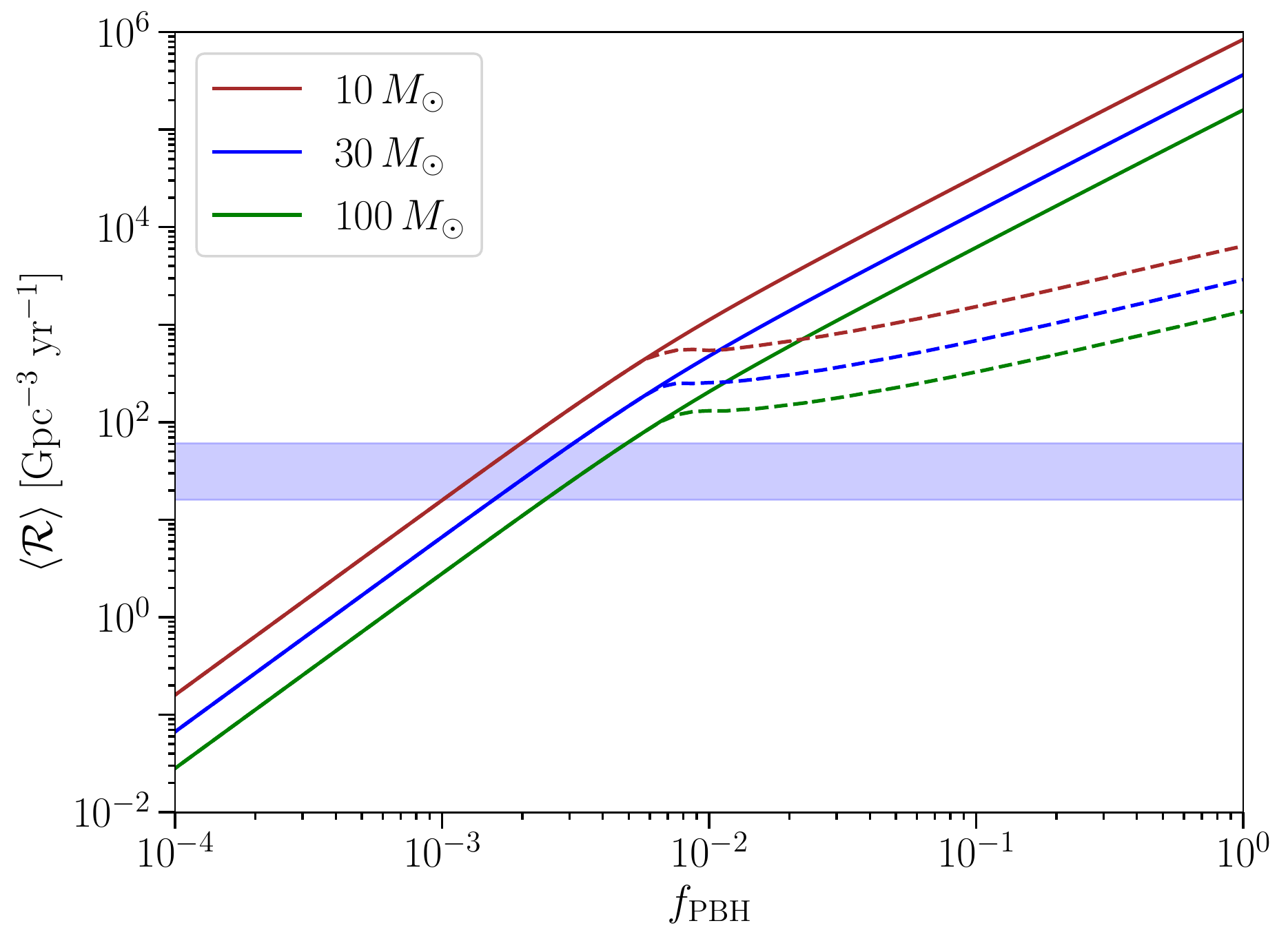}
\caption{Average merger rate density of PBH binaries formed in the early Universe, see \cref{eq:rate_avg}, as a function of \fPBH, for $M_{\rm PBH}=\SI{10}{\solarmass}$, $M_{\rm PBH}=\SI{30}{\solarmass}$ and $M_{\rm PBH}=\SI{100}{\solarmass}$. The dashed lines are obtained considering the disruption in clusters as described in \cref{sec:merger_rate_PBH}. The coloured band indicates for comparison the latest measure of the total binary BH merger rate as reported in \cite{LIGOScientific:2021psn}.}
\label{fig:MRclust}
\end{figure}

In \cref{fig:MRclust}, we plot the PBH merger rate for different values of $M_\mathrm{PBH}$ and $f_\mathrm{PBH}$, with and without the effects of clustering. The suppression factor increases going towards low redshifts, as clusters of increasingly larger size have sufficient time to undergo core collapse. As expected, it also increases with \fPBH. As we explain in \cref{sec:appendix_clustering}, for  small PBH fractions $\fPBH \lesssim 10^{-2}$, the effect becomes negligible, as most clusters in this case form late in the Universe and do not have sufficient time to undergo gravo-thermal collapse before today. As the disruption in clusters becomes greater with time, it causes an overall enhancement of the slope of the function $\mathcal{R}(z)$, making small values at $z=0 $ compatible with larger overall PBH abundances. Given that current bounds from GWs lie in the range $f_\mathrm{PBH} \sim 10^{-3} -10^{-2}$, where this effect may still be relevant, we update these to incorporate clustering in \cref{subsec:bound}.
    
\subsection{Astrophysical black holes}
\label{sec:merger_rate_ABH}

The ABH merger rate depends on the rate of formation of ABHs from stars via supernovae. We note that there are two channels through which ABH binaries can form: directly from binary stellar systems or from pairs of originally isolated black holes. In both scenarios, the ABH merger rate density~$\mathcal{R}_{\rm ABH}$ can be expressed as~\cite{Dvorkin:2016wac,Mukherjee:2021ags}
\begin{equation}
\mathcal{R}_{\rm ABH}(t, M_{\rm ABH}) = \mathcal{N}_{\rm ABH} \int_{\Delta t_{\rm min}}^{\Delta t_{\rm max}}
\diff \Delta t \; p(\Delta t) \, \mathcal{R}_{\rm birth}(t - \Delta t, M_{\rm ABH})\, ,
\label{eq:ABHmergerrate}
\end{equation}
where $\mathcal{R}_{\rm birth}(t,{M}_{\rm ABH})$ is the birth rate density of the stellar remnants as a function of redshift and mass, and $p(\Delta t)$ is the distribution function of the time delay between ABH formation and merger. In general, this time delay depends on the detailed rate of binary formation and on the orbital parameters of the resulting binaries (semi-major axis and eccentricity). In our model we follow the simplified approach of \cite{Dvorkin:2016wac} and express this distribution as $p(\Delta t) \propto 1/\Delta t$ for $\Delta t_{\rm min} \leq \Delta t \leq \Delta t_{\rm max}$ with $\Delta t_{\rm min} = \SI{50}{\mega\year}$ and $\Delta t_{\rm max}=H_0^{-1}$. This expression is motivated by the high-resolution numerical simulations of binary BH formation via the evolution of isolated binary stars presented in \cite{Belczynski:2016obo}. However, there are a number of uncertainties in the rate of ABH mergers~\cite{Mapelli:2018wys,Mapelli:2019bnp,Graziani:2020ekn,Santoliquido:2020bry,Mapelli:2020xeq,Mapelli:2021taw,Sedda:2021vjh}. We therefore include an overall normalising factor $\mathcal{N}_\mathrm{ABH}$, which we fix based on the observed low redshift merger rate, as we will describe in \cref{subsec:bound}.

We assume that the ABH birth rate density $\mathcal{R}_{\rm birth}(t,{M}_{\rm ABH})$ is proportional to the star formation rate (SFR) $\psi_{\rm SFR}(z)$. Galaxy surveys performed in the ultraviolet band and in the far-infrared band can directly measure the instantaneous star formation rate density at different redshifts, with the former range of wavelengths being a direct tracer of  short-lived massive stars, and the latter being a signature of UV light emitted by the same population of stars and subsequently absorbed and re-emitted by dust.
In-depth analyses of these data allow for modelling of the cosmic history of the SFR, unambiguously highlighting a rising trend at low redshift up to a peak of very intense star formation in the range $1 \leq z \leq 2$, followed by a fall-off at large redshift~\cite{SFR_Madau1998,SFR_MadauReview}.

However, the high redshift behaviour of the SFR may be affected by significant uncertainties and biases, mainly due to dust obscuration and to the fact that early star formation took place in very faint galaxies which are typically missed in existing surveys \cite{Kistler:2013jza}. Therefore we also consider an alternative tracer of the high redshift SFR, via the rate of gamma ray bursts (GRBs). Several studies (see e.g. \cite{Vangioni:2014axa} and references therein) based on this observable have indicated a higher SFR at high redshift compared to studies based on galaxy counts. With this method, the main uncertainty is the (model-dependent) relation between the rate of GRBs and the star formation rate itself. 

We follow~\cite{Nagamine:2003bd} and adopt the following parametrisation for the SFR,
\begin{equation}
\label{eq:SFR}
\psi_{\rm SFR}(z)
=
k \,
\frac{a \, \mathrm{e}^{b(z - z_{\rm m})}}
{a - b + b \, \mathrm{e}^{a(z - z_{\rm m})}} \ .
\end{equation}
The free parameters $k, a, b, z_{\rm m}$ in this expression are usually fitted by comparison to existing catalogues of different tracers of star formation. Aiming to bracket the uncertainties associated with the astrophysical rate, we consider two different choices for these parameters. The reference set of parameters corresponds to the \emph{fiducial model} quoted in \cite{Chen:2019irf} based on a fit to observations of bright galaxies. The values are
$k = \SI{0.178}{\solarmass\per\year\per\mega\parsec\cubed}$,
$z_{\rm m} = 2$,
$a = 2.37$,
$b = 1.8$.
We also consider a GRB-based fit as a \emph{maximal} model, with the values taken from~\cite{Vangioni:2014axa}:
$k = \SI{0.146}{\solarmass\per\year\per\mega\parsec\cubed}$,
$z_{\rm m} = 1.72$,
$a = 2.8$,
$b = 2.46$.

We stress that, unlike in the PBH case, the overall normalisation factor $\mathcal{N}_\mathrm{ABH}$ of the astrophysical rate is not determined {\it a priori}. Instead, we set this normalisation using the low redshift binary BH merger rate data provided by GW observatories currently in operation. Since in this work we are considering a hybrid scenario where a portion of the events is ascribed to a PBH population, each value of \fPBH actually corresponds to a different normalisation factor for the rate associated to the ABH population, so that the integrated rate in the redshift range $z\leq 1$ matches the observed one. The procedure to match the data naturally provides an upper limit on \fPBH, set by the requirement not to overshoot the rate measurement. The details of the procedure, and the corresponding bound on \fPBH, are described next.

\subsection{Normalisation of the redshift distributions}
\label{subsec:bound}

Given the models for the redshift evolution of the ABH and PBH rate density described above, we now aim to: {\it (i)} obtain an expression for the actual merger rate by convolving the theoretical rate estimates with the detector space-time sensitivity; {\it (ii)} compare the predicted merger rate for each value of \fPBH with the recent estimates provided by LVK. In this way, we will simultaneously set the normalisation of the astrophysical component for each value of \fPBH, and obtain an upper limit on this quantity.

\begin{figure}
\centering
\includegraphics[width=0.8\textwidth]{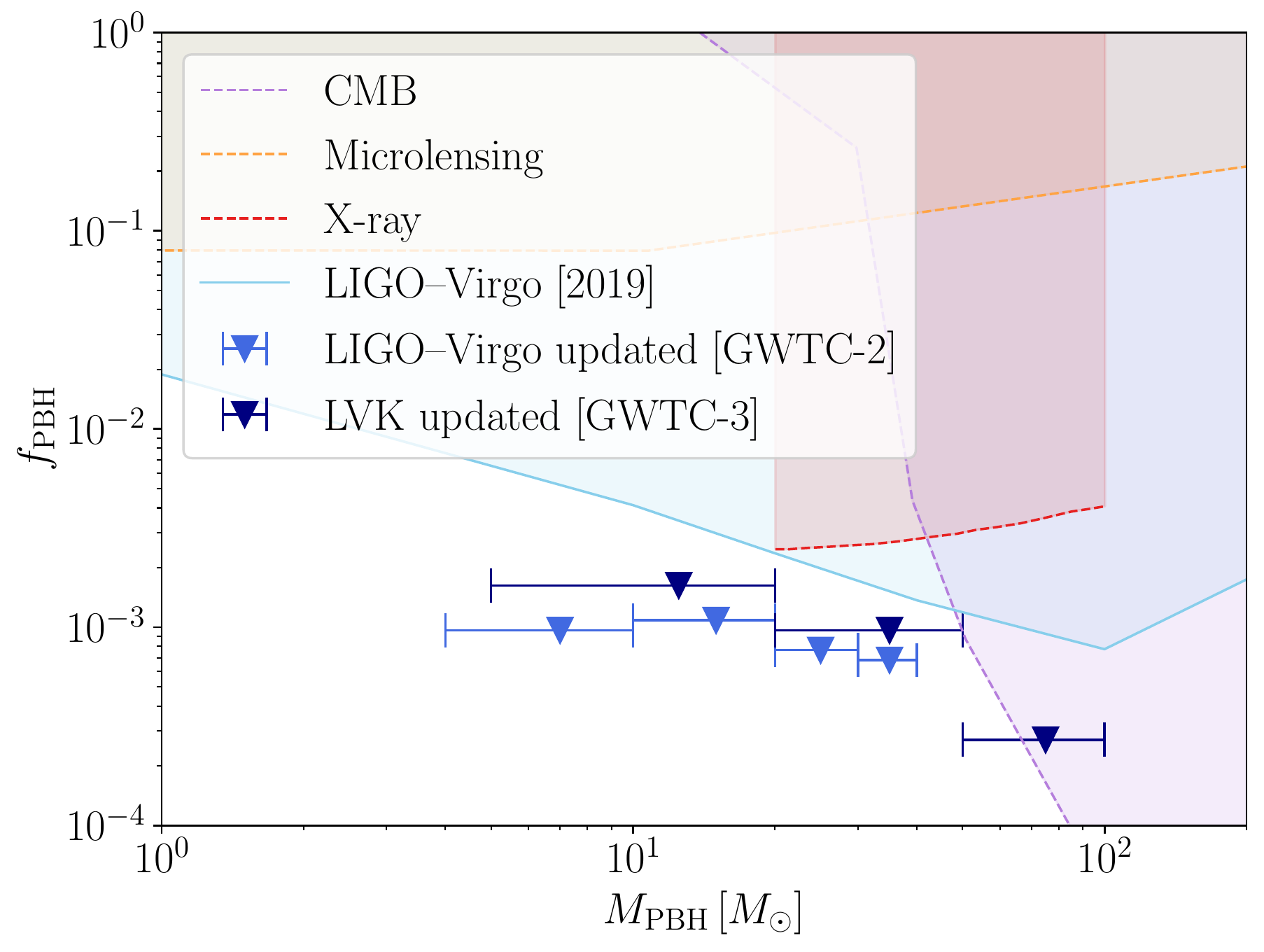}
\caption{Updated LIGO--Virgo and LVK bounds on $f_\mathrm{PBH}$, based on the observed low redshift merger rate reported in recent GW transient catalogues. The various bins in $M_\mathrm{PBH}$ correspond to the reported upper limits on the merger rate across different mass bins. For comparison, we also show bounds from the CMB  due to PBH accretion~\cite{Serpico:2020ehh}, microlensing constraints from the high redshift star Icarus~\cite{Oguri:2017ock}, and X-ray observations of the Milky Way~\cite{Manshanden:2018tze}. We obtained these bounds from the \href{github.com/bradkav/PBHbounds}{PBHbounds repository}~\cite{PBHbounds}.}
\label{fig:LIGObound}
\end{figure}

We compare the predicted PBH merger rate to the observational data from the second and third Gravitational-Wave Transient Catalogs (GWTC-2 and GWTC-3, produced by LIGO--Virgo and LVK respectively) for different PBH mass intervals and values of \fPBH. In particular, we adopt the  merger rate density estimates reported in Table 4 of \cite{LIGOScientific:2020kqk} for GWTC-2 and Table 4 of \cite{LIGOScientific:2021psn} for GWTC-3.\footnote{For the results presented in \cref{sec:cut_and_count} and \cref{sec:likelihood_based_method}, we fix the normalisation specifically using GWTC-3, from which the best estimate of merger rate density of BH binaries is $\mathcal{R} = 22 \,\mathrm{Gpc}^{-3}\,\mathrm{yr}^{-1}$.} These are reported assuming a non-evolving rate, while our models for the ABH and PBH merger rates evolve with redshift. We therefore compute the average merger rate density $\langle \mathcal{R} \rangle$, defined as
\begin{equation}
\label{eq:rate_avg}
    \langle \mathcal{R} \rangle \, =\, \dfrac{\int_0^{z_{\rm max}}  R_\mathrm{det}(z) \,\mathrm{d}z}{\int_0^{z_{\rm max}} \frac{f_\mathrm{det}(z)}{1+z}\frac{\mathrm{d}V_c}{\mathrm{d}z} \,\mathrm{d}z  } \, .
\end{equation}
This corresponds to the non-evolving merger rate density which would give rise to the same number of expected detectable events as our redshift-dependent model, over the redshift range $z \in [0, z_\mathrm{max}]$. For $f _{\rm det}(z)$ we use the selection function for aLIGO shown in \cref{fig:ET_selection_functions} and we fix $z_{\rm max}=2$.\footnote{In practice, this integral is largely insensitive to the precise value of $z_\mathrm{max}$, as the selection function drops rapidly above $z \sim 0.1$.} For each mass bin considered in these catalogues, we make the conservative choice to exclude the values of \fPBH that would correspond to an average merger rate density $\langle \mathcal{R} \rangle$ that exceeds the measured one by 2$\sigma$.
For values of \fPBH smaller than this bound, we adjust the ABH merger rate normalisation $\mathcal{N}_\mathrm{ABH}$ so that the sum of the averaged rates corresponds to the reported one, that is
    \begin{equation}
    \langle \mathcal{R}_{\rm PBH}(\fPBH)\rangle
    + \langle \mathcal{R}_{\rm ABH} (\mathcal{N}_{\rm ABH})\rangle
    = \mathcal{R}_{\rm obs}^{\rm LVK} \,. 
    \end{equation}
We have hence defined an allowed region for \fPBH and properly set the normalisation of the astrophysical rate for each value of \fPBH. We show the resulting new upper bounds we obtain on \fPBH in \cref{fig:LIGObound}. These bounds are compatible with previously reported bounds from GWTC-2 including the contribution of ABH mergers~\cite{Hutsi:2020sol}.

We find that the suppression due to late-time clustering, described in \cref{sec:merger_rate_PBH} and \cref{sec:appendix_clustering}, does not impact this upper limit. Its effect is relevant only for values of \fPBH larger than $\mathcal{O}(10^{-2})$, as shown in \cref{fig:MRclust}. Though the rate is significantly suppressed by clustering for large values of \fPBH, the predicted PBH merger rate would still exceed the merger rate observed by LVK, and large values of \fPBH remain excluded. Around our reported bound of $f_\mathrm{PBH} \lesssim 10^{-3}$, the suppression becomes negligible. Even so, we emphasise that the upper limit has to be taken {\it cum grano salis}, given the strong assumption of a monochromatic mass function. We cannot exclude scenarios featuring non-trivial mass functions and non-negligible initial clustering, which may potentially evade the bound. The study of such scenarios are beyond the scope of the current work.

\section{Generating mock GW catalogues}
\label{sec:generating_mock}

Having presented our merger rate calculation and the subsequent updated upper bounds on \fPBH from current data, we are ready to turn to the core of this work: assessing the ability of the ET to disentangle ABH mergers from PBH mergers using their redshift distributions. This section presents our method to produce mock data for the ET, accounting for the instrumental response and other observational effects such as lensing of the signal. Since we focus on the information contained in the redshift distribution of the GW events -- or, more accurately, their luminosity distance distribution which is the actual observable -- in the remainder of this article, a \emph{data set}~$\mathcal{D}$ will refer to $N_{\rm det}$ luminosity distance measurements~$D_i$ with their uncertainty~$\sigma_i$, $\mathcal{D}=\{(D_i, \sigma_i)\}_{i=1,\ldots, N_{\rm det}}$. Our mock generation code, \texttt{darksirens} \href{https://gitlab.com/matmartinelli/darksirens}{\faGitlab}, is publicly available and could be easily modified to include other observables.

\subsection{Parameters}

The parameters that must be set to produce a mock catalogue of GW distance measurements can be divided into four categories:
\begin{itemize}
    \item \textbf{Cosmology.} We assume a spatially flat homogeneous and isotropic \LCDM cosmological background. The parameters to be set are the Hubble--Lema\^{i}tre constant $H_0$ and the total matter density parameter~$\Omega_{\rm m}$.
    \item \textbf{Primordial black holes}. Following \cref{sec:merger_rate_PBH}, the PBH population is characterised by the masses of the objects, $M_{\rm PBH}$, the fraction of DM made up of PBHs, $f_{\rm PBH}$, and whether or not clustering of PBHs is considered (we remark once again that the effect of clustering is negligible within the current implementation).
    \item \textbf{Astrophysical black holes.} Following the model outlined in \cref{sec:merger_rate_ABH}, the ABH population is characterised by the masses of the objects~$M_{\rm ABH}$ and the SFR parameters $z_{\rm m}$, $a$ and $b$ which enter into \cref{eq:SFR}.
    \item \textbf{Specifications.} This class of parameters allows us to set the observational time $T_{\rm obs}$ of the survey and the SNR threshold, SNR$_{\rm min}$, which determines if a candidate event is detected or not. It also allows us to customise details of the mock generation, i.e.~to specify the redshift range over which calculations are done ($z\in\left[z_{\rm min},z_{\rm max}\right]$) and whether or not to include the effect of lensing.
\end{itemize}
We remind the reader that we assume for simplicity a single mass for PBHs, $M_\mathrm{PBH}$, and a single mass for ABHs, $M_\mathrm{ABH}$. Recall also that the normalisation of the ABH merger rate $\mathcal{N}_\mathrm{ABH}$ is not set as an external parameter but is fixed by comparison with low redshift GW observations, as detailed in \cref{subsec:bound}.

For the results presented in \cref{sec:cut_and_count,sec:likelihood_based_method}, we fix the parameters used to generate the mock data to the fiducial values listed in \cref{tab:fiducials}, unless otherwise specified in the text. We report the survey specifications assumed for the ET observations in  \cref{tab:specs}.

\begin{table}[ht]
	\centering{}%
	\begin{tabular}{|c|c||Cc|c|c|c||c|c|}
		\hline
		\multicolumn{2}{|Cc||}{Cosmology} & \multicolumn{4}{Cc||}{ABH parameters} & \multicolumn{2}{Cc|}{PBH parameters} \\
		\hline
		 $\Omega_{\rm m}$  & $H_0$ [\si{\kilo\meter\per\second\per\mega\parsec}] & $M_{\rm ABH}$ [$M_\odot$]  & $z_{\rm m}$  & $a$    & $b$ & $M_{\rm PBH}$ [$M_\odot$] & clustering \\
         \hline
         $0.315$ & $67.4$ & $7$ & $2$ & $2.37$ & $1.8$ & $10$ & yes \\
		 \hline
		  
	\end{tabular}\protect
	\caption{Cosmological and BH related parameters accessible in \texttt{darksirens} \href{https://gitlab.com/matmartinelli/darksirens}{\faGitlab}. The values shown here represent our baseline settings used throughout the article. We do not report a value of $f_{\rm PBH}$ as this will be changed depending on the analysis done with the data.}\label{tab:fiducials}
\end{table}

\begin{table}[ht]
	\centering{}%
	\begin{tabular}{|Cc|c|c|c|c|}
		\hline
		\multicolumn{5}{|Cc|}{Specifications} \\
		\hline
		 $T_{\rm obs}$ [yrs]  & SNR$_{\rm min}$  & lensing & $z_{\rm min}$  & $z_{\rm max}$\\
         \hline 
         $1$ & $8$ & yes & $0.001$ & $100$\\
		 \hline
		  
	\end{tabular}\protect
	\caption{Parameters available in \texttt{darksirens} \href{https://gitlab.com/matmartinelli/darksirens}{\faGitlab} to specify the characteristics of the survey and the redshift range over which the mock is constructed. The values shown here represent our baseline settings used throughout the article.}\label{tab:specs}
\end{table}

\subsection{Sketch of the generation algorithm}

Once the parameters are chosen, \texttt{darksirens} produces a mock catalogue~$\mathcal{D}$ of GW distances with their uncertainties as they would be measured by ET. For $\fPBH \neq 0$, the data is a mix of ABH and PBH mergers, which takes into account the potential clustering of the latter, the effect of lensing on distance measurements, and ET's instrumental uncertainties. We now explain exactly how the mock data are generated.

The very first step consists in computing the total number $N_{\rm tot}$ of mergers occurring in the redshift range $[z_\mathrm{min}, z_\mathrm{max}]$. We draw a random number from a Poisson distribution whose mean, $\bar{N}_{\rm tot}$, is given by \cref{eq:N_bar_tot}, in which we consider the total merger rate~$R(z)=R_{\rm ABH}(z)+R_{\rm PBH}(z)$ as described in \cref{sec:merger_rates}. Each data point~$i = 1,\ldots,N_{\rm tot}$ is then produced as follows:
\begin{enumerate}
\item Randomly draw the ``true'' redshift~$z_i$ of the merger from the redshift distribution $p(z)$ given in \cref{eq:prob_redshift}, once again considering the total merger rate.
\item Convert $z_i$ into the ``true'' unlensed luminosity distance~$\bar{D}_i\equiv \bar{D}(z_i)$ of the merger, using the background cosmological model and \texttt{CAMB}.
\item Compute the unlensed signal-to-noise ratio (SNR)~$\bar{\rho}_i$ of the event. The position, polarisation and inclination of the event are drawn randomly and the SNR computed based on these quantities along with the specifications of the ET. The full details of this calculation are given in \cref{sec:appendix_mocks}.
\item Determine the weak gravitational lensing magnification~$\mu_i$ of the event by randomly drawing it from the theoretical PDF $p(\mu)$ -- see \cref{sec:appendix_lensing} for details on lensing and its statistics. The lensing magnification enhances (or reduces) the SNR as $\rho_i = \sqrt{\mu_i} \bar{\rho}_i$, and reduces (or increases) the luminosity distance as $\tilde{D}_i = \bar{D}_i/\sqrt{\mu_i}$.
\item Determine whether the SNR is large enough for the event to be properly detected: if $\rho_i < 8$, then the event is considered to be too faint to be a true GW candidate and is removed from the catalogue. The choice of discarding events with an SNR smaller than eight follows the approach taken by the LIGO collaboration as an estimate for the detection threshold~\cite{Abbott:2016xvh}.
\label{step:SNR_cut}
\item Compute the measured luminosity distance,
$
D_i = \tilde{D}_i + \Delta D_i,
$
where $\Delta D_i$ represents the instrumental error on the measurement. The latter is drawn from a Gaussian distribution $\mathcal{N}(0, \sigma^{\rm inst}_i)$, with error inversely proportional to the SNR, $\sigma^{\rm inst}_i = 2 \tilde{D}_i/\rho_i$~\cite{Li:2013lza}. 
\item Compute the total uncertainty~$\sigma_i$ on this data point as the quadratic sum of the instrumental and lensing uncertainties,
\begin{equation}
\sigma_i^2
= \left(\sigma_i^{\rm inst}\right)^2 + \left(\sigma_i^{\rm lens}\right)^2
= \left( \frac{2\tilde{D}_i}{\rho_i} \right)^2
+ \sigma_\kappa^2(z_i) \, \bar{D}_i^2 \ ,
\label{eq:distance_error}
\end{equation}
where $\sigma_\kappa^2(z)$ is the variance of the weak-lensing convergence -- see \cref{sec:appendix_lensing} for details.
\end{enumerate}

\begin{figure}[ht!]
	\centering
	\includegraphics[width=0.8\columnwidth]{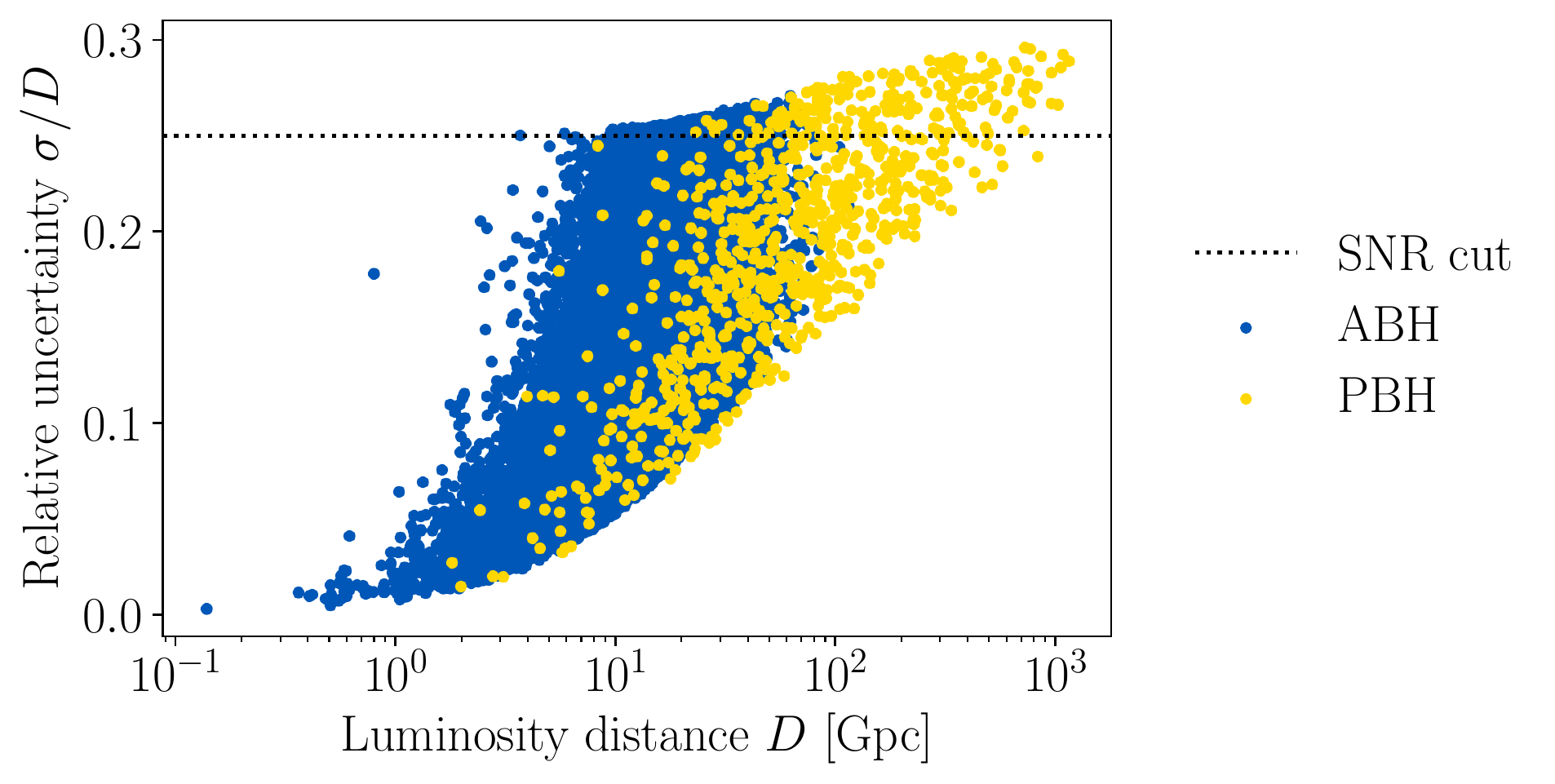}
	\caption{Data set generated using \texttt{darksirens} \href{https://gitlab.com/matmartinelli/darksirens}{\faGitlab} with the baseline parameters of \cref{tab:fiducials} and \cref{tab:specs}, with $f_{\rm PBH}=10^{-5}$. The blue dots show the observed ABH events, while the yellow dots represent the observed PBH events. At a fixed distance~$D$, the typical uncertainty is smaller for PBHs than for ABHs due to our choice of a larger PBH mass ($M_\mathrm{PBH} = 10\,M_\odot$, $M_\mathrm{ABH} = 7\,M_\odot$). The dotted horizontal line shows the level of instrumental uncertainty corresponding to our SNR cut of $\rho_i = 8$. The events that lie above this line still satisfy the SNR condition we impose, 
	but receive an extra contribution to the error from lensing.}
	\label{fig:mockplot}
\end{figure}

The end product is a catalogue of distance measurements with their uncertainties, $\mathcal{D}=\{(D_i, \sigma_i)\}_{i=1,\ldots, N_{\rm det}}$, where $N_{\rm det}\leq N_{\rm tot}$ is the number of events that survive the SNR cut of step~\ref{step:SNR_cut}. We have checked that $N_{\rm det}$ agrees on average with the theoretical expectation~$\bar{N}_{\rm det}$. \Cref{fig:mockplot} shows an example of a catalogue that can be obtained using this approach. The figure shows the distance $D_i$ of the events that survive the SNR cut,\footnote{A plot including the events which would not survive the SNR cut is shown in \cref{fig:SNRplot}.} together with their relative error, for both ABH (blue) and PBH (yellow). The data shown here are obtained setting $f_{\rm PBH}=10^{-5}$.

\subsection{Distance uncertainty}

Correctly modelling uncertainties in the luminosity distance of the GW events is crucial to disentangling high redshift and low redshift populations of mergers. A full analysis pipeline based on real GW data would provide as an output a probability distribution for the true luminosity distance $P(\bar{D})$ of the merger. Instead, in our catalogues, the distance uncertainty is described by only a single number $\sigma_i$, given in \cref{eq:distance_error}. We therefore model the probability distribution for the luminosity distance as a Gaussian, given explicitly as:
\begin{equation}
    p(\bar{D}|D_i) = \frac{1}{\sqrt{2 \pi} \sigma_i} \exp\left[- \frac{(\bar{D} - D_i)^2}{2\sigma_i^2}\right]\,.
    \label{eq:Gaussian_error}
\end{equation}

In the statistical modelling which we will describe in \cref{sec:cut_and_count} and \cref{sec:likelihood_based_method}, it is important to distinguish between $P(D_i|\bar{D})$ and $p(\bar{D}|D_i)$. In particular, some of the calculations presented in those sections are simplified if we identify the right hand side of \cref{eq:Gaussian_error} as $p(D_i|\bar{D})$, rather than $p(\bar{D}|D_i)$. However, this does not follow an intuitive definition of what is typically meant by the measurement error. We would typically consider that $D_i$ is our best estimate of the luminosity distance $\bar{D}$, with $\sigma_i$ parametrising our uncertainty on $\bar{D}$, matching the definition given in \cref{eq:Gaussian_error}.
We also find that interpreting the right hand side of \cref{eq:Gaussian_error} as $p(D_i|\bar{D})$ leads to inconsistencies. Consider the case where both $D_i$ and $\sigma_i$ are large, in which case the right hand side of \cref{eq:Gaussian_error} is a relatively flat function of $\bar{D}$. If we interpret this as $p(D_i|\bar{D})$, it would imply that a wide range of $\bar{D}$ values would all give rise to the same observed $D_i$ with similar probabilities. However, we know that the measurement error becomes smaller as we reduce $\bar{D}$, in which case large $D_i$ values should be very unlikely. The only consistent definition for the error is therefore given by \cref{eq:Gaussian_error}.

\medskip

We will now present two different analysis methods of the mock data we have generated:
an intuitive binned approach, and a more complete Bayesian analysis of the data, in \cref{sec:cut_and_count} and \cref{sec:likelihood_based_method} respectively. With the first method we aim to estimate the detectability threshold of a PBH population, while with the second method we also assess the capability of the ET to measure the PBH fraction. We refer to these methods as ``cut-and-count'' and ``likelihood-based'' respectively.

\section{Detecting PBHs: the cut-and-count method}
\label{sec:cut_and_count}

In this section, we present the ``cut-and-count'' method which we use to determine the smallest PBH fraction that could be detected by the ET. The idea is simple and intuitive: since we expect no ABHs to be formed beyond some high redshift (as there is a significant delay between the beginning of the Universe, the birth and death of the first stars and hence the formation and merging of ABHs), any sufficiently high redshift GW event produced by a BH merger should be the result of merging \textit{primordial} BHs. 

In practice, however, large distances are also the most uncertain ones, since the GW signal is typically much fainter than closer events. Hence the question we aim to answer is better phrased as: what is the lowest value of \fPBH that would still produce a sufficient number of large-distance events so that they would be statistically distinguishable from $\fPBH=0$ when observing with the ET? We hence \textit{cut} the data set into two subsets containing the small- and large-distance events, \textit{count} the number of events in the large-distance subset, and compare to the expected number for $\fPBH=0$. This method only exploits a fraction of the available information in the data set, but it is nevertheless a natural starting point.

\subsection{Description of the method}
\label{subsec:cut_and_count_method}

Let $D_* \equiv \bar{D}(z_*)$ be an arbitrary distance threshold ($z_*$ is the corresponding background redshift threshold). Given a data set~$\mathcal{D}$, we divide it into two subsets: a small-distance subset $\mathcal{D}_\leq$ on the one hand, made of the events with distances beneath the threshold ($D\leq D_*$); and a large-distance subset~$\mathcal{D}_>$ on the other hand, with events beyond the threshold ($D> D_*$). We call $N_>$ the cardinal of $\mathcal{D}_>$, i.e., the number of events above the threshold~$D_*$; the larger \fPBH, the larger the expected~$N_>$.

To be more specific, the computation of $N_>$ does not simply consist in counting the number of events whose best-fit distance is above the threshold. Due the measurement errors, it may happen that an event truly lies beneath the distance threshold ($\bar{D}_i < D_*$) but is actually measured beyond it ($D_i > D_*$), or vice versa. In order to account for this, we calculate the probability that the true distance lies beyond the arbitrary distance threshold as
\begin{equation}\label{eq:probdens}
P_i
=
\frac{1}{\sqrt{2\pi}\sigma_i}
\int_{D_*}^\infty{{\rm d}\bar{D}_i \; \exp\left[-\frac{(\bar{D}_i-D_i)^2}{2\sigma_i^2}\right]}\, ,
\end{equation}
using the definition of the uncertainty from \cref{eq:Gaussian_error}.
We can therefore obtain the value of $N_>$ by summing over all the events the probability of falling within the bin being considered,
\begin{equation}\label{eq:stat_counts}
    N_> \equiv \sum_{i=1}^{N_{\rm det}}{P_i}\,.
\end{equation}

The uncertainty on $N_>$ is twofold. First, since the data set is discrete, it inevitably comes with a Poisson uncertainty with variance
\begin{equation}
\sigma^2_{\rm P} = \bar{N}_> \ ,
\end{equation}
where $\bar{N}_{>}$ is the theoretical expectation value of $N_>$.\footnote{In practice,
we estimate $\bar{N}_>$ by generating a number of mock data sets and taking the mean of the $N_>$ values obtained for each.} On the other hand, the number~$N_>$ of large-distance events may be seen as the sum of $N_{\rm det}$ independent Bernoulli variables, the $i^{\rm th}$ one having a probability $P_i$ of being equal to $1$, and a probability $1-P_i$ of being equal to zero. This observational contribution to the variance of $N_>$ is the sum of the individual Bernoulli variances,
\begin{equation}
\sigma_{\rm B}^2 = \sum_{i=1}^{N_{\rm det}}{P_i(1-P_i)}\ .
\end{equation}
Summing the two sources of uncertainty in quadrature yields the total error on $N_>$,
\begin{equation}\label{eq:binerror}
\sigma_{>}
= \sqrt{\sigma^2_{\rm P} + \sigma_{\rm B}^2} \ .
\end{equation}

\begin{figure}[ht!]
	\centering
	\includegraphics[width=0.6\columnwidth]{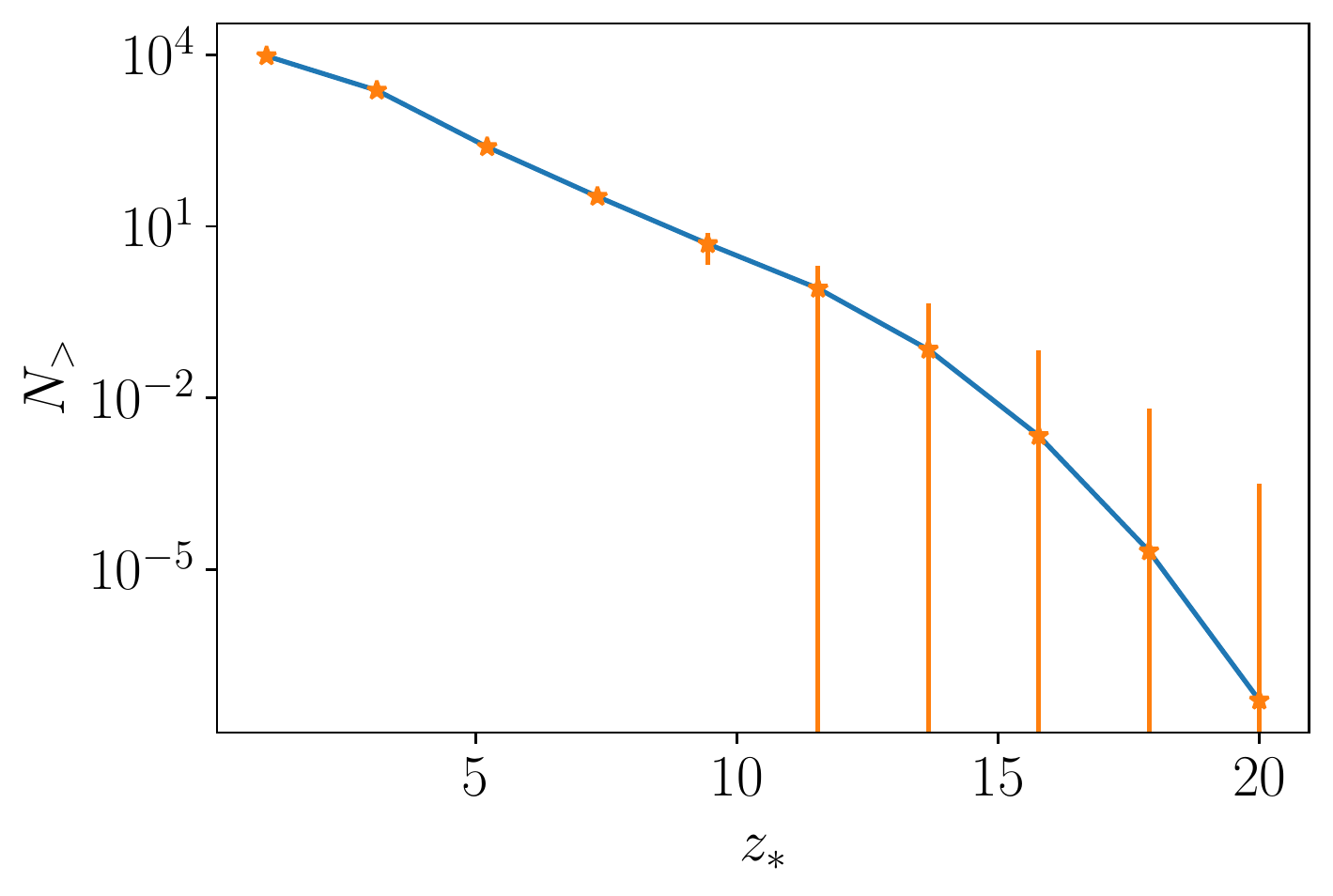}
	\caption{Value of $N_>$ as a function of the redshift threshold~$z_*$, for a data set with a negligible number of PBH. The error bars represent the uncertainty~$\sigma_>$ on $N_>$.}
	\label{fig:N_>_no_PBH}
\end{figure}

As an example, we consider the null case of a data set~$\mathcal{D}_0$ containing no PBH ($\fPBH \to 0$). \Cref{fig:N_>_no_PBH} shows the expected value of $N_>$ as a function of the redshift threshold~$z_*$, to which the $N_>$ of an observed data set should be compared. The error bars of that figure are computed following the approach described aobve, and they represent the uncertainty~$\sigma_>$ on $N_>$. 

\subsection{Optimal distance threshold and smallest detectable PBH fraction}
\label{subsec:cut_and_count_results}

Let us now determine what is the smallest detectable PBH fraction~\fPBH that could be detected using the cut-and-count method. This will also be the occasion to determine the optimal choice for the arbitrary distance threshold, that is, the value of $D_*$ leading to the maximal sensitivity in \fPBH.

For that purpose, we generate mock data sets~$\mathcal{D}_{\fPBH}$ for $100$ values of \fPBH ranging from $10^{-6}$ to $10^{-2}$. For each of these data sets, we apply the cut-and-count method for 10 different values of the redshift threshold~$z_*$, taken as equispaced in the range $[5, 50]$. This yields 
a number $N_>(\mathcal{D}_{\fPBH}, z_*)$ for each mock data set and each value of $z_*$. Those numbers come with their own uncertainty~$\sigma_>(\mathcal{D}_{\fPBH}, z_*)$, computed as described in \cref{subsec:cut_and_count_method} for the null case~$\mathcal{D}_0$. We estimate the statistical significance of the detection of a non-zero \fPBH as
\begin{equation}\label{eq:detfac}
\mathcal{S}(\mathcal{D}_{\fPBH}, z_*)
\equiv
\frac{\left| N_>(\mathcal{D}_{\fPBH}, z_*) - N_>(\mathcal{D}_0, z_*) \right|}
{\sqrt{\sigma^2_{>}(\mathcal{D}_{\fPBH}, z_*) + \sigma^2_{>}(\mathcal{D}_0, z_*)}}\ ,
\end{equation}
which represents the ``number of sigmas'' with which we could claim a detection.

\begin{figure}[ht!]
	\centering
	\begin{tabular}{cc}
	     \includegraphics[height=5.6cm]{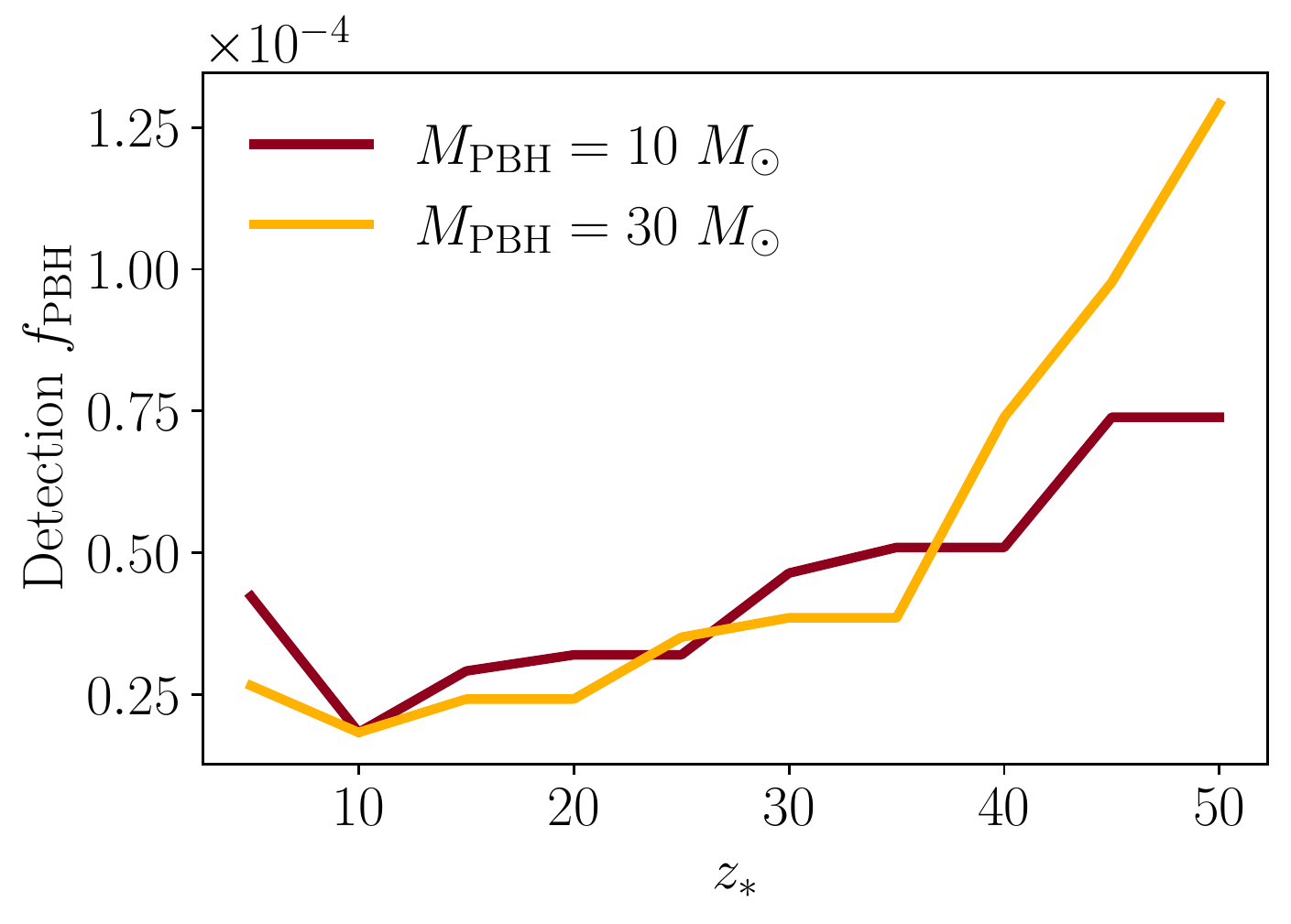} &  
	     \includegraphics[height=5.6cm]{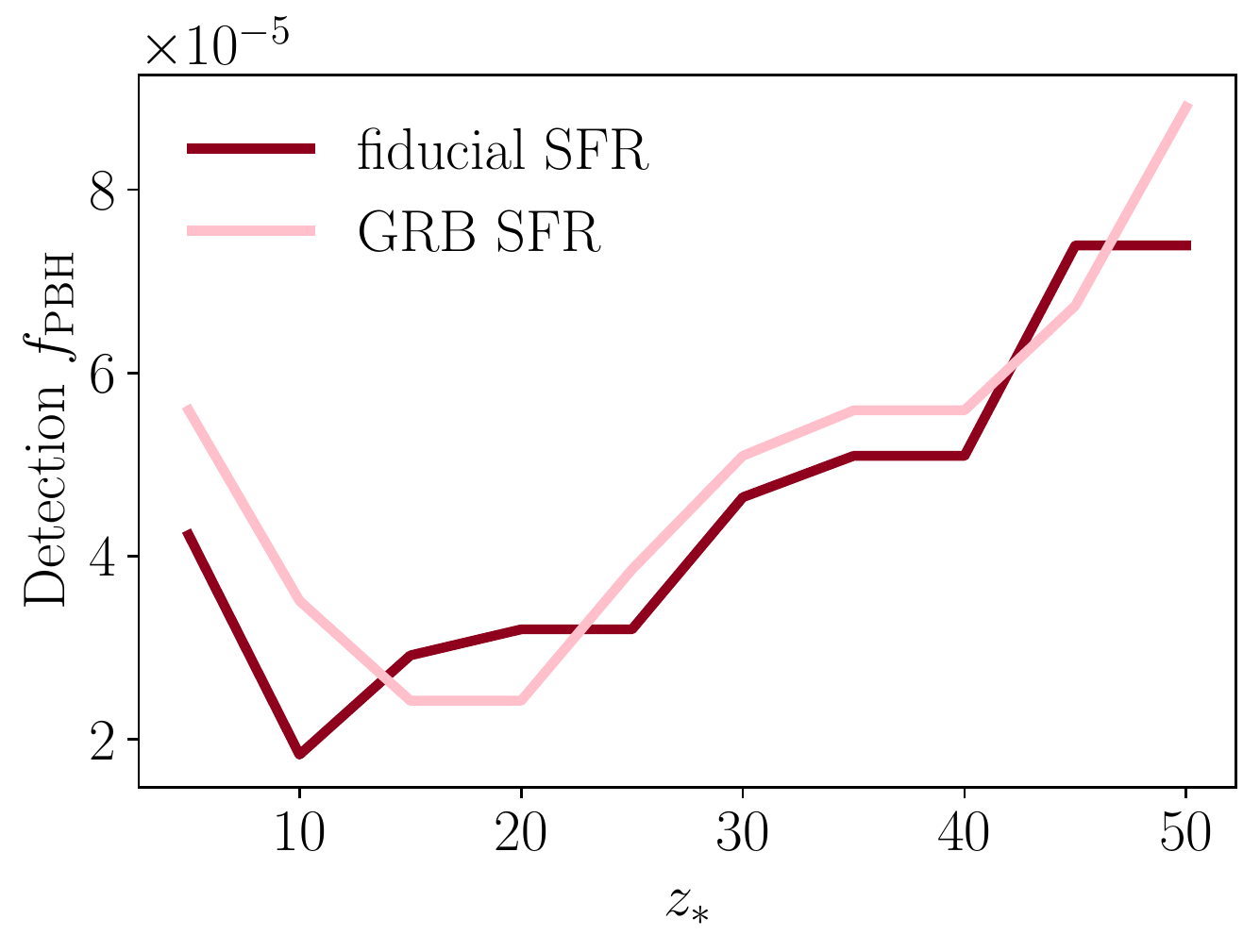}\\
	     \multicolumn{2}{c}{\includegraphics[height=5.6cm]{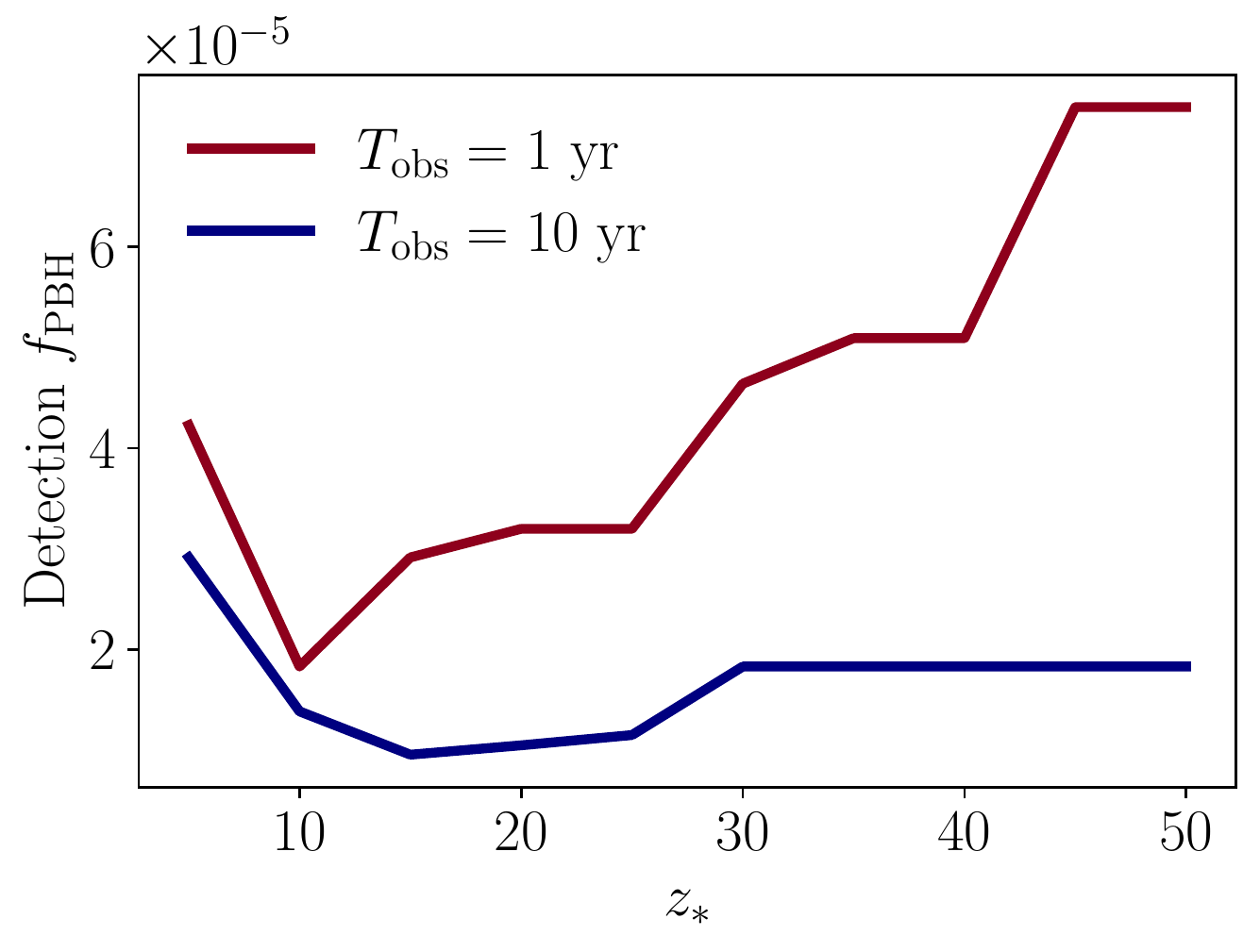}}
	\end{tabular}
	\caption{
	{\bf Top left}: Lowest $3\sigma$-detectable $f_{\rm PBH}$ as a function of the redshift~$z_*$ discriminating between the low redshift and high redshift subsets, for $M_{\rm PBH}=10\,M_\odot$ (red lines) and $M_{\rm PBH}=30\,M_\odot$ (yellow lines). 
	{\bf Top right}: Lowest $3\sigma$-detectable $f_{\rm PBH}$ as a function of $z_*$, for $M_{\rm PBH}=10\,M_\odot$ for the fiducial ABH merger rate (red line) and for the one obtained using the GRB SFR (pink line).
	{\bf Bottom}: Lowest $3\sigma$-detectable $f_{\rm PBH}$ as a function of the chosen $z_*$, for $M_{\rm PBH}=10\,M_\odot$ for an observation time $T_{\rm obs}=\SI{1}{\year}$ (red line) and for $T_{\rm obs}=\SI{10}{\year}$ (blue line).}
	\label{fig:detection_zT}
\end{figure}

In the following, we consider a ``significant'' detection to be one made at $3\sigma$; in other words, for each value of $z_*$, the smallest detectable \fPBH is determined by finding the data set $\mathcal{D}_{\fPBH}$ such that $\mathcal{S}(\mathcal{D}_{\fPBH}, z_*)>3$. In the top left panel of \cref{fig:detection_zT}, both curves (for $M_{\rm PBH} = \SI{10}{\solarmass}$ in red and $M_{\rm PBH} = \SI{30}{\solarmass}$ in yellow) show the evolution of the smallest detectable \fPBH with $z_*$. The optimal value for the redshift threshold, leading to the best sensitivity in \fPBH, is found to be $z_*\approx 10$, corresponding to a distance of $D_*\approx\SI{106}{\giga\parsec}$.

The existence of an optimal value for $z_*$ arises from a trade-off between two effects: on the one hand, increasing $z_*$ reduces the contamination due to ABH events in the high redshift subset; on the other hand, reducing $z_*$ increases the number of events in that high redshift subset and hence the significance of the detection. 
Our results suggests that choosing $z_*<10$ implies too much ABH contamination to easily assess whether or not PBHs are present in the data set, while $z_*>10$ worsens the statistics and only provides a detection for large values of \fPBH.
These conclusions are valid for both $M_{\rm PBH}=\SI{10}{\solarmass}$ and $M_{\rm PBH}=\SI{30}{\solarmass}$, although the exact value of the optimal $z_*$ does change when the mass of the PBHs is changed.

We expect the SFR model to affect the results, as it determines how fast the probability of ABH mergers vanishes with redshift, see \cref{fig:MergerRate}. We verify the impact of this uncertainty by repeating our analysis using the merger rate obtained assuming the GRB SFR model.
In the top right panel of \cref{fig:detection_zT}, we compare the GRB SFR results with the baseline case ($M_{\rm PBH}=10\ M_\odot$, fiducial SFR and $T_{\rm obs}=1$ yr). We can see how in the GRB case, the optimal threshold redshift~$z_*$ increases due to the ABH merger probability being non-vanishing up to higher redshifts than in the fiducial case. We also explore the impact of the observation time of the survey on these results; we compare our baseline results with an extended survey time for ET, setting $T_{\rm obs}=10$ yr. As can be seen in the bottom panel of \cref{fig:detection_zT}, increasing the time of the survey generally leads to a lower detection threshold for $f_{\rm  PBH}$, while also decreasing the dependency of such a threshold on the choice of the binning strategy. With a greater survey time, more high redshift events are observed, meaning that a large detection significance is still achievable even as the high redshift bin is moved to larger values of $z_*$. 

Overall, we find that a fraction $\fPBH\approx2\times10^{-5}$ is the lowest value detectable by the ET with this method for $T_{\rm obs}=1$ year, with a weak dependency on the mass of the progenitor systems and on the SFR model chosen. In our baseline case, for the optimal $z_*\approx10$, such a fraction of PBH corresponds to $N_>=16\pm4.6$, while the no PBH case yields $N_>=1\pm1.7$, a result that puts this \fPBH over the $3\sigma$ threshold we consider for detection.

\begin{figure}[ht!]
	\centering
	\includegraphics[width=0.49\textwidth]{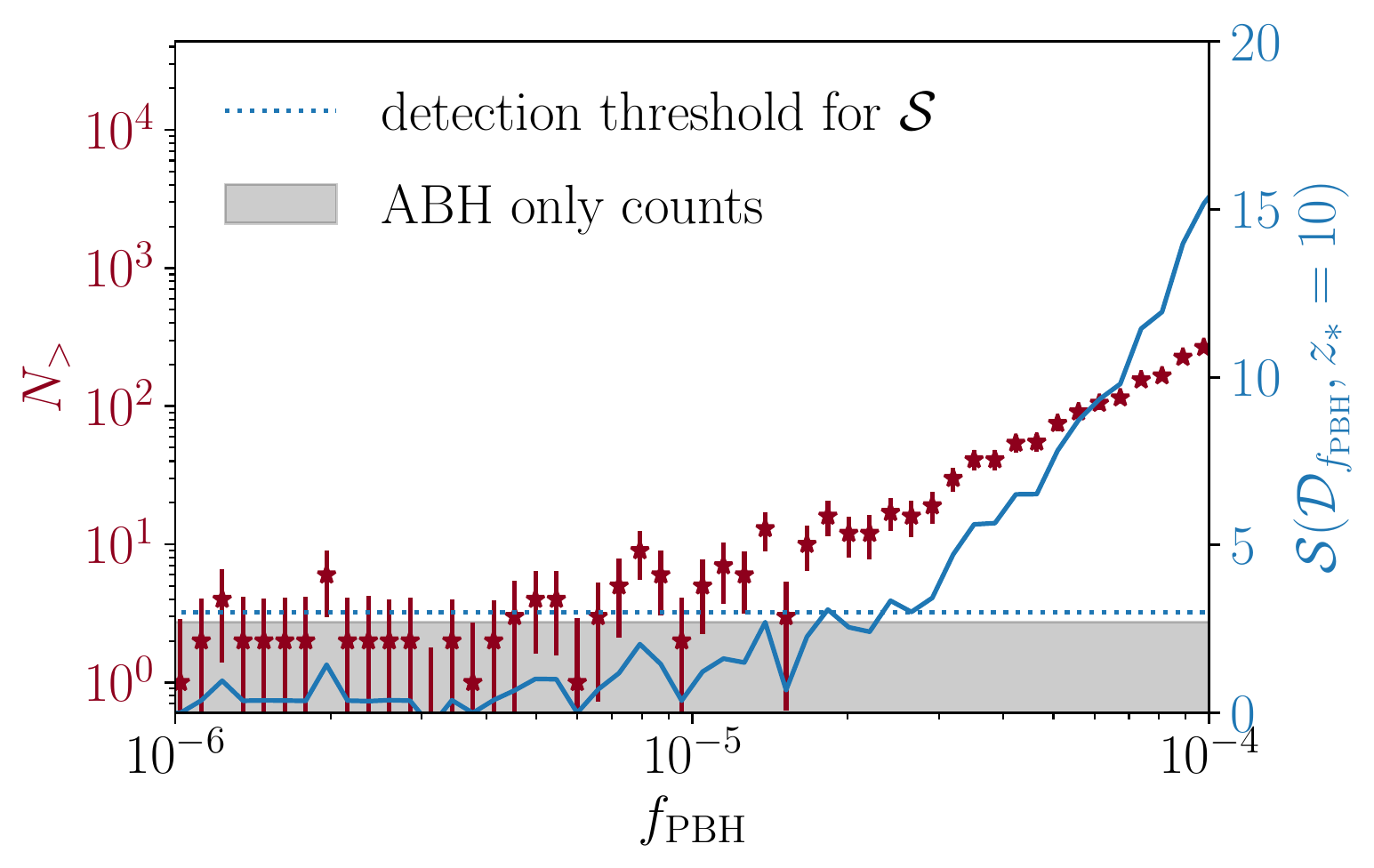}
	\hfill
	\includegraphics[width=0.49\textwidth]{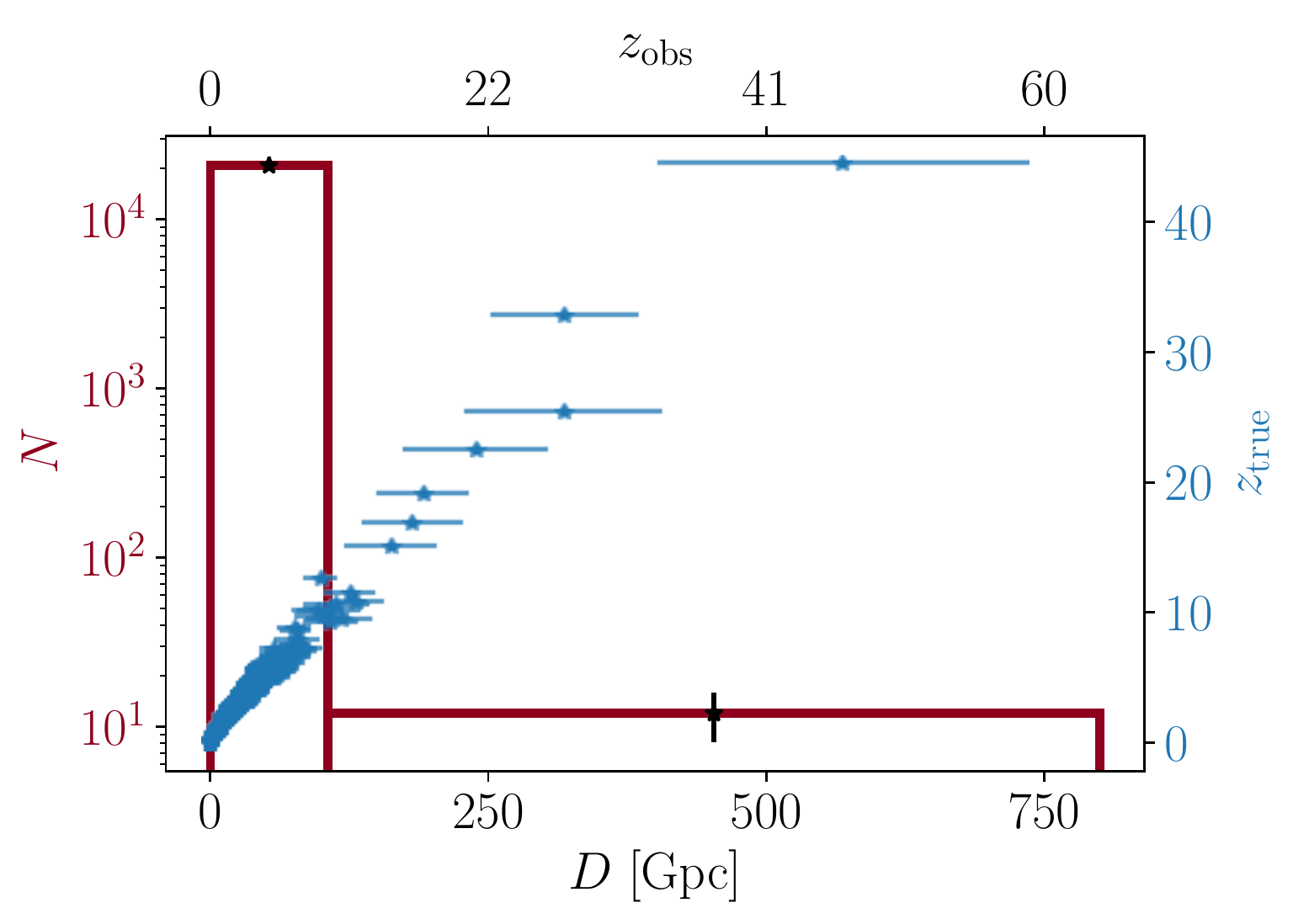}
	\caption{{\bf Left}: Number $N_>$ of events in the high redshift subset, $z>z_*=10$, for each value of $f_{\rm PBH}$ in the baseline case (see \cref{tab:fiducials} and \cref{tab:specs}) (red points). The grey band shows the uncertainty for $N_>$ in the no-PBH case. The blue line shows instead the trend of the detection significance for a departure from the no-PBH case as a function of $f_{\rm PBH}$. 
	{\bf Right}: Counts in the small- and large-distance bins are shown by the red bar plot for the baseline case, with $z_*=10$ and $f_{\rm PBH}=2\times10^{-5}$, together with the errors we determine on the value of the counts (left $y$-axis). In blue, the distances and their errors for all the events contained in the data set are shown as a function of the true redshift $z_{\rm true}$ (right $y$-axis), while the top $x$-axis shows the redshift $z_{\rm obs}$ that could potentially be inferred from the distance measurements.}\label{fig:best_baseline}
\end{figure}

\Cref{fig:best_baseline} shows more details for the results obtained in our baseline case, when taking $z_*=10$. The left panel shows in red the results for $N_>$ for the sampled values of \fPBH, together with their error bars and the uncertainty on $N_>$ in the no PBH case (grey band). In this plot, we also show the trend of the detection significance $\mathcal{S}$ with \fPBH (blue line). The right panel shows the counts in the two redshift bins for $\fPBH=2\times10^{-5}$ in red, i.e. the first value for which the counts in the large-distance bin result in $\mathcal{S}\geq3$. Together with the counts we also show  the distance and its error for the events contained in this mock data set, as a function of the event redshifts (which would not be observed in reality).

The conclusions we have found for our baseline case using this cut-and-count approach are compatible with those that can be obtained using the approach of \cite{Ng:2021sqn}, where the discriminant for the presence of PBH in the observed data set is the observation of events at $z>30$, where the contribution of ABHs is negligible. With our simulated data we indeed find that $f_{\rm PBH}\approx2\times10^{-5}$ is the lowest value in the baseline case for which we have at least one event above $z=30$ with $99.7\%$ confidence level.

\section{Quantifying the PBH fraction: a likelihood-based method}
\label{sec:likelihood_based_method}

Having followed our simple cut-and-count method to quantify the lowest value of \fPBH that would be statistically distinguishable from \fPBH$=0$ (i.e. the ability of the ET to \textit{detect} PBHs), we now utilise a more powerful, likelihood-based method to assess the ET's potential capacity to both detect PBHs \textit{and} measure the \fPBH associated with a detection. The analysis that follows assumes fixed values of the cosmological parameters $\Omega_{\rm m}$ and $H_0$, considering them as priors set from independent experiments. We thus neglect the potential degeneracies that could exists between \fPBH and the cosmological model. This approach is justified by the fact that even future GW experiments are expected to yield much looser cosmological constraints than established probes such as \emph{Planck}.

\subsection{Description of the method}

We describe an unbinned likelihood-based approach to comparing the data to a given model for the merger rate, which may include a contribution from both ABH and PBH mergers. To do this, we must compute the likelihood $\mathcal{L}(\fPBH)$, which is the probability to observe the data set $\mathcal{D}$ given the PBH fraction $f_\mathrm{PBH}$.

We write the probability distribution for the \textit{true} luminosity distances of ABH and PBH mergers as  $p_\mathrm{ABH}(\bar{D})$ and  $p_\mathrm{PBH}(\bar{D}|\fPBH)$ respectively, where we have indicated explicitly that the distribution for PBH mergers depends on \fPBH. The probability that a BH merger has a true luminosity distance in the range $[\bar{D}, \bar{D} + \diff \bar{D}]$ is given by
\begin{equation}
    p(\bar{D}|\fPBH) \, \diff \bar{D}
    = \frac{N_\mathrm{ABH}}{\bar{N}_\mathrm{obs}} \, p_\mathrm{ABH}(\bar{D}) \, \diff \bar{D} + \frac{N_\mathrm{PBH}}{\bar{N}_\mathrm{obs}} \, p_\mathrm{PBH}(\bar{D}|\fPBH) \, \diff \bar{D} \,, 
    \label{eq:P_full}
\end{equation}
where $\bar{N}_\mathrm{obs}(\fPBH) = N_\mathrm{ABH} + N_\mathrm{PBH}(\fPBH)$ is the total \textit{expected} number GW events and $N_\mathrm{ABH}$ and $N_\mathrm{PBH}$ are the \textit{expected} numbers of ABH and PBH mergers respectively. These are the numbers of events expected to be observed above the SNR threshold, taking into account the detection efficiency of the observatory, as detailed in \cref{subsec:preliminary_definitions}. 

We do not observe the true distance $\bar{D}$ but instead an estimate of the luminosity distance $D$. The probability that we observe an event with estimated luminosity distance $D$ can be written as
\begin{equation}
    p(D|f_\mathrm{PBH}) = \int p(D|\bar{D})\,p(\bar{D}|f_\mathrm{PBH})\,\mathrm{d}\bar{D}\,.
    \label{eq:measurementerror}
\end{equation}
The distribution of ``true'' luminosity distance $p(\bar{D}|f_\mathrm{PBH})$ is given by the theoretical expectation in \cref{eq:P_full}. The term $p(D|\bar{D})$ can be obtained from the measurement uncertainty in \cref{eq:Gaussian_error} via Bayes' theorem~\cite{Bayes:1764vd},
\begin{equation}
    p(D|\bar{D}) = \frac{p(\bar{D}|D)}{\tilde{p}(\bar{D})}\,
    \tilde{p}(D)\,.
\end{equation}
Here, $\tilde{p}(D)$ and $\tilde{p}(\bar{D})$ are the \textit{overall} probability distributions of $\bar{D}$ and $D$, where by \textit{overall} we mean that they are marginalised over all theory parameters (which in our case consists of only $f_\mathrm{PBH}$).
The term $\tilde{p}(D)$ enters as an overall normalisation which can be pulled out of the integral in \cref{eq:measurementerror} and does not depend on the theory parameters. This can therefore be safely neglected in the likelihood. The final task is then to compute $\tilde{p}(\bar{D})$, for which we need a prior on the PBH fraction $\mathrm{Pr}(f_\mathrm{PBH})$:
\begin{equation}
    \tilde{p}(\bar{D}) = \int_{0}^{1} p(\bar{D}|f_\mathrm{PBH}) \, \mathrm{Pr}(f_\mathrm{PBH})\,\mathrm{d}f_\mathrm{PBH}\,.
\end{equation}
We assume an uninformative log-flat prior,  $\mathrm{Pr}(f_\mathrm{PBH}) \propto 1/f_\mathrm{PBH}$, with $f_\mathrm{PBH}\in[10^{-9}, 10^{-3}]$.

For a sample of $N_\mathrm{obs}$ observed merger events, the likelihood can then be written as
\begin{equation}
    \mathcal{L}(\mathcal{D}|\fPBH) = \frac{\bar{N}_\mathrm{obs}(f_\mathrm{PBH})^{N_\mathrm{obs}} \mathrm{e}^{-\bar{N}_\mathrm{obs}(f_\mathrm{PBH})}}{N_\mathrm{obs}!} \times \prod_{i = 1, N_\mathrm{obs}} p(D_i|\fPBH)\,.
    \label{eq:likelihood}
\end{equation}
The first term is the Poisson probability to observe $N_\mathrm{obs}$ merger events, given that we expect to observe $\bar{N}_\mathrm{obs}(f_\mathrm{PBH})$. The second term accounts for the contribution of each observed event to the likelihood, where $D_i$ are the estimated luminosity distances. Combining the results above, the probability of observing a merger at an estimated distance $D_i$ is given by
\begin{equation}
    p(D_i|f_\mathrm{PBH}) \propto \int \frac{p(\bar{D}_i|D_i)}{\tilde{p}(\bar{D}_i)}\,p(\bar{D}_i|f_\mathrm{PBH})\,\mathrm{d}\bar{D}_i\,,
\end{equation}
which depends on the uncertainty $\sigma_i$ through $p(\bar{D}_i|D_i)$ in \cref{eq:Gaussian_error}. 

We then adopt a Bayesian approach in order to construct the posterior distribution function $p(f_\mathrm{PBH}|\mathcal{D})$ and thus determine projected constraints on the PBH fraction,
\begin{equation}
\label{eq:PDF}
    p(f_\mathrm{PBH}|\mathcal{D}) \propto \mathcal{L}(\mathcal{D}|f_\mathrm{PBH}) \mathrm{Pr}(f_\mathrm{PBH})\,.
\end{equation}
We sample $\log_{10}f_{\rm PBH}$ from a uniform prior $\log_{10}f_{\rm PBH}\in\left[-9,-3\right]$ using the public cosmological sampling code \texttt{Cobaya} \cite{Torrado:2020dgo}, while fixing all other parameters to the values used to generate mock data.
We analyse the results using \texttt{GetDist} \cite{Lewis:2019xzd} to obtain the bounds achievable on $f_{\rm PBH}$ using this approach.

\subsection{Future constraints on \texorpdfstring{$f_{\rm PBH}$}{fPBH}}
\label{subsec:likelihood-based_results}

\begin{figure}[th!]
	\centering
	\begin{tabular}{cc}
	     \includegraphics[height=5.cm]{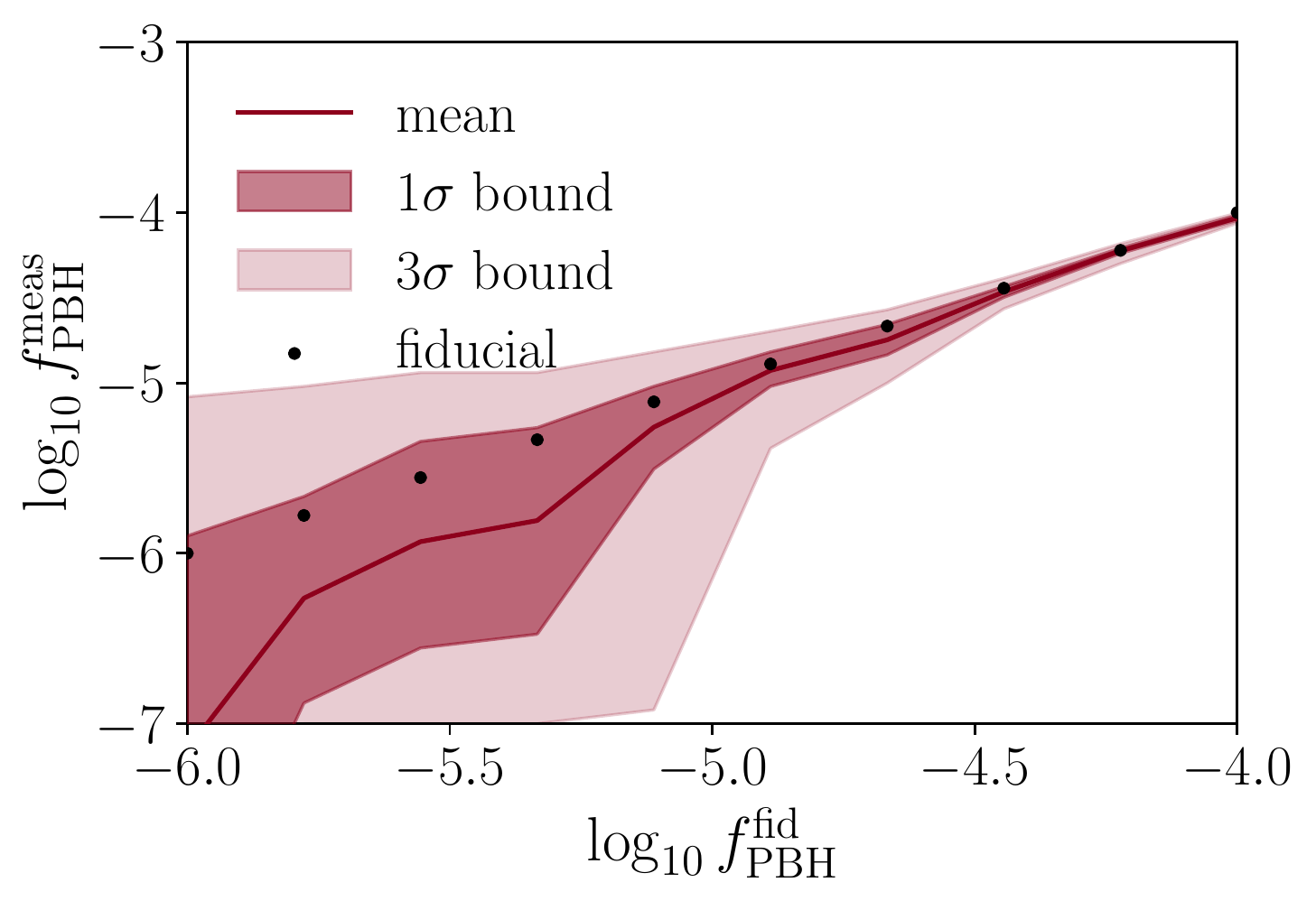} &
	     \includegraphics[height=5.cm]{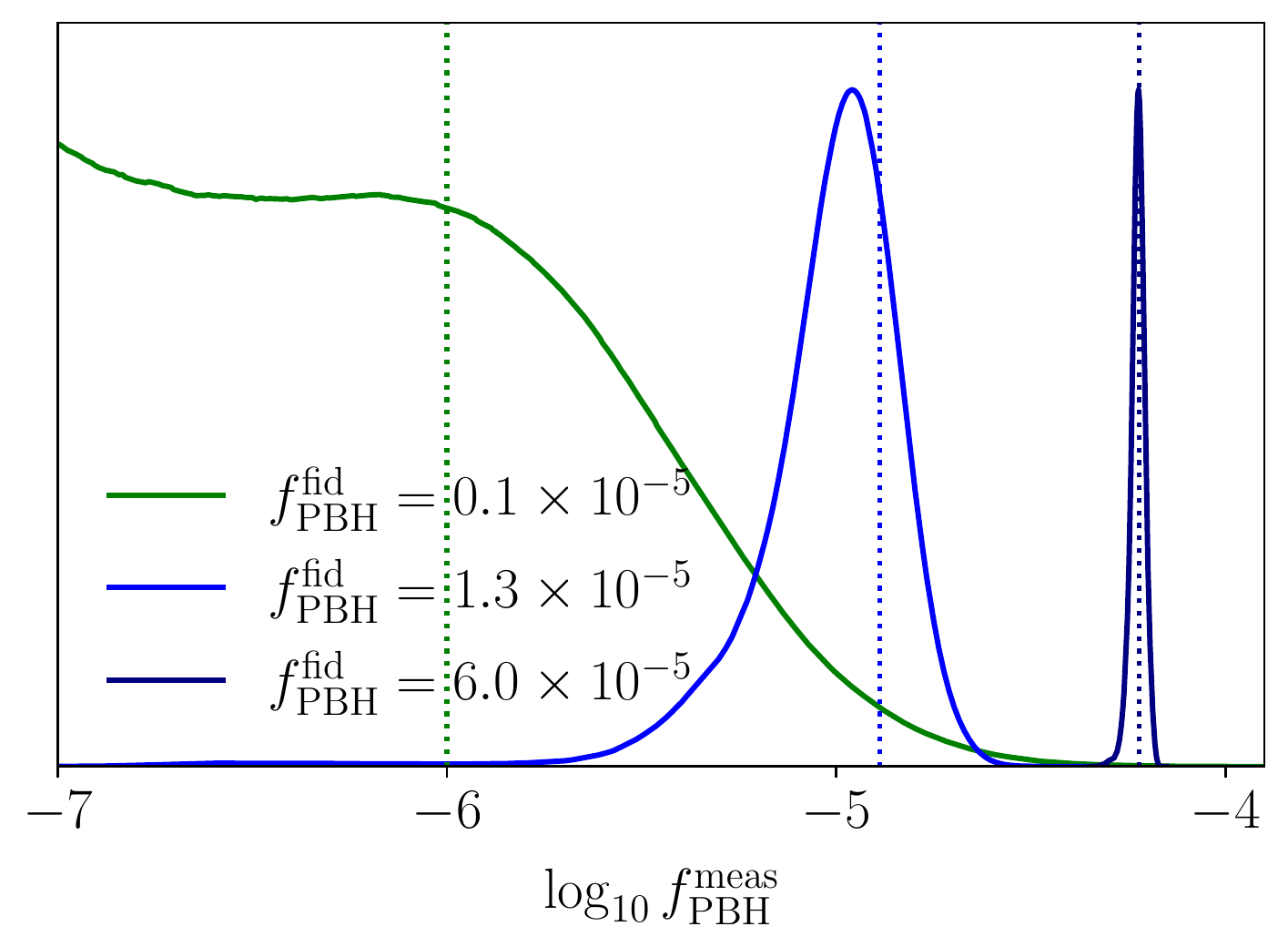}\\
	     \multicolumn{2}{c}{\includegraphics[height=5.cm]{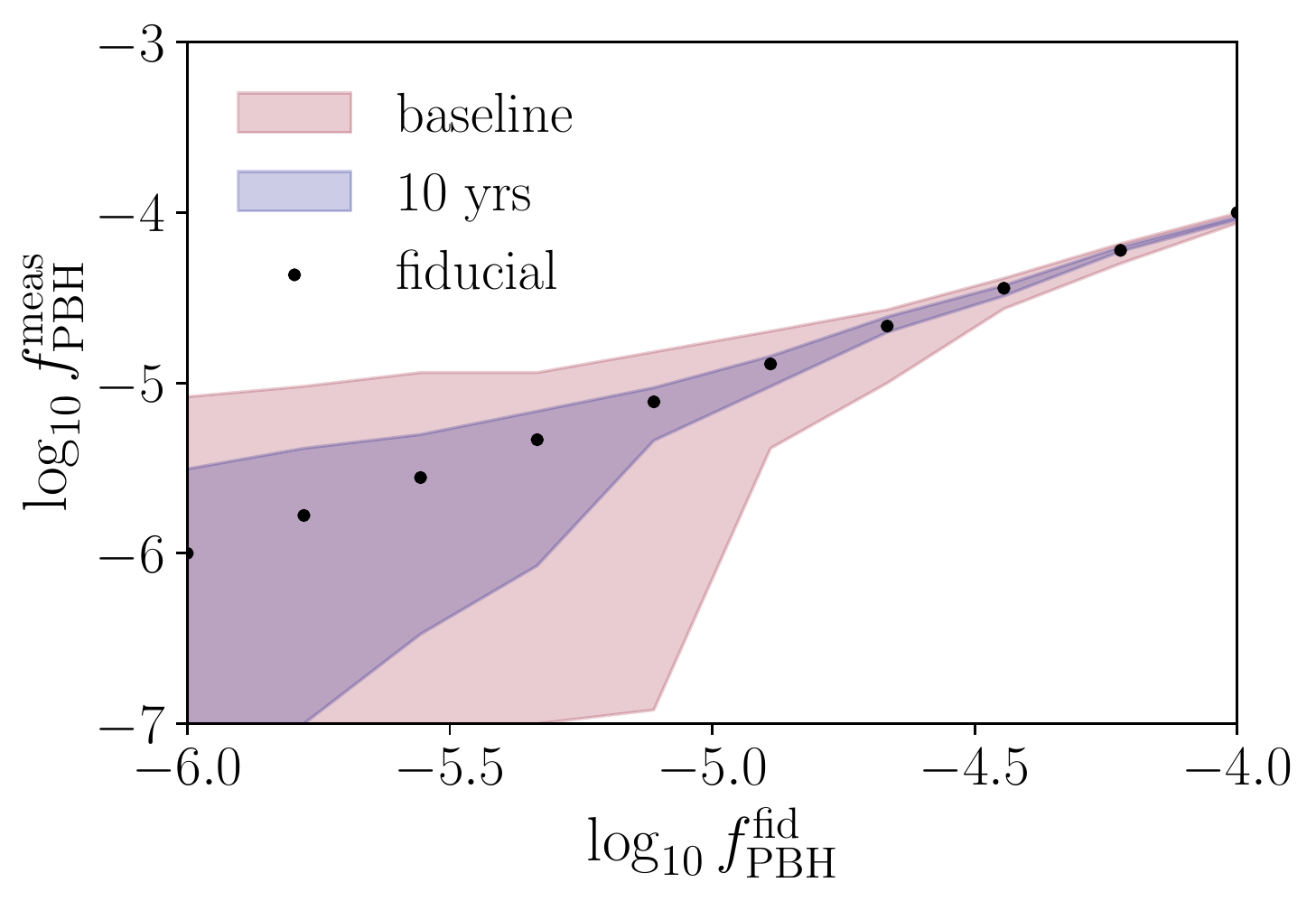}}
	\end{tabular}
	\caption{{\bf Left:} recovered mean value $f_{\rm PBH}^{\rm meas}$ (solid line) and $68\%$ and $99.7\%$ confidence level limits (red bands) as a function of the fiducial $f_{\rm PBH}^{\rm fid}$ (black dots). {\bf Right:} posterior distributions obtained on $f_{\rm PBH}^{\rm meas}$ for low (green), intermediate (blue) and high (navy) values of $f_{\rm PBH}^{\rm fid}$, with the vertical dotted line showing the value of $f_{\rm PBH}^{\rm fid}$ corresponding to each color. {\bf Bottom:} Comparison between the $99.7\%$ ($3\sigma$) confidence regions for the baseline case (outer band) and for an observation time of $T_{\rm obs}=10$ yrs (inner band).}\label{fig:bradleyplot}
\end{figure}

In this section we discuss  the bounds on $f_{\rm PBH}$ that ET can potentially obtain, as a function of the ``true'' value of this parameter, using the likelihood-based method.
In order to estimate the constraints, we generate 10 mock data sets with fiducial values $f^{\rm fid}_{\rm PBH}$ logarithmically distributed in the range $\left[10^{-6},10^{-4}\right]$. 
For each fiducial value of $f_{\rm PBH}^{\rm fid}$, we compute the likelihood associated to each data set, and obtain the posterior distribution function applying \cref{eq:PDF}. We define the mean of the distribution as the ``measured value'' $f^{\rm meas}_{\rm PBH}$. We show the results of this analysis in \cref{fig:bradleyplot}.
In the top left panel, we visualise the $68\%$ and $99.7\%$ confidence level bounds (i.e. $1\sigma$ and $3\sigma$) on the measured PBH fraction for each of the chosen $f^{\rm fid}_{\rm PBH}$ in our baseline settings, together with the mean value obtained for $f_{\rm PBH}^{\rm meas}$. We interpolate between the results obtained for our ten values of $f_{\rm PBH}^{\rm fid}$, in order to visualise the trend of these bounds.

We notice how the qualitative behaviour of these results is similar to that of the previous method; the analysis highlights how for low fiducial values ($f_{\rm PBH}\lesssim10^{-5}$) the ET is not able to detect the presence of PBHs, and only an upper bound can be placed on $f_{\rm PBH}$. Choosing a threshold of $3\sigma$ for a non-vanishing value of $f_{\rm PBH}^{\rm meas}$ to be considered a detection, as we did in \cref{sec:cut_and_count}, we find that ET will be able to detect the presence of PBHs for $f_{\rm PBH}^{\rm fid}\approx7\times10^{-6}$. Such a value is the result of interpolating between the $99.7\%$ confidence level lower bounds obtained for the different data sets, and then finding for which value of $f_{\rm PBH}^{\rm fid}$ this function would exclude the no PBHs case (set to be at $f_{\rm PBH}=10^{-7}$). Such a value is roughly three times lower than what we found with the cut-and-count method, highlighting how using the full amount of information present in the data set helps to boost the survey sensitivity.

For higher values of $f^{\rm fid}_{\rm PBH}$, a detection is possible and the constraining power of ET increases with higher fiducial values.
In the right panel of \cref{fig:bradleyplot}, we show the posterior distribution of $f_{\rm PBH}^{\rm meas}$ for three different $f_{\rm PBH}^{\rm fid}$; here we highlight how for low $f_{\rm PBH}^{\rm fid}$, the posterior distribution is extremely flat at low $f_{\rm PBH}$, as the ET cannot distinguish between fractions of PBHs that produce a very low number of events. Moving towards higher $f_{\rm PBH}^{\rm fid}$, the posterior becomes increasingly peaked, showing how these PBH abundances could be measured with very high precision.

These results highlight how, should the Universe contain a high enough number of PBHs, the method proposed in this work will be able to constrain the value of $f_{\rm PBH}$. For example, considering $f_{\rm PBH}^{\rm fid}=1.3 \times 10^{-5}$, i.e. the first generated data set above the detection threshold of $f_{\rm PBH}\approx7\times10^{-6}$ found above, we find that one can obtain a measurement of $f_{\rm PBH}^{\rm meas}$ with a precision of $\SI{35}{\percent}$ at $\SI{68}{\percent}$ confidence level.

Finally, we compare the baseline results with what can be achieved increasing the observation time to $T_{\rm obs}=\SI{10}{yrs}$. This comparison is depicted in the bottom panel of \cref{fig:bradleyplot}, where the baseline bound is shown in red, while the result for the extended observation time is shown in navy. Our results confirm that extending the survey time allows the detection of lower fractions of PBH; specifically, the smallest detectable $f_{\rm PBH}^{\rm fid}$ in the ten year case is about five times smaller than in the one year case. For a given detectable value of $\fPBH$, the uncertainty on its measurement is also reduced; for instance, with $f_{\rm PBH}^{\rm fid}=1.3\times10^{-5}$ we find that $f_{\rm PBH}^{\rm meas}$ can be obtained with a precision of $\SI{11}{\percent}$ with $\SI{68}{\percent}$ confidence in the ten year case.

\section{Discussion}
\label{sec:discussion}

The rationale of the two statistical frameworks presented here is to present complementary approaches, and emphasise different aspects of the analysis of GWs from BH merger events as a tool to detect and measure a PBH population. In particular, the cut-and-count method is especially designed to intuitively capture and visualise the essential aspect of the problem, namely the potential discovery of an anomalous excess of high redshift events beyond some reference distance. On the other hand, the likelihood-based approach exploits the capability of a Bayesian framework while using all the available information encoded in the redshift distribution of the events. 
In both cases, the characterisation of the instrument response plays a crucial role, and the main results we obtained revealed the capacity of the ET to disentangle astrophysical and primordial BH merger events on solid statistical grounds.

However, when searching for signatures of new physics against a highly uncertain astrophysical background, a key question is whether such signatures remain undetected simply due to the limitations of instrumental sensitivity or instead due to poor understanding of either the background or the signal itself, and the \emph{systematic} uncertainties associated with each. 
In the search for PBHs we are considering here, the modelling of both the background and the signal poses a clear challenge.

Beginning with the signal, our model relies on a set of simplifying assumptions which includes: \emph{(i)} a statistically \emph{uniform} initial spatial distribution of PBHs (i.e. no initial clustering); and \emph{(ii)} a \emph{monochromatic} mass function (i.e. all the PBHs have the same mass). Regarding \emph{(i)}, as pointed out in \cite{DeLuca:2021hde}, initial clustering may significantly change the picture: the authors of that study aimed to find the minimum PBH abundance testable by future detectors and concluded that abundances as small as $f_{\rm PBH} \sim 10^{-10}$ can be probed, if PBHs are highly clustered at formation; in the case of no initial clustering, they estimated that values of $f_{\rm PBH} \sim 10^{-5}$ can provide at least one event per year at the ET. Our results based on the mock data generation and careful statistical analysis are compatible with the latter estimate. We leave the assessment of the role of the initial clustering, as well as a more refined analysis of the late-time clustering, to future work.

Regarding \emph{(ii)}, realistic PBH production scenarios generally predict an extended mass function rather than a monochromatic one as assumed here. Assessing the role of the mass function requires us to understand the role of the PBH mass itself in the merger rate. On the one hand, the larger the chirp mass of a PBH binary, the higher the amplitude of the corresponding GW signal,
which may result in a higher SNR, and hence an increased detectability of the PBH population. However, changing the chirp mass also affects the frequency range of the GW emitted by the merger. Our choice for the PBH mass, $M_{\rm PBH}=\mathcal{O}(\SI{10}{\solarmass})$, roughly coincides with an optimal detectability of high-redshift events, given the ET noise curve.\footnote{Note however that the optimal PBH mass depends on the precise value of the event's redshift.} On the other hand, for a fixed \fPBH, the larger the PBH mass, the fewer PBHs, and hence the rarer the merger events. Therefore, if we consider a broad mass function, we expect that a high-mass tail would be associated with rare but (potentially) high-SNR events, while a low-mass tail would correspond to abundant but low-SNR events.
Overall, for a relatively narrow mass function we do not expect our main conclusions to dramatically change, but we leave the study of more complex mass functions for future work.

Related to the above point, we chose to keep the masses of the ABH and PBH populations fixed in the statistical analysis. We have not used any information about the inferred masses of the mergers in this work, instead focusing on how the distance information can disentangle the two populations. Of course, including mass information is likely to enhance our ability to detect a PBH population, since detecting a deviation from the conventional ABH mass distribution model would be a hint at a possible population of PBHs, but that would require further assumptions about the ABH mass distribution.

The possibility to detect a subdominant population of PBHs by analysing the mass function of binary BH mergers has progressively gained momentum with the detection of many BH merger events with masses around $\SI{30}{\solarmass}$ \cite{LIGOScientific:2020kqk, LIGOScientific:2021psn}.
For example, the authors of \cite{DeLuca:2021wjr} claim that there is decisive evidence in favour of a two-population model from these data, with a second, subdominant, population peaked around $\SI{30}{\solarmass}$. This population displays a significant high-mass tail, hence its presence is  further supported by the detection of a BH with mass in the {\it pair-instability gap} region \cite{LIGOScientific:2020iuh} (several astrophysical models of BH formation in the pair-instability mass gap also exist, see e.g. \cite{LIGOScientific:2020ufj}). 

Motivated by these considerations, we emphasise that an exploration of the impact of broad mass distributions for the PBH population (possibly with multiple peaks on different mass scales~\cite{Carr:2019kxo}) on the PBH detection prospects with the ET is an important future goal in this research line. 
This would require a more careful treatment of late-time clustering than that presented in this work, given the complex phenomenology that may arise (for instance, mass segregation in clusters \cite{Trashorras:2020mwn}). 

We have assumed throughout this work that PBH binaries formed in the early Universe dominate the merger rate. Dense environments, such as PBH clusters, can potentially lead to the disruption of early Universe binaries. We have considered  the effect of late-time clustering (see \cref{sec:appendix_clustering}), which seems to be negligible for the small PBH abundances considered in this work; the impact of initial PBH clustering is left for future work. PBH binaries could also be disrupted by interactions with other objects; while it is unlikely that they will be disrupted by stellar encounters in the disks of typical galaxies~\cite{Sasaki:2016jop}, it seems feasible that they may be disrupted in sufficiently dense environments, such as close to the centre of the host galaxy. This is unlikely to affect the high redshift signals we study here, with the first haloes having little effect on early-formed binaries~\cite{Ali-Haimoud:2017rtz}. On the other hand, dense environments can lead to the formation of a population of late-time PBH binaries. This is typically expected to be subdominant~\cite{Bird:2016dcv,Ali-Haimoud:2017rtz}, in particular for the small values of \fPBH considered in this work. These late-time binaries may be more difficult to distinguish from ABH ones, as they are expected to form only at low redshift, in which case alternative approaches are necessary, such as a measurements of the clustering bias~\cite{Canas-Herrera:2021qxs}.
However, as long as the population of early-time binaries remains relatively unperturbed, high redshift observations would be able to probe it, and hence detect the presence of PBHs, even if this population were to be subdominant with respect to the late-time one. Regarding the measurability discussion, a sizable population of late-time binaries merging at relatively high redshifts could lead us to overestimate the abundance of PBHs. In neglecting this, we rely on the expectation that the merger rate due to such binaries is only relevant at low redshifts and for large PBH abundances.


There are also a number of ways in which the modelling of the astrophysical background could alter the results presented here. In particular, we emphasise that our model for the ABH evolution, characterised by a peak at $z \sim 2$ and a monotonic decrease at higher redshifts, is designed to capture the rate associated to binary systems made of second- and third-generation stars (usually called Population I and II). These systems may form via different channels, such as binary stellar evolution in galactic fields, or  dynamical formation through multi-body interactions in star clusters.

However, a significant contribution to the binary BH merger rate, including a peak at high redshift, $z  \sim \mathcal{O}(10)$, can be expected from the first-generation (Population III) stars, which formed out of the pristine gas left over after cosmological nucleosynthesis, and generated the first heavy elements in the Universe \cite{Vangioni:2014axa,deSouza2011Aaanda}. Taking into account this additional component would certainly require, again, to address the mass dependence in more detail. In fact, this population would be modelled as a high (or intermediate) mass model of star formation at high redshift, to be added to the background model~\cite{Ng:2020qpk}. An even more detailed treatment would imply a marginalisation over the parameters that describe the metallicity dependence of the SFR, treated as nuisance parameters. 
We leave this more detailed treatment to future studies, and emphasise that the numerical and statistical tools presented here are the ideal framework to consistently address uncertainties in both the signal and background models.

As we were finalising this work, Ng et al.~\cite{Ng:2022agi} proposed an independent analysis on the same topic. Reference~\cite{Ng:2022agi} explores how a network consisting of two third generation detectors (ET and CE) can be used to distinguish between ABH and PBH populations using high redshift mergers. Their analysis differs from ours in that they consider exclusively population~III ABHs and restrict their analysis to events above $z > 8$, as well as working directly with measurement errors on the event redshift, rather than on the luminosity distance, as we do here. 
Their results are broadly consistent with ours, highlighting that a PBH population with $f_\mathrm{PBH} \sim \mathcal{O}(10^{-5})$ should be within reach of third generation detectors.  We find a one year sensitivity estimate for $f_\mathrm{PBH}$ roughly a factor of three stronger, which may be partly due to differences in the distance uncertainties and population models assumed. This emphasises the importance of careful modelling of these details in quantifying the PBH detection and measurement potential of future GW detectors. 

\section{Conclusions}
\label{sec:conclusion}

In this article we have assessed the capability of the forthcoming Einstein Telescope (ET) gravitational wave (GW) observatory to detect and measure a subdominant population of primordial black hole (PBH) mergers using a novel statistical framework. 

We have described a procedure that computes the redshift evolution of both the expected background associated with the astrophysical black hole (ABH) merger events, and the signal under consideration associated with the PBH population (assumed to be characterised by masses of the same order of magnitude). In our modelling, we paid particular attention to the impact of late-time clustering of PBHs. We set the relative normalisation of the two contributions by comparison with the latest (low redshift) data released by the LIGO, Virgo and KAGRA collaborations. This procedure naturally provided an updated upper limit on the fraction of dark matter in the form of PBHs (quantified using the parameter \fPBH), based on the third Gravitational Wave Transient Catalog (GWTC-3) -- see \cref{fig:LIGObound}.  

The key feature of our merger rate models is the different behaviour at high redshift of signal (i.e.\ PBH mergers) and background (i.e.\ ABH mergers), with the former monotonically increasing with increasing distance, and the latter steadily decreasing. Motivated by this qualitative aspect, and taking into account the expected high sensitivity of ET, we presented two methods to assess the potential of discovering a distant PBH population, which both accurately take into account the experimental sensitivity. Both methods are based on the generation and analysis of a set of mock data catalogues, associated to the null hypothesis (no PBHs) and to different values of the fraction of DM in the form of PBHs.

The first method (dubbed ``cut-and-count'') relies on a two-redshift-bin data analysis strategy that highlights the PBH contribution as a significant excess in the high redshift bin. We demonstrated that a PBH fraction as low as $f_{\rm PBH}\approx2\times10^{-5}$ can be detected with a 3$\sigma$ significance, given a proper choice of the cut position -- see \cref{fig:detection_zT}.

The second method takes a Bayesian approach to the problem, and fully exploits the information on the redshift dependence of both the signal and background. Within this framework, we provided an assessment of the posterior probability density function for different fiducial values of \fPBH. We found that fractions of DM in the form of PBHs as low as $\mathcal{O}(10^{-5})$ can be measured by ET with one year of data taking, with a precision of $\sim\SI{35}{\percent}$ at $\SI{68}{\percent}$ confidence level -- see \cref{fig:bradleyplot}. These results demonstrate a well-defined avenue towards a discovery or, in the absence of a discovery, the prospect of setting a significantly improved upper limit on the existence of PBHs in this mass window.

Furthermore, in the process of this work, we have developed the \texttt{darksirens} \href{https://gitlab.com/matmartinelli/darksirens}{\faGitlab} code, which allows for the fast generation of realistic mock GW catalogues sourced by ABH and PBH merger events. We have made the code publicly available for community use, with extensive scope for development and application to other investigations.

The final message of this work is clear: the possibility of detecting and measuring a distant population of PBHs is well within the capabilities of future GW observatories such as the ET. The exciting prospect that these exotic objects could constitute a fraction of the dark matter, a substance that despite its fundamental importance in the Universe we still know so little about, makes the pursuit of this goal nothing short of vital. In the rigorous analysis we have here presented, based on the twin pillars of robust statistics and a thorough treatment of the modelling of both signal and noise, we have shown that the ET will be able to measure a fraction of PBHs as low as $\mathcal{O}(10^{-5})$, through observations of their luminosity distances alone. If a high redshift population of PBHs exists in our Universe, the detection of these distant dark sirens is therefore something we no longer need to hope for, but can begin to expect.

\section*{CRedIT statement} 

\textbf{Matteo Martinelli}: Conceptualization; Methodology; Software; Writing -- original draft; Writing -- review \& editing; Supervision; Funding acquisition.
\textbf{Francesca Scarcella}: Software; Writing -- original draft; Writing -- review \& editing.
\textbf{Natalie B. Hogg}: Software; Data Curation (code documentation); Writing -- original draft; Writing -- review \& editing.
\textbf{Bradley J. Kavanagh}: Methodology; Software (likelihoods); Writing -- original draft; Writing -- review \& editing; Project administration.
\textbf{Daniele Gaggero}: Conceptualization; Formal analysis (astrophysical background, PBH upper limit); Software (astrophysical background, PBH upper limit); Writing -- original draft, review \& editing; Supervision; Funding acquisition.
\textbf{Pierre Fleury}: Conceptualization; Formal analysis (lensing); Software (lensing); Writing - review \& editing; Supervision; Funding acquisition.

\acknowledgments
We are very grateful to Evan Hall for making the \href{https://git.ligo.org/evan.hall/gw-horizon-plot}{gw-horizon} code public and open. We thank Tjonnie Li, Michele Maggiore and Bangalore Sathyaprakash for useful discussions, and Cristina Fern\'andez Su\'arez for contributions in the early stages of this work.

MM acknowledges funding by the Agenzia Spaziale Italiana (ASI) under agreement n. 2018-23-HH.0.
FS was supported by the Spanish Agencia Estatal de Investigación through the grants IFT Centro de Excelencia Severo Ochoa CEX2020-001007-S and PGC2018-095161-B-I00, and, during the early stages of this work, through the grants Severo Ochoa SEV-2016-0597 and Red Consolider MultiDark FPA2017-90566-REDC.
FS has received financial support through la Caixa Banking Foundation (grant  n.~LCF/BQ/LI18/11630014) during the early stages of the project.
NBH is supported by a postdoctoral position previously funded through two ``la Caixa'' Foundation fellowships (ID00010434), with codes LCF/BQ/PI19/11690015 and LCF/BQ/PI19/11690018 respectively, and currently funded by the French Commissariat à l’énergie atomique et aux énergies alternatives (CEA).
BJK thanks the Spanish Agencia Estatal de Investigaci\'on (AEI, MICIU) for the support to the Unidad de Excelencia Mar\'ia de Maeztu Instituto de F\'isica de Cantabria, ref. MDM-2017-0765.
DG has received financial support through the Postdoctoral Junior Leader Fellowship Programme from la Caixa Banking Foundation (grant n.~LCF/BQ/LI18/11630014) during the early stage of the project.
DG was also supported by the Spanish Agencia Estatal de Investigaci\'{o}n through the grants PGC2018-095161-B-I00, IFT Centro de Excelencia Severo Ochoa SEV-2016-0597, and Red Consolider MultiDark FPA2017-90566-REDC during the early stages of the project.
DG acknowledges funding from the “Department of Excellence” grant awarded by the Italian Ministry of Education, University and Research (MIUR) in October-December 2021.
DG also acknowledges support from the INFN grant “LINDARK,” and the project “Theoretical Astroparticle Physics (TAsP)” funded by the INFN in October-December 2021.
DG acknowledges support from Generalitat Valenciana through the plan GenT program (CIDEGENT/2021/017) starting from 01/01/2022.
In the early stages of this work, MM and PF received the support of a fellowship from ``la Caixa'' Foundation (ID 100010434). The fellowship codes are LCF/BQ/PI19/11690015 and LCF/BQ/PI19/11690018 for MM and PF respectively.

\appendix 

\section{Merger rate suppression due to PBH clustering }
\label{sec:appendix_clustering} 

In this appendix, we discuss the suppression of the PBH merger rate due to PBH clustering. We follow the numerical implementation given in Ref.~\cite{Vaskonen:2019jpv} but here we provide additional details which are useful for understanding why the suppression of the merger rate is negligible for the ranges of $f_\mathrm{PBH}$ allowed by present bounds.

\subsection{PBH cluster formation}

Even if PBHs are uniformly distributed in the early Universe (i.e.\ according to a Poisson distribution), their discrete nature means that the mean PBH density in a given region can be subject to large fluctuations. Regions with overdensities of PBHs begin to collapse early, forming PBH \textit{clusters}~\cite{Chisholm:2005vm,Chisholm:2011kn,Inman:2019wvr}. Here, we briefly review this mechanism, before commenting on the relevance for the survival of PBH binaries. 

At matter--radiation equality, $t_{\rm i} = t_\mathrm{eq}$, we consider a spherical region of radius $r_{\rm i}$, containing an overdensity $\delta_{\rm i} = (\rho_{\rm i} - \rho_\mathrm{eq})/\rho_\mathrm{eq}$. This overdense region will evolve as an isolated matter-dominated Universe with effective curvature~\cite{Gunn:1972sv,1984ApJ...281....1F,1985ApJS...58...39B},
\begin{equation}
\mathcal{K}
= \frac{8 \pi G \rho_\mathrm{eq} \delta_{\rm i} r_{\rm i}^2}{3}
= \frac{2 G \bar{M}_{\rm i} \delta_{\rm i}}{r_{\rm i}}\,,
\end{equation}
where $\bar{M}_{\rm i} = (4 \pi/3) G \rho_\mathrm{eq} r_{\rm i}^3$ would be the initial mass of the region in the unperturbed background. The evolution of the radius of this region $r(t)$ can be solved parametrically as
\begin{align}
r(\theta) &= \frac{G M_{\rm i}}{\mathcal{K}}(1-\cos \theta)\,, \\
t(\theta) &= \frac{G M_{\rm i}}{\mathcal{K}^{3 / 2}}(\theta-\sin \theta)\,.
\end{align}
The region collapses then at
\begin{align}
    t_\mathrm{coll} = t(\theta = 2\pi) = \frac{2\pi G M_{\rm i}}{\mathcal{K}^{3/2}}\,,
\end{align}
or, using the fact that $z \propto t^{-2/3}$ in a matter-dominated Universe:
\begin{align}
    z_\mathrm{coll} = z_\mathrm{eq} \left(\frac{t_\mathrm{eq}}{t_\mathrm{coll}}\right)^{2/3} = z_\mathrm{eq} \mathcal{K} \left(\frac{t_\mathrm{eq}}{2\pi G M_{\rm i}}\right)^{2/3}\,.
\end{align}
We can relate $t_\mathrm{eq}$ and $r_{\rm i}$ by assuming that the size of the region corresponds to its comoving size at matter--radiation equality. With this, we find the redshift of collapse as
\begin{equation}
    z_\mathrm{coll} = \frac{2}{(18\pi^2)^2} \, z_\mathrm{eq} \,
    \frac{\delta_{\rm i}}{(1+\delta_{\rm i})^{2/3}}
    \approx 0.36 \, z_\mathrm{eq} \, \delta_{\rm i}\,,
\end{equation}
for $\delta_{\rm i} \ll 1$. The redshift at which the expansion of the region turns around is a little larger, $z_\mathrm{ta} \approx 0.56 z_\mathrm{eq} \delta_{\rm i}$.

Consider now a region of the Universe which we expect to contain, on average, $N$ PBHs with total PBH mass $M = N M_\mathrm{PBH}$. If a cluster forms from this collection of PBHs, it is likely to form from a typical overdensity in PBHs of $\delta_\mathrm{PBH} = \sqrt{N}/N$, due to Poisson fluctuations. If PBHs make up only a fraction $f_\mathrm{PBH}$ of the total DM density, then the total overdensity will be $\delta_{\rm i} = f_\mathrm{PBH}/\sqrt{N}$ (assuming that the DM fluctuations are subdominant). Dropping some order one factors, we therefore expect that the PBH cluster to form at a redshift
\begin{equation}
    z_\mathrm{coll} \approx z_\mathrm{eq} f_\mathrm{PBH}/\sqrt{N}\,.
    \label{eq:z_coll}
\end{equation}
While this expression is a good estimate of collapse time for large values of $N$, it fails for smaller values. In our numerical analysis, we rely on the more complete approach discussed in \cite{Inman:2019wvr}.
However, in the following sections we will make use of \cref{eq:z_coll} to estimate the typical size of clusters which are disrupted at late times.

\subsection{Relaxation time}

We will assume that a PBH binary is completely disrupted if the PBH cluster it resides in undergoes core collapse due to gravo-thermal instability and that all other PBH binaries are unperturbed. We take the characteristic core-collapse timescale to be $t_{\rm cc} \sim 18 \, t_{\rm r}$~\cite{Quinlan:1996bw}, where $t_{\rm r}$ is the relaxation time. 

The relaxation time $t_{\rm r}$ for a PBH cluster containing $N$ PBHs. For a system with density $\rho$ and component masses $m$, this relaxation time can be estimated as~\cite{Quinlan:1996bw}
\begin{equation}
    t_{\mathrm{r}}=0.065 \, \frac{ v^{3}}{ G^2 m \rho \ln \Lambda}\,,
\end{equation}
where $\ln\Lambda \approx \ln(N/f_{\rm PBH})$ is the Coulomb logarithm associated with interactions between component masses. Assuming that the fraction of PBHs in clusters matches the mass fraction in the Universe, then the total mass of each cluster is $M = m N/f_\mathrm{PBH}$.

The mean density of a cluster will be
\begin{equation}
\bar{\rho}
\equiv \frac{3 M}{4\pi R^3}
\approx 18 \pi^2\rho_c a^{-3}
= 18 \pi^2 \rho_{\rm c} \, a_\mathrm{eq}^{-3}\,\frac{f_\mathrm{PBH}^{3}}{N^{3/2}}\,,
\label{eq:MeanDensity}
\end{equation}
where the second equality follows from the theory of spherical collapse~\cite{1980lssu.book.....P} and from setting $a = 1/z_\mathrm{coll}$ (the scale factor at which the cluster forms). Here, $\rho_{\rm c}$ is the critical density.

Next, we set the PBH velocity $v$ in the cluster equal to some typical velocity dispersion $v \approx \sigma_v \approx \sqrt{GM/R}$. Using the definition of $\bar{\rho}$ in Eq.~\eqref{eq:MeanDensity}, we then have
\begin{align}
\sigma_v^3
= \left(\frac{GM}{R}\right)^{3/2}
= \sqrt{\frac{4\pi}{3}} G^{3/2} M \sqrt{\bar{\rho}}\,.
\end{align}

Assuming that there is no substantial growth or evaporation of the cluster after formation, the relaxation timescale can then be written as
\begin{align}
t_r &= 0.065 \, \frac{\sigma_v^3}{G^2 m \bar{\rho}\ln\Lambda}
    = 0.065 \, \sqrt{\frac{4\pi}{3}}
            \frac{M}{\sqrt{G \bar{\rho}} \, m \ln \Lambda}\\
&= 0.065 \, \sqrt{\frac{2}{27\pi}}
        \frac{1}{\sqrt{G\rho_{\mathrm{m, eq}}}}  \frac{N^{7/4}}{f_\mathrm{PBH}^{5/2}\ln\Lambda}
= \SI{2.1}{\kilo\year}\,\frac{N^{7/4}}{f_\mathrm{PBH}^{5/2}\ln\Lambda}\,.
        \label{eq:relaxationtime}
\end{align}
Here, we have defined $\rho_{\mathrm{m,eq}} = \rho_{\rm c} a_\mathrm{eq}^{-3}$ (the matter density at matter--radiation equality). We have also taken the numerical values $\rho_{\rm c} = 2.78 \times 10^{11} h^2~M_\odot~\si{\per\mega\parsec\cubed} $~\cite{Zyla:2020zbs} and $H_0 = h\times \SI{100}{\kilo\meter\per\second\per\mega\parsec}$, with $h = 0.673$~\cite{Aghanim:2018eyx}. This value for the  relaxation time $t_{\rm r}$ matches the result given in Eq.~(7) of \cite{Vaskonen:2019jpv}.

\subsection{Cluster collapse}

\begin{figure}[th!]
    \centering
    \includegraphics[width=0.49\textwidth]{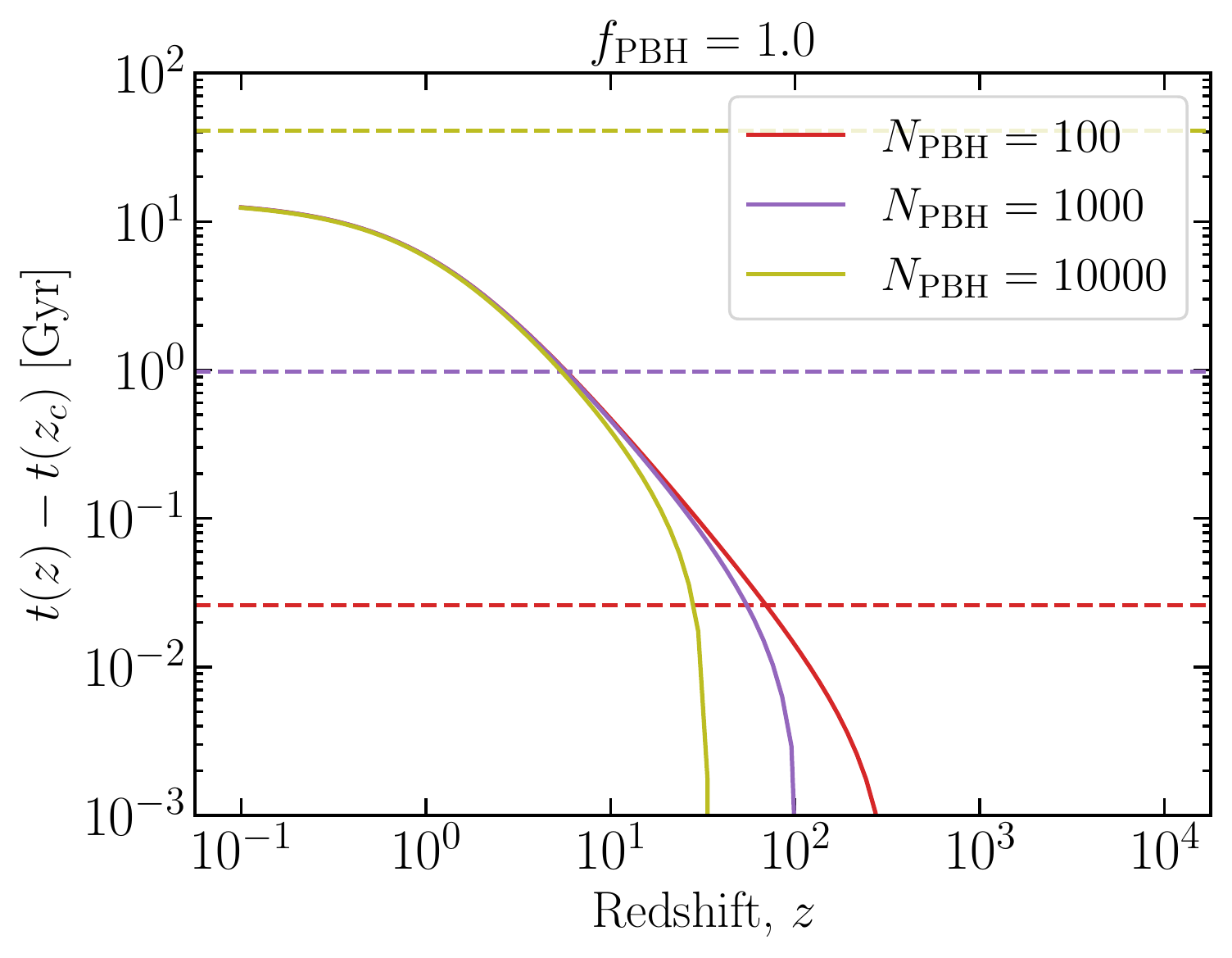}
    \hfill
    \includegraphics[width=0.49\textwidth]{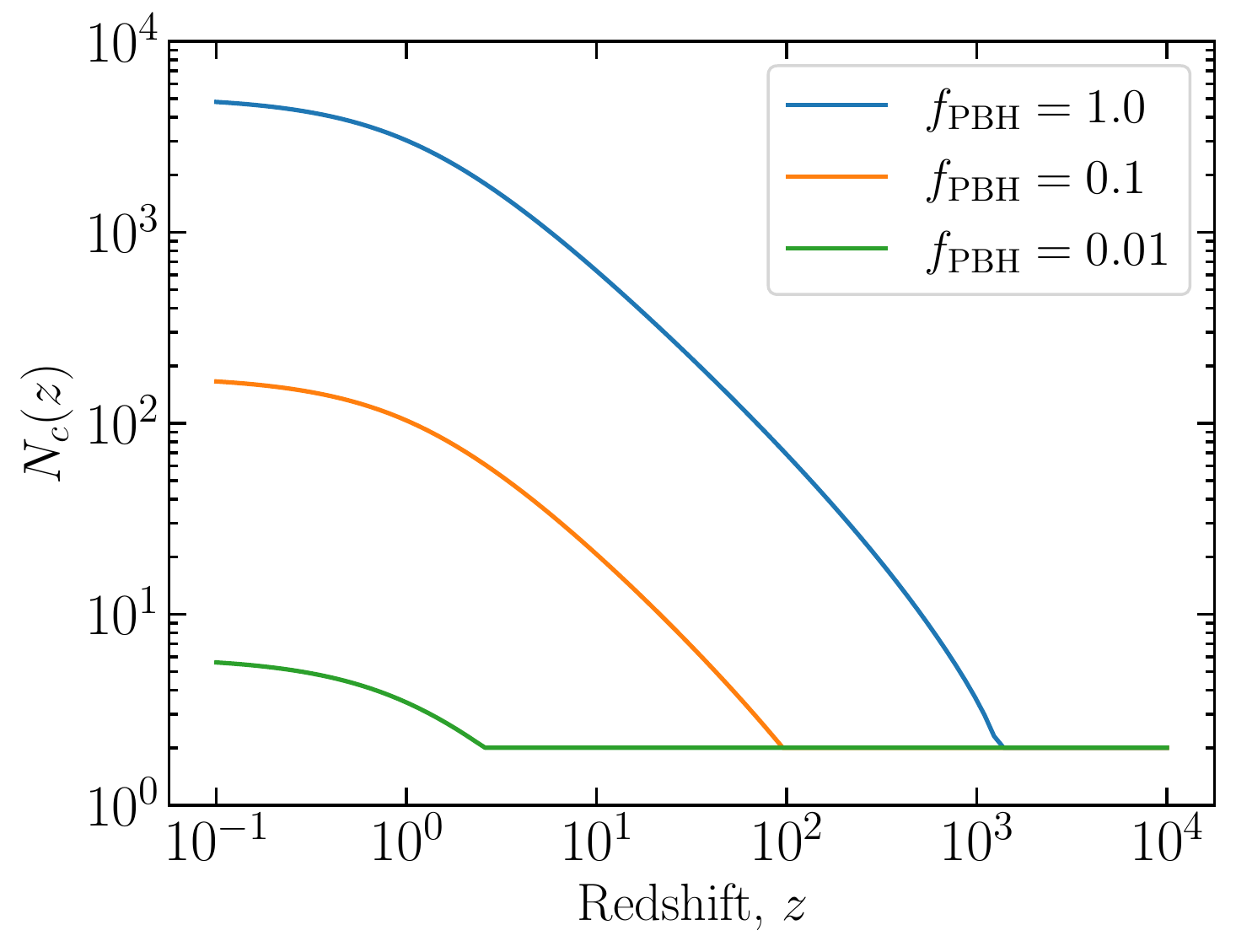}
    \caption{\textbf{Left:} Time-scales associated with PBH clusters. Solid lines show the time between the redshift of cluster formation $z_{\rm c}$ and a given redshift $z$, for clusters containing different numbers of PBHs $N_\mathrm{PBH}$. Horizontal dashed lines show the core-collapse timescale for clusters with $N_\mathrm{PBH}$ members. Where the solid and dashed lines (of a given colour) cross then $N_\mathrm{PBH} = N_{\rm c}(z)$. \textbf{Right:} Critical number of PBHs per cluster $N_{\rm c}$ as a function of redshift $z$. Clusters with $N < N_{\rm c}(z)$ will be disrupted due to core-collapse before redshift $z$.}
    \label{fig:Nc}
\end{figure}

In order to undergo core collapse before redshift $z$, a PBH cluster must be formed with fewer than a critical number of members $N_{\rm c}(z)$. This critical number $N_{\rm c}(z)$ is obtained by equating the core-collapse timescale to the time between formation and collapse,
\begin{equation}
\label{eq:Nc}
    18 \, t_{\rm r}(N_{\rm c}) = t(z) - t(z_{\rm c})\,,
\end{equation}
where $z_{\rm c}$ is the redshift at which clusters of size $N_{\rm c}$ are formed. Here, $t_{\rm r}(N_{\rm c})$ is the relaxation time for clusters of size $N_{\rm c}$, defined in \cref{eq:relaxationtime}. Recall also from the previous sections that
\begin{equation}
    z_{\rm c} \approx z_\mathrm{eq} \, \frac{f_\mathrm{PBH}}{\sqrt{N_{\rm c}}}\,.
\end{equation}
With these definitions, we can calculate $N_{\rm c}(z)$ from Eq.~\eqref{eq:Nc}. This is illustrated in Fig.~\ref{fig:Nc}. There, we see that for $f_\mathrm{PBH} = 1$, the critical size of clusters is $N_{\rm c} \sim 5000$ at $z = 0$. This decreases with decreasing $f_\mathrm{PBH}$. Below $f_\mathrm{PBH} \sim 0.01$, core-collapse becomes more or less irrelevant, as the critical number of PBHs tends to one.

\subsection{Merger rate suppression factor}

Following Ref.~\cite{Vaskonen:2019jpv}, the suppression factor is expressed as the probability of a binary not belonging to a cluster that undergoes gravo-thermal collapse. Estimating this factor requires knowledge of the PBH halo mass function, which we describe using the analytical model discussed in \cite{Inman:2019wvr}. We indicate the probability of a PBH belonging to a cluster of $N$ elements at given $z$ and \fPBH with $P(N | z, \fPBH)$. We thus have $P(N | z, \fPBH) \propto  \exp[-N/N^*(z, \fPBH)]$, where $N^*(z, \fPBH)$ is the characteristic number of PBHs in clusters forming at redshift $z$. 

\begin{figure}[th!]
    \centering
    \includegraphics[width=0.49\textwidth]{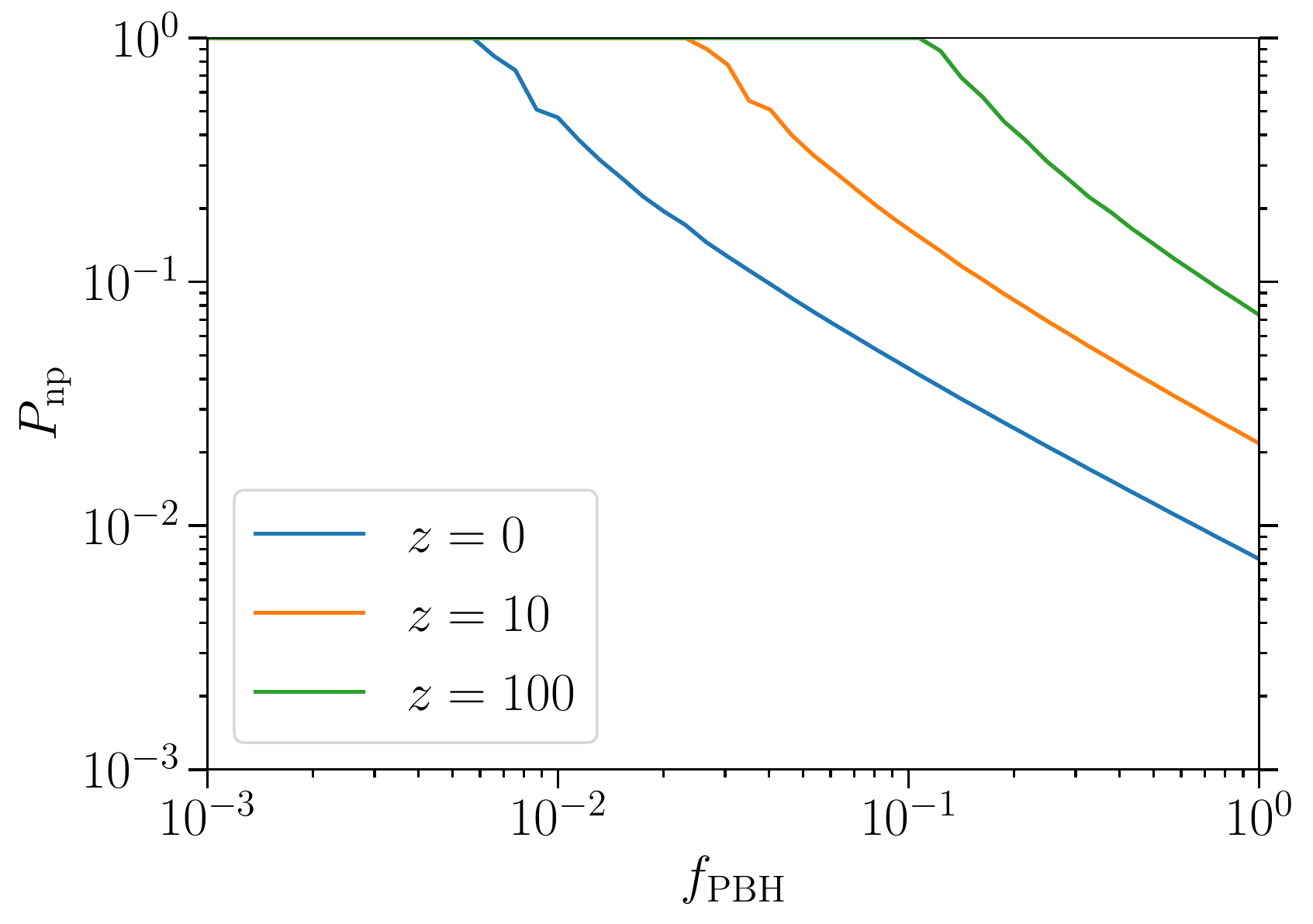}
    \hfill
    \includegraphics[width=0.49\textwidth]{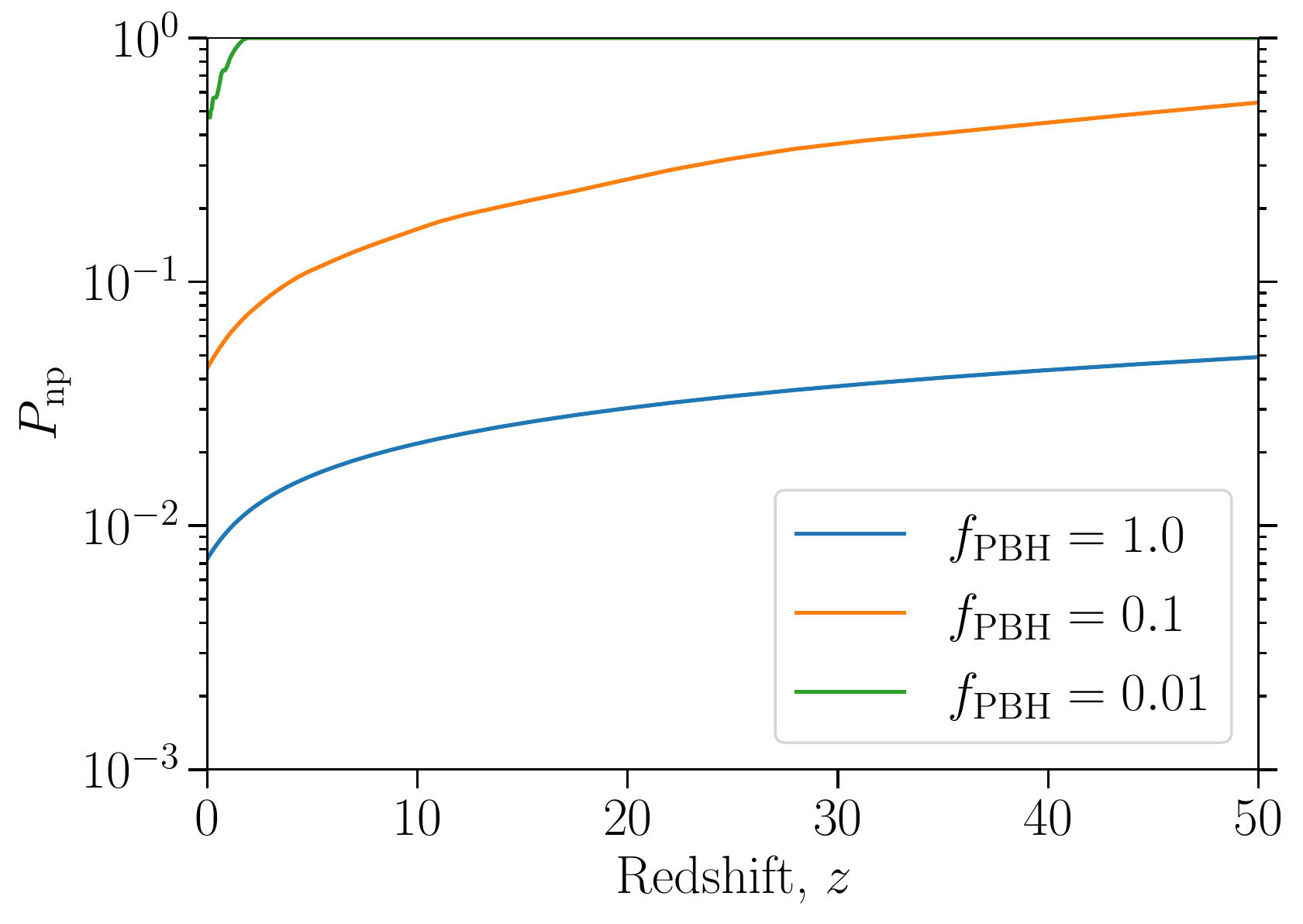}
    \caption{ Merger rate suppression factor due to interactions in gravo-thermally unstable clusters, computed following \cite{Vaskonen:2019jpv}. \textbf{Left:} Probability of  a binary \textit{not} being disrupted as a function of \fPBH for different redshifts. Notice that, depending on the value of $z$, there is a threshold value of \fPBH below which no binaries are perturbed.   \textbf{Right:} Same probability as a function of redshift, for different values of \fPBH.  }
    \label{fig:Pnp}
\end{figure}

The probability of a binary being disrupted by redshift $z$ is given, according to \cite{Vaskonen:2019jpv}, by the sum of two terms:
\begin{enumerate}
\item The probability of the binary belonging to a cluster which has reached instability before redshift $z$, i.e. a cluster with $N \leq N_{\rm c} (z)$,
\begin{equation}
  P_{\rm p}^{(1)} = \sum\limits_{N=3}^{N_{\rm c}} P_{\rm bin}(N|z_{\rm c}) \;,
\end{equation}
where the probabilities of binaries belonging to a cluster $P_{\rm bin}(N)$ are taken to be $\propto P(N)$ but normalised to exclude $P(N = 1)$:
\begin{equation}
\sum\limits_{N=2}^{\infty} P_{\rm bin}(N) =  1 .
\end{equation}

\item The probability of a binary belonging to a sub-cluster of $N \leq N_{\rm c}$ within a larger cluster ($N > N_{\rm c}$), 
\begin{equation}
P_{\rm p}^{(2)} = \sum\limits_{N>N_{\rm c}} \left( \sum\limits_{N'=3}^{N_{\rm c}} P_{\rm sub}(N'|z_{\rm c}) \right) P_{\rm bin}(N|z_{\rm c}) \; ,
\end{equation}
\end{enumerate}
where the probabilities of belonging to a sub-cluster $P_{\rm sub}(N)$ are taken to be proportional to $P(N)$, but normalised summing up to the size of the cluster that contains them
\begin{equation}
\sum\limits_{N' = 2}^{N} P_{\rm sub}(N') =  1 .
\end{equation}
All the probabilities in the expressions above are evaluated at $z=z_{\rm c}$, which is the typical redshift of formation of the clusters with $N_{\rm c}$ elements. Then $z_{\rm c}$ is obtained by requiring that $N^*(z_{\rm c}) = N_{\rm c}$. Finally, the probability of a binary not being perturbed is given by $ P_\mathrm{np} = 1 - P_{\rm p}^{(1)}- P_{\rm p}^{(2)} $. This suppression factor is shown in Fig.~\ref{fig:Pnp} for different values of $z$ and \fPBH.

\section{Obtaining the signal-to-noise ratio of an event} \label{sec:appendix_mocks}

In this appendix, we provide a detailed description of how our mock catalogues are constructed. We start by introducing GW signals in \cref{subsec:appendix_gw}, move to a discussion of the luminosity distance from GWs in \cref{subsec:appendix_dl} and finally the computation of the uncertainty on the luminosity distance in \cref{subsec:appendix_sigmadl}.

\subsection{Gravitational waves}\label{subsec:appendix_gw}
The merging of massive compact objects results in the emission of GWs. These can be detected by the strain they produce in GW interferometers, housed in observatories like LIGO and Virgo. Let us note the different terms we use in the following discussion: the term GW \textit{detector} refers to the equipment used to measure the strain itself; the term GW \textit{interferometer} refers to the combination of two separate vacuum tubes or ``arms'' housing the laser and mirrors with a detector at the junction of the arms; and the term GW \textit{observatory} refers to an entire facility, which may contain multiple detectors, or as in the case of LIGO, multiple interferometers at different physical locations (LIGO Livingston and LIGO Hanford).

The GW strain is given in the transverse traceless gauge as 
\begin{equation}\label{eq:hpatterns}
    h(t) = F_+ (\theta, \phi, \psi) \, h_+(t) + F_\times (\theta, \phi, \psi) \, h_\times (t),
\end{equation}
where $h_{+,\times}$ are the plus and cross polarisations of the metric perturbation $h_{\alpha\beta}$, $F_{+,\times}$ are the corresponding antenna pattern functions which describe the angular dependence of the sensitivity of the interferometers. The angles $(\theta, \phi)$ are the spherical polar coordinates of the wave source in the celestial sphere, $\theta=\pi/2$ being the detector plane; $\psi$ denotes the angle between the two-dimensional basis of the plane orthogonal to the line of sight with respect to which the plus and cross polarisations are defined, and the natural basis~$(\vect{e}_\theta, \vect{e}_\phi)$ associated with the celestial coordinates~$(\theta, \phi)$~\cite{Maggiore2007}. More simply put, $\psi$ is the polarisation angle.

For the results presented in this work, we generate mock catalogues of GW events based on the configuration and sensitivity of the ET, a proposed third-generation GW observatory. Specifically, we consider the ET-D configuration, which involves three interferometers arranged in an equilateral triangle shape and two detectors at each vertex of the triangle, one sensitive to higher frequencies and one to lower, i.e. six detectors in total. The antenna patterns for a single interferometer with a $60$ degree opening angle are given by
\begin{align}
    F_+ &= \frac{\sqrt{3}}{2}\left[\frac12 (1 + \cos^2 \theta) \cos(2\phi) \cos (2\psi) - \cos\theta \sin(2\phi) \sin(2\psi) \right],\\
    F_\times &= \frac{\sqrt{3}}{2}\left[\frac12 (1 + \cos^2\theta) \cos(2\phi) \sin(2\psi) + \cos\theta \sin(2\phi) \cos(2\psi) \right].
\end{align}
The antenna patterns for the two other interferometers in the ET-D configuration are then given by $F_{+, \times}(\theta, \phi + 2\pi/3, \psi)$ and $F_{+, \times}(\theta, \phi + 4\pi/3, \psi)$.

We assume that the basic information available from the detection of GWs from merging compact objects by the ET will be the luminosity distance to the merger event, $D$, and the uncertainty on that luminosity distance, $\sigma$, which has contributions from the instrumental noise and from the uncertainty due to weak lensing of the GW.

\subsection{Luminosity distance} \label{subsec:appendix_dl}

To obtain the luminosity distance for a given simulated event, we firstly draw a redshift from a probability distribution defined based on the merger rate of the progenitor system being considered (binary PBHs or binary PBHs), which we previously described in \Cref{sec:merger_rates}.

However, unlike binary neutron star mergers which produce an electromagnetic counterpart, the redshift of an individual\footnote{There have been proposals to cross-correlate GW events with galaxy catalogues (see e.g. \cite{Canas-Herrera:2019npr,Mukherjee2021}), allowing the redshift of the event to be estimated, but this introduces a good deal of uncertainty, especially with the current small number of events and relatively poor sky localisation.} binary BH merger is always unknown -- hence the name \textit{dark siren}. We therefore convert the simulated redshift to a luminosity distance in the standard $\Lambda$CDM cosmology using \CAMB. All of the analysis presented in the main text is based solely on the luminosity distance information, rather than the redshift information, in order to accurately simulate the actual analysis that could be done with a catalogue which contains only binary BHs.

Finally, we rescale the luminosity distance to account for weak lensing of the signal. The full computation is shown in \Cref{sec:appendix_lensing}.

\subsection{Uncertainty on the lensed luminosity distance} \label{subsec:appendix_sigmadl}

The second step in the generation of the mock catalogue is then to compute the uncertainty on the lensed luminosity distance for a given event. As we mentioned, this has two components: the instrumental noise and the noise due to weak lensing of the GW. The instrumental noise depends on the signal-to-noise ratio for a given event.

The unlensed signal-to-noise ratio for a single detector is given by 
\begin{equation}
   \rho_{\rm det} = \frac{\int^\infty_{-\infty} \mathrm{d}f \; h(f) \, K^*(f)}{\sqrt{\int^\infty_{-\infty} \mathrm{d}f \; \frac12 \, S_n(f) \, |K(f)|^2}},
\end{equation}
where $h(f)$ is the waveform of the GW (the strain as a function of the frequency~$f$), $K(f)$ is the filter used in the \textit{matched filtering} process which maximises the signal-to-noise ratio and $S_n$ is the noise power spectral density \cite{Li:2013lza}. Asterisks denote the complex conjugate. The optimal unlensed signal-to-noise ratio is obtained using a Wiener filter, and is given by
\begin{equation}
    \rho_{\rm opt} =  \left[4 \int^{f_{\rm upper}}_{f_{\rm lower}} \mathrm{d}f \, \frac{h(f) h^*(f)}{S_n(f)} \right]^{\frac12},
\end{equation}
where $f_{\rm upper}$ and $f_{\rm lower}$ are the cutoff frequencies for the strain, beyond which it is assumed to be zero. The waveform $h(f)$ is the Fourier transform of the strain $h(t)$ given by \cref{eq:hpatterns}. We randomly draw the angles $\phi$ and $\psi$ from a uniform distribution between $0$ and $2\pi$. However, the angles $\theta$ and $\iota$ (the inclination of the event i.e. the angle between the source plane and the detector plane) should have their cosine uniformly distributed, meaning that for these quantities we draw their values randomly from a uniform distribution between $-1$ and $1$ and then take the arccosine of the result. The inclination of the event enters into the function used to weight the final signal-to-noise ratio sum, as we will see in a moment.

We use the publicly available package \texttt{PyCBC}\footnote{\url{https://pycbc.org/}.} to generate a mock waveform $h(f)$ for a given event, using the \texttt{IMRPhenomD} waveform model, and by inputting the masses and simulated luminosity distance of the event. We keep the spins fixed to zero. We also generate the frequency range, and hence $f_{\rm upper}$ and $f_{\rm lower}$, using \texttt{PyCBC}.

Assuming that the noise can be described by a Gaussian stochastic process, the power spectral density is given by 
\begin{align}
S_n(f) &= 2 \int^\infty_{-\infty} \mathrm{d}\tau \; R(\tau) \, \mathrm{e}^{\mathrm{i} 2 \pi f \tau},
\intertext{where}
R(\tau) &= \langle {n}(t + \tau)\; {n}(t) \rangle
\end{align}
is the autocorrelation of the noise ${n}(t)$ \cite{Li:2013lza}. In practice, we use the publicly available power spectral density data for ET-D\footnote{\url{http://www.et-gw.eu/index.php/etsensitivities}.}.

With the waveform and power spectral density in hand, the signal-to-noise ratio for an event seen in a single interferometer can be computed. The total signal-to-noise ratio for an event $i$ seen by the three-armed ET-D observatory is obtained by summing the squares of the individual interferometer signal-to-noise ratios, 
\begin{align}
\bar{\rho}_i &= \sqrt{w \left(\rho^2_{\text{$i$ opt, \nth{1} arm}} + \rho^2_{\text{$i$ opt, \nth{2} arm}} + \rho^2_{\text{$i$ opt, \nth{3} arm}} \right)},
\intertext{where the weighting factor $w$ is given by}
w &= F_+^2(\theta, \phi, \psi) (1+ \cos^2\iota)^2 + 4F_\times^2(\theta, \phi, \psi) \cos^2\iota,
\end{align} 
where $F_+, F_\times$ are the antenna patterns and $\iota$ is the inclination.
Lastly, following \cite{Li:2013lza}, 
we approximate the instrumental uncertainty on the lensed luminosity distance $\tilde{D}_i$ for a given event $i$ as
\begin{equation}
\sigma^{\rm inst}_{i} \approx \frac{2 \tilde{D}_{i}}{\rho_i} \ ,
\qquad
\rho_i = \sqrt{\mu_i} \, \bar{\rho}_i \ .
\label{eq:inst_error}
\end{equation}
Here, $\bar{\rho}_i$ is the unlensed signal-to-noise ratio for the event determined by the observatory in question and $\mu_i$ is the magnification of the signal-to-noise ratio due to weak lensing. 
The factor $2$ in \cref{eq:inst_error} accounts for the contribution of the uncertainty on the inclination $\iota$ to the instrumental noise \cite{Li:2013lza}. Once the rescaling of the unlensed signal-to-noise ratio by the lensing magnification has been carried out, it is then trivial to compute the instrumental uncertainty for the event from \cref{eq:inst_error}. 

However, we note that a precise modelling of the uncertainty on the distance measurement would require access to the full parameter estimation pipeline of the experiment. Furthermore, the uncertainty actually depends on the full network of observatories that may be operating in collaboration with the ET, see \cite{Vitale:2016icu}. Extending the uncertainty computation in this way is beyond the scope of this work.

Furthermore, our analysis also rests on the assumption that the expression for the instrumental uncertainty given in \cref{eq:inst_error} is correct. The signal-to-noise ratio in this expression depends on the noise being stationary and Gaussian. In a real detector this is not exactly the case. Depending on the specific template used in the matched filtering process, so-called glitches -- artifacts with very high signal-to-noise ratios -- can be seen in the detector \cite{Nuttall2018}. The identification and removal of glitches and other noise artifacts through proper characterisation of the detectors is an important part of  current gravitational wave analysis pipelines \cite{LIGO:2021ppb}. However, since we are working solely with mock data, we generate our catalogues using the assumption of stationary, Gaussian noise. This is effectively equivalent to using a real catalogue of events with glitches removed. 

\begin{figure}[ht!]
	\centering
	\includegraphics[height=5.6cm]{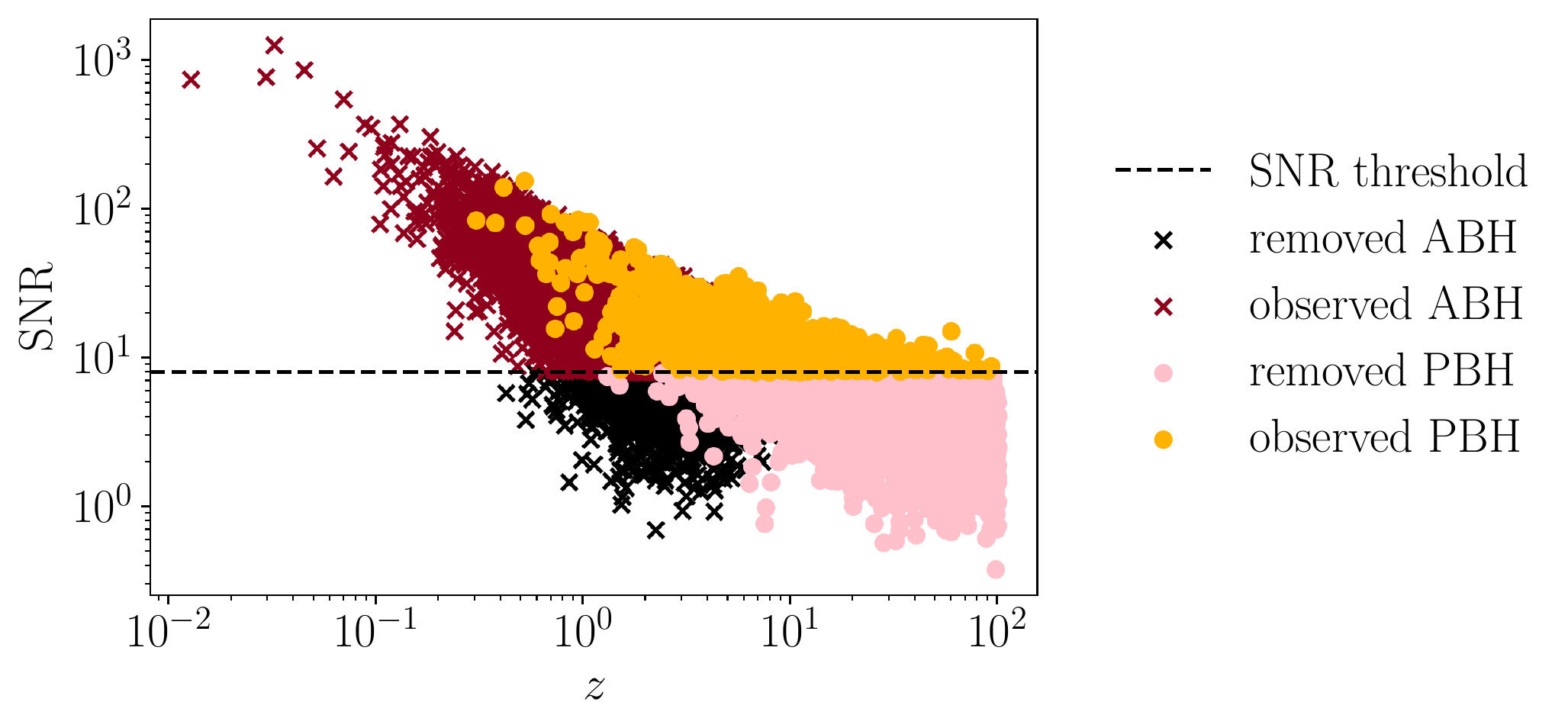}
	\caption{Trend of the signal-to-noise ratio with event redshift for ABHs (crosses) and PBHs (dots). Events that are above the signal-to-noise ratio threshold (black dashed line) are considered to be properly observed GW events (red for ABHs and yellow for PBHs), while the black and pink points are respectively the ABHs and PBHs with signal-to-noise ratios that make them too faint to be observed.}\label{fig:SNRplot}
\end{figure}

Finally, as mentioned in \cref{sec:generating_mock}, after computing the \textit{lensed} signal-to-noise ratio for all the events, we remove all the events with a lensed signal-to-noise ratio of less than eight from the catalogue. This serves to exclude events which may or may not be true detections of GWs. The signal-to-noise ratio threshold that we use follows that of the LIGO collaboration  \cite{Abbott:2016xvh}. In \cref{fig:SNRplot} we show the computed signal-to-noise ratio as a function of redshift for ABH and PBH events, highlighting the significant amount of events which are discarded due to the cut in the signal-to-noise ratio.

\section{Lensing}
\label{sec:appendix_lensing}

This appendix provides details about our modelling of the weak lensing of GW signals. \Cref{subsec:magnification_GWs} is a short theoretical reminder where we state our assumptions and define the relevant quantities to be used. \Cref{subsubsec:variance_magnification} indicates how the variance of the magnification is estimated. \Cref{subsubsec:lognormal_PDF_magnification} deals with our model for the magnification PDF in the mocks.

\subsection{Magnification of gravitational waves}
\label{subsec:magnification_GWs}

Consider a GW modelled as a small perturbation~$h_{\mu\nu}$ of an arbitrary background geometry~$g_{\mu\nu}$. If the wave is propagating in vacuum, in the sense that we can neglect its direct interaction with matter, then its equation of motion reads~\cite{Straumann:2013spu}
\begin{equation}
\label{eq:EoM_GW}
\nabla_\rho \nabla^\rho h_{\mu\nu} = 0 \ ,
\end{equation}
in the linear regime, harmonic transverse-traceless gauge, and assuming that the GW's wavelength is much shorter than the typical curvature radii of the background spacetime geometry, whose covariant derivative is denoted with $\nabla_\rho$ in \cref{eq:EoM_GW}.

Introducing the wave ansatz $h_{\mu\nu}=H_{\mu\nu}\exp(\rmi w)$, where $H_{\mu\nu}$ and $w$ respectively denote the amplitude and the phase of the GW, \cref{eq:EoM_GW} implies (i) that the wave follows null geodesics of the background spacetime; and (ii) that its amplitude satisfies
\begin{equation}
(1+z)^{-1} D H_{\mu\nu} = \text{const.} ,
\end{equation}
where $z$ is the source's redshift and $D$ is the electromagnetic luminosity distance -- see, e.g., \cite{Dalang:2019rke} for further details. It follows that $H_{\mu\nu}\propto D^{-1}$, that is, the energy of a GW ``dilutes'' just like the energy of an electromagnetic wave as it propagates. Since this is valid in any background spacetime, we conclude that GWs experience the same gravitational lensing effects as light.

In the following, we only consider \emph{weak-lensing} effects, in the sense that we neglect the possibility that a GW source might be multiply imaged. We also assume that the luminosity distance can be computed from the propagation of an infinitesimal beam of null geodesics. In that framework, the distortions of the beam with respect to the homogeneous-isotropic FLRW case are customarily encoded in the so-called distortion matrix. If $\vect{\theta}$ is the observed incoming direction of GWs and $\vect{\beta}$ the direction in which they would be observed in FLRW, then the distortion matrix is defined as the Jacobian matrix of $\vect{\theta}\mapsto\vect{\beta}(\vect{\theta})$,
\begin{equation}
\vect{\mathcal{A}}
\equiv
\frac{\diff\vect{\beta}}{\diff\vect{\theta}}
=
\begin{bmatrix}
1 - \kappa - \gamma_1 & -\gamma_2 + \omega\\
-\gamma_2 - \omega & 1 - \kappa + \gamma_1
\end{bmatrix} ,
\end{equation}
where $\kappa$ is called the convergence of the beam, $\gamma=\gamma_1+\rmi\gamma_2$ its shear distortion, and $\omega=\mathcal{O}(\gamma^2)$ its rotation.

The correction to the observed luminosity distance, $D$, relative to the FLRW case, $\bar{D}$, is quantified by the \emph{magnification}~$\mu$, which is related to the distortion matrix as follows,
\begin{equation}
\label{eq:magnification}
\mu^{-1}
\equiv
\left(
\frac{D}{\bar{D}}
\right)^2
=
\frac{\diff^2\vect{\beta}}
{\diff^2\vect{\theta}}
=
\det\vect{\mathcal{A}}
=
(1-\kappa)^2 - |\gamma|^2 + \omega^2 .
\end{equation}
In particular, at lowest order in $\kappa, \gamma, \omega$, the magnification only depends on the convergence, $\mu \approx 1 + 2\kappa$. We shall use this approximation for the estimate of the variance of the magnification -- see next subsection. It follows from \cref{eq:magnification} that the absolute uncertainty on the luminosity distance due to lensing may be estimated as
\begin{equation}
\sigma_{\rm lens}
= \frac{1}{2} \, \bar{D} \, \sigma_\mu \ ,
\end{equation}
where $\sigma_\mu$ denotes the standard deviation of the magnification.

\subsection{Variance of the magnification in weak lensing}
\label{subsubsec:variance_magnification}

We aim to model the PDF of the magnification of the observed GW signals in a realistic inhomogeneous universe. As a first step, we shall estimate the variance of that distribution. Previous analyses, both in the context of supernova cosmology~\cite{2014A&A...568A..22B} or GW cosmology~\cite{Sathyaprakash:2009xt, Zhao:2010sz, Cai2016, Du:2018tia, Jin:2020hmc, Hogg:2020ktc}, considered a dispersion of magnitude due to that grows linearly with redshift~\cite{2010MNRAS.405..535J}, $\Delta m = 0.055\,z$, which corresponds to a magnification dispersion of $\sigma_\mu=4\ln(10)\Delta m/5=0.10\,z$. While this linear approximation may be valid at low redshift, we expect it to fail at the high redshifts ($z\sim 10$ to $100$) considered here. The intuition is that at high redshift the Universe is increasingly homogeneous, thereby reducing the growth of the lensing dispersion.

In order to get a more accurate estimate of $\sigma_\mu$ across a wide range of redshifts, we shall use perturbation theory at second order. In that framework, we first notice that
\begin{equation}
\sigma_\mu^2 
\equiv
\langle \mu^2 \rangle - \langle \mu \rangle^2
=
4\sigma_\kappa^2
+ \mathcal{O}(\kappa^4) \ ,
\end{equation}
where $\sigma_\kappa^2$ is the variance of the convergence, because $\langle\kappa\rangle = 0$ at linear order and $\omega\sim\kappa^2\sim\gamma^2$. Hence, we may estimate $\sigma_\mu$ at second order from the linear-order results on $\sigma_\kappa$.

In the flat-sky and Limber approximations, the variance of the convergence for a source at redshift $z$ is related to the convergence power spectrum~$P_\kappa(\ell, z)$ as
\begin{equation}
\sigma_\kappa^2(z)
= \int_0^\infty \frac{\ell\diff\ell}{2\pi} \; 
                P_\kappa(\ell, z) \ ,
\end{equation}
which is itself related to the power spectrum~$P_W$ of the Weyl potential $W=(\Phi+\Psi)/2$, $\Phi$ and $\Psi$ being the Bardeen potentials, via
\begin{equation}
\label{eq:P_kappa}
P_\kappa(\ell, z)
= \ell^2(\ell+1)^2 \int_0^r
    \diff r'
    \left(
        \frac{r - r'}{r}
    \right)^2
    \left(
        \frac{1}{\ell+1/2}
    \right)^4
    P_W \left(\eta_0-r', \frac{\ell+1/2}{r'} \right) ,                          
\end{equation}
where $\eta_0$ denotes today's conformal time, and $r=r(z)$ is the comoving distance to the source.

\begin{figure}
\centering
\includegraphics[width=0.49\columnwidth]{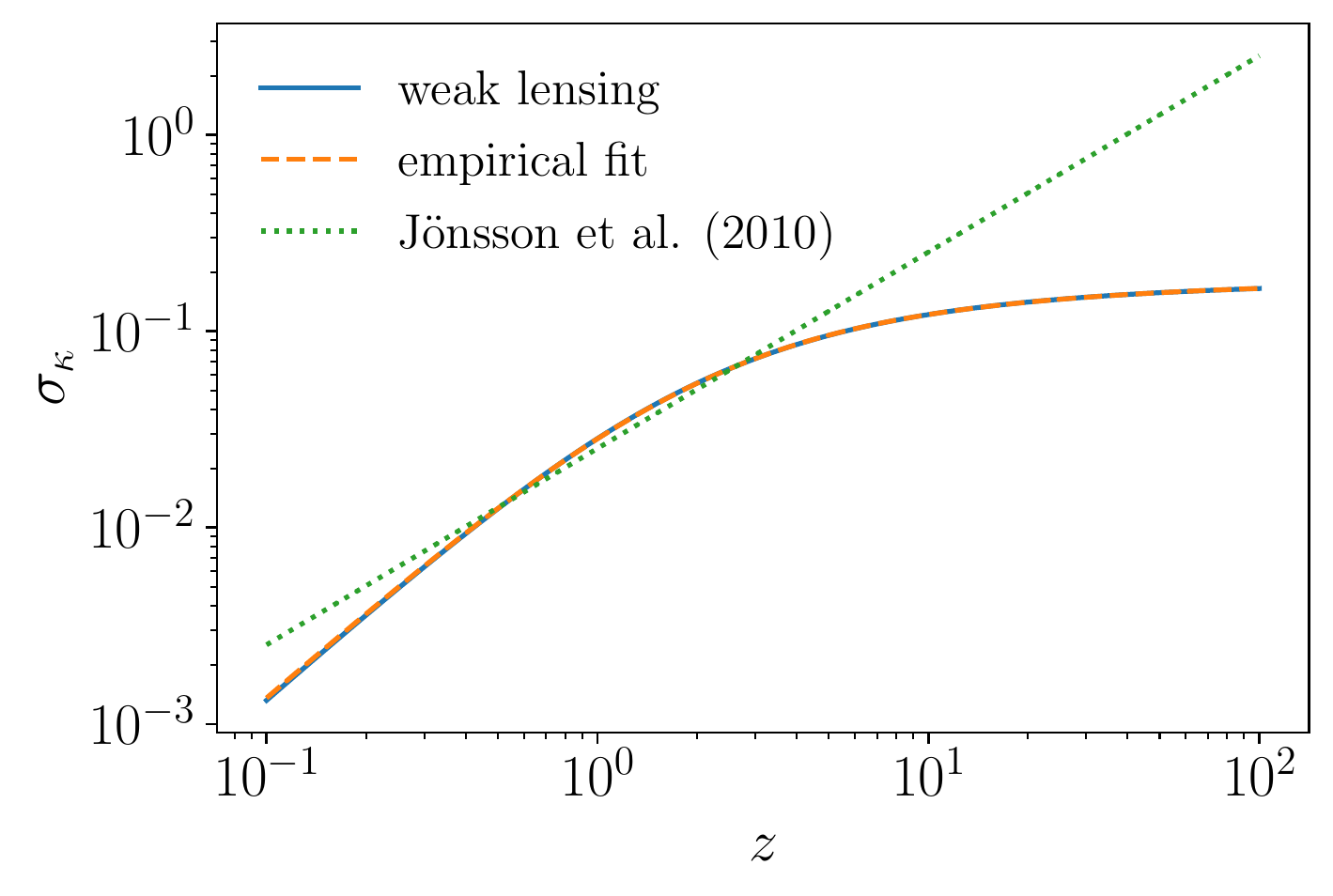}
\hfill
\includegraphics[width=0.49\columnwidth]{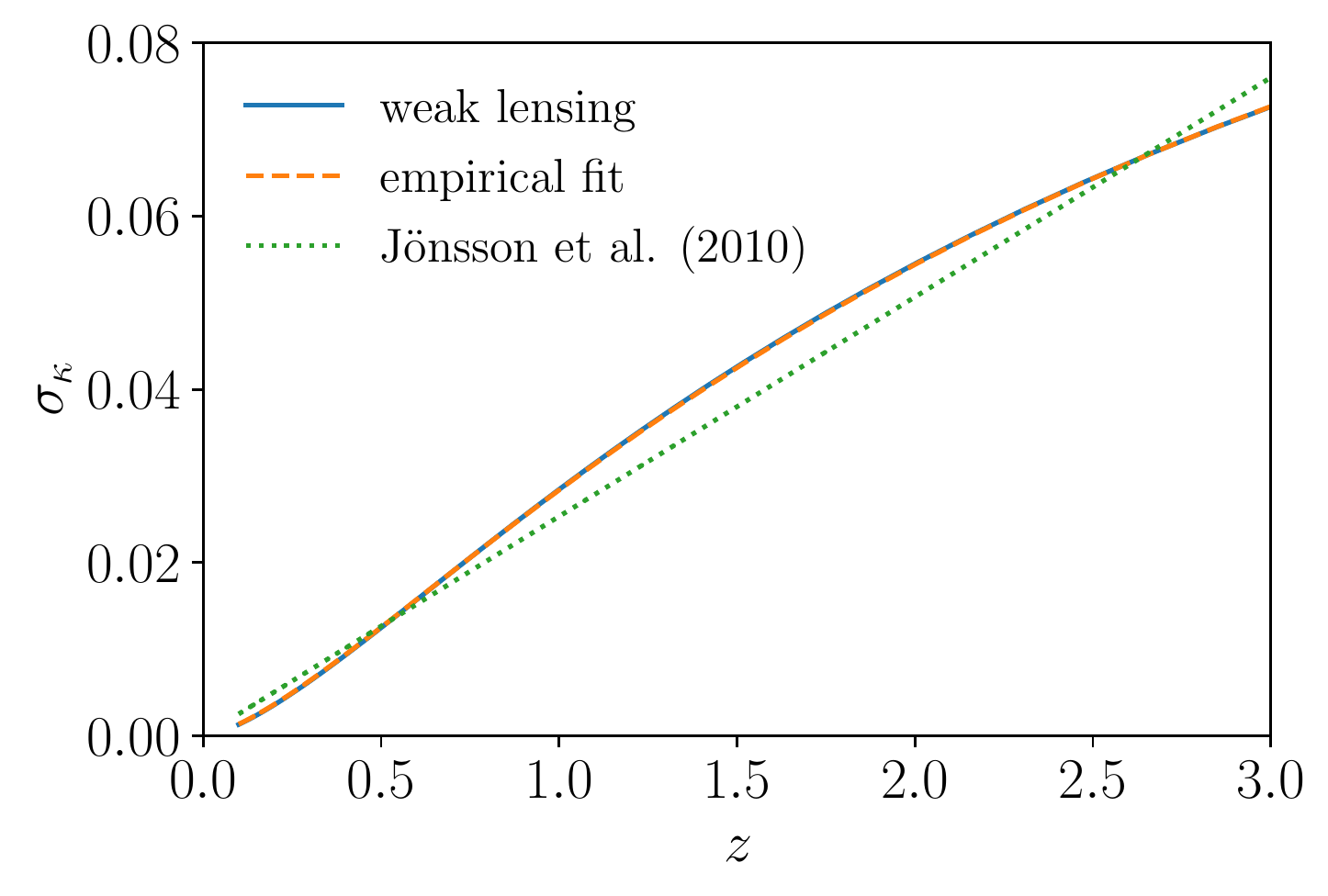}
\caption{Standard deviation of the convergence~$\sigma_\kappa$ as a function of the redshift~$z$ of the source. Blue solid lines indicate the numerical results from \CAMB; dashed orange lines indicate the empirical fitting function~\eqref{eq:empirical_fit_sigma_kappa}; green dotted lines show the linear prescription of \cite{2010MNRAS.405..535J}. The left panel shows the entire redshift range $z\in[0, 100]$, while the right panel focuses on $z<3$.}
\label{fig:sigma_kappa}
\end{figure}

We compute $P_W$ and the associated quantities using \CAMB. For a \LCDM cosmology with $H_0=\SI{67.4}{\kilo\meter\per\second\per\mega\parsec}, \Omega_{\rm m} = 0.315, \Omega_{\rm b} = 0.05,  A_{\rm s}=2\times 10^{-9}, n_{\rm s}=0.965$, we find that the standard deviation of the magnification as a function of redshift is very well fit (with sub-percent accuracy) by
\begin{equation}
\label{eq:empirical_fit_sigma_kappa}
\sigma_\kappa(z)
= a \arctan\left[\left(1 + b \, z^c\right)^d - 1\right]
\end{equation}
with $a=0.116, b=1.26, c=1.46, d=0.268$.\footnote{These numerical values depend on the cosmology; our code includes a notebook \href{https://gitlab.com/matmartinelli/darksirens/-/blob/master/notebooks/variance_convergence.ipynb}{\faGitlab} capable of generating such a fitting function for other values of the cosmological parameters.} The result is depicted in \cref{fig:sigma_kappa}, where we also indicate the linear ansatz of \cite{2010MNRAS.405..535J} for comparison. The latter is a reasonably good approximation of $\sigma_\kappa(z)$ up to $z\approx 3$, but it highly overestimates it at high $z$. At $z=100$, we find $\sigma_\kappa \approx 16\%$, which is more than an order of magnitude below the prediction of the linear ansatz. This emphasises the importance of a careful modelling of lensing at high redshift.

\subsection{Lognormal probability distribution of the magnification}
\label{subsubsec:lognormal_PDF_magnification}

\begin{figure}
\centering
\includegraphics[width=0.49\columnwidth]{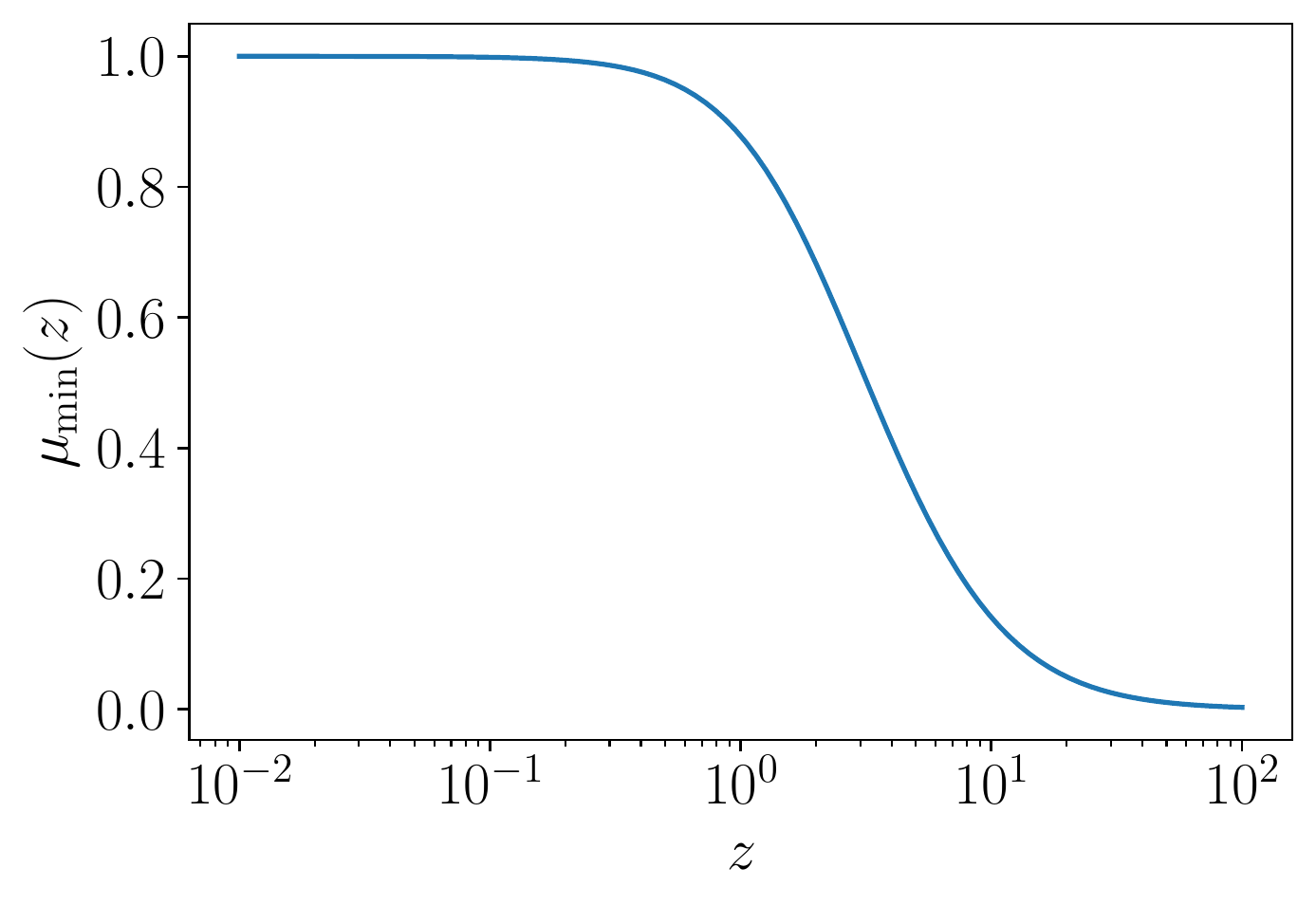}
\hfill
\includegraphics[width=0.49\columnwidth]{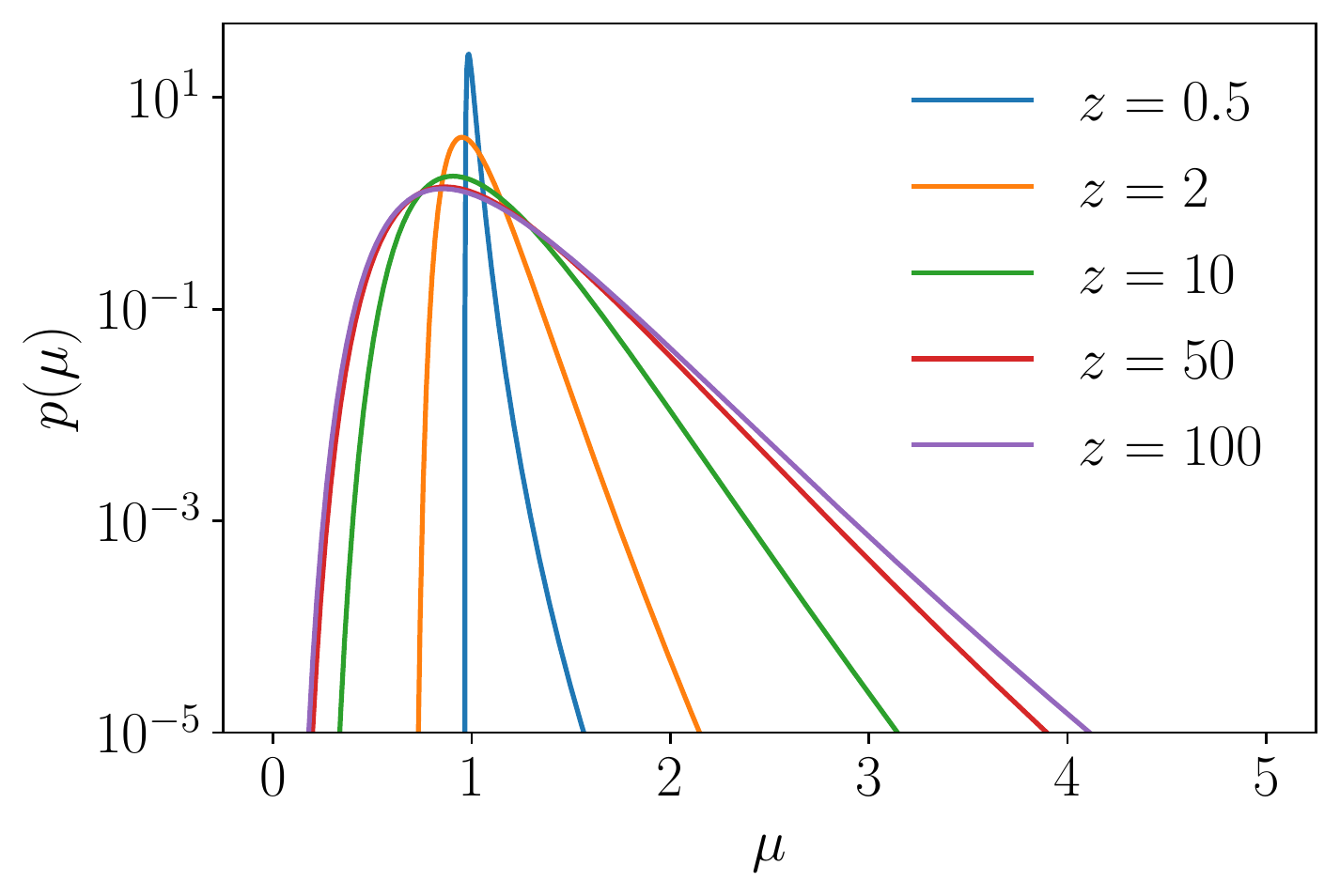}
\caption{\textit{Left}: minimum magnification corresponding to Zel'dovich's empty-beam case, as defined in \cref{eq:empty_beam}. \textit{Right}: PDF of the magnification determined by the shifted lognormal model~\eqref{eq:model_PDF_mu} for different values of the source redshift~$z$.}
\label{fig:mu_min_PDF_mu}
\end{figure}

We now turn to the full PDF~$p(\mu)$ of the magnification. This distribution must satisfy three physical requirements:
\begin{enumerate}
\item It must vanish for $\mu<\mu_{\rm min}$, where $\mu_{\rm min}$ corresponds to the (de)magnification of Zel'dovich's empty beam~\cite{1964SvA.....8...13Z}. This is due to the fact that a bundle of null geodesics cannot be less focused than a beam propagating through pure vacuum. In that case, the angular diameter distance coincides with the affine parameter~$\lambda$ along the bundle; the minimum magnification thus reads
\begin{equation}
\label{eq:empty_beam}
\mu_{\rm min}(z) = \left[ \frac{\bar{D}(z)}{(1+z)^2\lambda(z)}\right]^2 < 1 \ ,
\qquad
\lambda(z) = \int_0^z \frac{\diff\zeta}{(1+\zeta)^2 H(\zeta)} \ .
\end{equation}
The evolution of $\mu_{\rm min}$ with redshift $z$ is depicted in the left panel of \cref{fig:mu_min_PDF_mu}.
\item The magnification averaged over sources must be unity~\cite{2008MNRAS.386..230W}
\begin{equation}
\label{eq:magnification_theorem}
\langle \mu \rangle = \int_{\mu_{\rm min}}^\infty \diff \mu \; \mu \, p(\mu) = 1 \ .
\end{equation}
This magnification theorem assumes that sources are homogeneously distributed in space. Note that the underlying averaging procedure -- over sources rather than over random directions in the sky -- is essential to this result. See \cite{2021A&A...655A..54B} and references therein for detailed discussions.
\item The variance of the magnification should coincide with the one evaluated in \cref{subsubsec:variance_magnification},
\begin{equation}
\label{eq:magnification_variance_requirement}
\langle \mu^2 \rangle - 1
= \int_{\mu_{\rm min}}^\infty \diff \mu \; \mu^2 \, p(\mu) - 1
= 4\sigma_\kappa^2 \ .
\end{equation}
\end{enumerate}

There are, of course, many models that would satisfy the above three requirements. We adopt, for simplicity, the following shifted lognormal model,
\begin{equation}
\label{eq:model_PDF_mu}
p(\mu) =
\frac{1}{\sqrt{2\pi} \sigma (\mu - \mu_{\rm min})} \,
\exp\left\{- \frac{[\ln(\mu - \mu_{\rm min}) - m]^2}{2\sigma^2} \right\} ,
\end{equation}
where $\mu_{\rm min}$ is given by \cref{eq:empty_beam}, and the two free parameters $m, \sigma$ are fixed by the two conditions~\eqref{eq:magnification_theorem} and \eqref{eq:magnification_variance_requirement},
\begin{align}
\sigma &= \sqrt{\ln\left[1 + \frac{4\sigma_\kappa^2}{(1-\mu_{\rm min})^2}\right]} \ , \\
m &= \ln(1 - \mu_{\rm min}) - \frac{\sigma^2}{2} \ .
\end{align}

The resulting magnification PDF is depicted in the right panel of \cref{fig:mu_min_PDF_mu} for various values of the source redshift. Despite its simplicity, the shifted lognormal model mimics important properties of the expected magnification distribution in the inhomogeneous Universe. On the one hand, $p(\mu)$ peaks at $\mu<1$, thereby encoding the fact that most lines of sight have a lower column density than average, because voids occupy more volume in the Universe. On the other hand, $p(\mu)$ exhibits a longer tail towards high magnifications, which encodes that, albeit rare, overdensities can produce large magnifications. Note however that our model tends to underestimate the probability of those high magnifications compared to what is obtained with ray tracing in $N$-body simulations.

\clearpage

\bibliographystyle{JHEP}
\bibliography{PBHpaper}

\end{document}